\documentclass{article}     

\usepackage{amsmath, amsfonts,amsthm,amssymb}
\usepackage{dsfont}

\usepackage[english]{babel}

\usepackage[left=1.75cm, top=2.7cm, bottom=2.50cm,right=1.75cm]{geometry}

\usepackage{verbatim}
\usepackage{mathrsfs}
\usepackage{bm}
\usepackage{color}
\usepackage{graphicx}

\usepackage{url}

\usepackage{bm}

\usepackage{colortbl}

 \newtheorem{thm}{Theorem}[section]
 
 \newtheorem{lem}[thm]{Lemma}

 \theoremstyle{definition}

 \newtheorem{rem}[thm]{Remark}
 \numberwithin{equation}{section}
\newcommand{\be}{\begin{equation}}
\newcommand{\ee}{\end{equation}}
\newcommand{\bq}{\begin{eqnarray}}
\newcommand{\eq}{\end{eqnarray}}

\newcommand{\half}{\frac{1}{2}}

    \def\ld{\;{\stackrel{\frak {D}}{\longrightarrow}}\;}

    \def\ld{{\stackrel{\frak {D}}{\longrightarrow}}}

    \def\bbl{{\mathbb L}}

    \def\calF{{\mathcal F}}

    \def\calR{{\mathcal R}}

    \def\bbr{{\mathbb R}}
    \def\bbe{{\mathbb E}}
    \def\bbp{{\mathbb P}}
    \def\bbc{{\mathbb C}}
    \def\bbn{{\mathbb N}}

    \def\ld{\;{\stackrel{\frak {D}}{\longrightarrow}}\;}

    \def\a{{a}}
    \def\b{{b}}

    \def\q{{q}}
    
    \def\s{{s}}
    \def\t{{t}}

    \def\B{{B}}
    \def\C{{C}}
    \def\D{{D}}

    \def\M{{M}}
    \def\N{{N}}

    \def\R{{R}}
    \def\S{{S}}
    
    \def\U{{U}}
    \def\V{{V}}
    \def\W{{W}}
    \def\X{{X}}
    \def\Y{{Y}}
    \def\Z{{Z}}

  \definecolor{Red}{rgb}{0,0,0}
    
    \definecolor{DRed}{rgb}{0.7, 0.3, 0.00}
    
    \definecolor{Green}{rgb}{0.2, 0.5, 0.2}
    
    \definecolor{Blue}{rgb}{0.00, 0.00, 1.00}
    
    \definecolor{PaleGrey}{rgb}{.6, .6, .6}

\title{{Short-time expansions for close-to-the-money options under a L\'evy jump model with stochastic volatility}}
\author{{Jos\'e E. Figueroa-L\'opez\thanks{Department of Statistics, Purdue University, West Lafayette, IN 47907, USA ({\tt figueroa@purdue.edu}). Research supported in part by the NSF Grants: DMS-1149692.}\,\, {and} Sveinn \'Olafsson\thanks{Department of Statistics, Purdue University, West Lafayette, IN 47907, USA ({\tt sveinno@purdue.edu}).}}}
\date{\today}

\begin{document}
\maketitle
\begin{abstract}
In {Figueroa-L\'opez et al.~(2013) [High-order short-time expansions for ATM option prices of exponential L\'{e}vy models]}, a second order approximation for at-the-money (ATM) option prices is derived for a large class of exponential L\'evy models, with or without a Brownian component. The purpose of this article is twofold. First, we relax the regularity conditions imposed in {Figueroa-L\'opez et al.~(2013)} on the L\'evy density to the weakest possible conditions for such an expansion to be well defined. Second, we show that the formulas extend both to the case of ``close-to-the-money" strikes and to the case where the continuous Brownian component is replaced by an independent stochastic volatility process with leverage.

\vspace{0.2 cm}

\vspace{0.2 cm}

\vspace{0.2 cm}

\end{abstract}

\section{Introduction}
In recent years, a great deal of effort has been put into the study of the asymptotic behavior of option prices and implied volatility in a variety of asymptotic regimes. For a recent review of the topic, the reader is referred to \cite{Andersen}. Despite the attention that the problem has received in the literature, there are still important open problems, such as the lack of accurate (i.e., high-order) asymptotics for an ample class of models, general enough to incorporate several stylized features of asset prices, and the determination of a suitable asymptotic regime when including ``close-to-the-money" options near expiration. These two key issues are addressed in the present work. 

In the presence of jump risk, there were, until recently, no available high-order asymptotics for at-the-money (ATM) options near expiration, let alone for close-to-the-money options. In \cite{LopGonHou:2013}, a second order approximation for ATM option prices for a certain class of exponential L\'evy models is derived. More specifically, the asset price at time $t$ is given by $S_t:=S_0e^{X_t}$, where $X:=(X_t)_{t\geq 0}$ is a L\'evy process with L\'evy density $s:\bbr\backslash\{0\}\to [0,\infty)$ of the form
\begin{align}\label{DstyFrm0}
	s(x)= |x|^{-Y-1}{C\left(\frac{x}{|x|}\right)\bar{q}(x)},
\end{align}
{for $Y\in(1,2)$, 
constants $C(1),C(-1)\in(0,\infty)$, 
and a bounded measurable function $\bar{q}:\bbr\backslash\{0\}\to [0,\infty)$ such that}
\begin{equation}\label{FirstCdn0}
	{\lim_{x\to{}0} \bar{q}(x)=1.}
\end{equation}
Conditions (\ref{DstyFrm0}) and (\ref{FirstCdn0}) entail 
that the ``small" jumps of the process $X$ behave like those of a $Y$-stable process, while the function $\bar{q}$ allows for a tempering of large jumps in order for $X$ to have, say, finite exponential moments, which is needed for the price process to be a martingale. A {pure-jump L\'evy process having a L\'evy density $s$ satisfying (\ref{DstyFrm0})-(\ref{FirstCdn0})} will hereafter be called a \emph{tempered stable-like process}. {It is worth pointing out that most of the standard L\'evy models used in finance admit L\'evy densities of the form (\ref{DstyFrm0}). These include the CGMY model and the normal tempered stable processes as defined in \cite{CT04}, the Meixner processes (\cite{Schoutens}), and the generalized hyperbolic class (\cite{Schoutens}). Note, however, that in the later two cases $Y=1$. The restriction of $Y\in(1,2)$ is actually motivated by recent econometric studies based on high-frequency financial data (see Remark 2.2 in \cite{LopGonHou:2013} and references therein).}

{Additionally to (\ref{DstyFrm0})-(\ref{FirstCdn0})}, in \cite{LopGonHou:2013}, $\bar{q}$ was also assumed to satisfy the following {rather technical} conditions for some constants $M,G>0$:
\begin{align}\label{Eq:StndCnd1a}
  &{\rm (i)}\; {1-\bar{q}(x)\sim Mx},\;x\searrow{}0;
	&{\rm (ii)}&\; {1-\bar{q}(x)\sim -G x},\;x\nearrow{}0;\nonumber\\
	&{\rm (iii)}\; \bar\q(x)\leq e^{-x},\;\text{ for {all} }\;x>0;
	&{\rm (iv)}&\; \bar\q(x)\leq 1,\;\text{ for {all} }\;x<0;\\
	&{\rm (v)}\;\limsup_{|x|\to\infty} \frac{|\ln \bar\q(x)|}{|x|}<\infty;
	&{\rm (vi)}&\;\inf_{|x|<\epsilon} \bar\q(x)>0,\;\text{ for any }\epsilon>0\nonumber.
\end{align}
A natural and important question is how necessary the regularity conditions (\ref{Eq:StndCnd1a}) are for the validity of the second-order short-term expansion in \cite{LopGonHou:2013}. In what follows, we show that they are mostly superfluous and all what is needed is the following integrability condition
\begin{align}\label{MinCond0}
	\int_{|x|\leq 1}|x|^{-Y}\left|1-\bar{q}(x)\right|dx<\infty, 
\end{align}
which, as will be shown below, is the minimal possible condition under which the second-order expansion is well defined. Let us briefly outline the strategy of our proof. First, using arguments similar to those in \cite{LopGonHou:2013}, we show the validity of the result for a tempered stable-like process $\widetilde{X}$, whose $\bar{q}$ function satisfies the following conditions:
\begin{align}\label{NewAssumEq}
  {\rm (i)}\; |1-\bar\q(x)| = O(|x|),\quad x\to{}0;\qquad
	{{\rm (ii)}}\;\limsup_{|x|\to\infty} \frac{|\ln \bar\q(x)|}{|x|}<\infty;\qquad
	{{\rm (iii)}}&\;\inf_{|x|<\epsilon} \bar\q(x)>0,\;\text{ for any }\epsilon>0.
\end{align}
Second,  using an approach similar to the one introduced in \cite{MuhNut:2009}, we show that the option prices corresponding to a process satisfying only (\ref{MinCond0}) can be closely approximated, up the second order, by those corresponding to a process satisfying all three conditions in (\ref{NewAssumEq}).

It is well known that exponential L\'evy models fail to capture accurately the time dynamics of volatility surfaces, and that they do not account for some stylized features of asset prices such as volatility clustering and the leverage effect. A natural remedy to that is to replace the (constant volatility) Brownian component of the L\'evy process with an independent stochastic volatility process of the form
\begin{align}
		dV_t & = \mu(Y_t)dt+\sigma(Y_t)\left(\rho dW_t^1+\sqrt{1-\rho^2}dW_t^2\right),\qquad V_0=0, \label{modelV0}\\
		dY_t & = \alpha(Y_t)dt+\gamma(Y_t)dW_t^1,\qquad Y_0=y_0,\label{modelY0}
\end{align}
where $(W_t^1)_{t\geq 0}$ and $(W_t^2)_{t\geq 0}$ are independent standard Brownian motions. This framework includes the most common stochastic volatility models, such as the mean reverting Heston and Ornstein-Uhlenbeck processes. We will then consider the asset price process $S_t:=S_0e^{X_t+V_t}$, where $X$ is a pure-jump tempered stable-like process as described above, and show that under some mild conditions on the drift and volatility parameters in (\ref{modelV0})-(\ref{modelY0}), the second order expansion in \cite{LopGonHou:2013} is still valid, but with the volatility of the Brownian component, $\sigma$, replaced by the spot volatility $\sigma(y_0)$. The steps in doing so are in spirit similar to the ones used in the pure-jump case. First, by conditioning on the stochastic factor $Y$ and utilizing properties of the Gaussian distribution, a strategy similar to the one in \cite{LopGonHou:2013} can be employed to show that the expansion is valid when $\sigma(\cdot)$ is assumed to be bounded away from $0$ and $\infty$. Then, an approach similar to the one introduced in \cite{MuhNut:2009} can be used to show that the second order expansion extends to the case of potentially unbounded $\sigma(\cdot)$.

The {{aforementioned}} asymptotic expansions are concerned with ATM option prices, i.e. $\bbe\left(S_t-S_0e^{\kappa}\right)^{+}$ with $\kappa=0$. As mentioned above, a problem of practical importance is then raised by the fact that, as maturity approaches $0$, the most liquid options have strike prices that are close to being ATM, i.e. $\kappa\approx 0$ (cf. \cite{TankMij}). However, in the presence of jumps, the implied volatility explodes for out-of-the-money options ($\kappa\neq 0$) as maturity decreases (cf. \cite{FigForde:2012,Tankov:2010}), while for ATM options it converges to a finite value (cf. \cite{RoperRutkowski:2007,Tankov:2010}). It is therefore of interest to see whether the at-the-money expansions can be extended to include options whose strike prices are ``close-to-the-money". To formalize that idea, we follow the lines of \cite{TankMij} and consider option prices of the form $\bbe\left(S_t-S_0e^{\kappa_t}\right)$, where the log-strike $\kappa_t$ is now a deterministic function such that $\kappa_t\to 0$, as $t\to 0$. In other words, the strike is allowed to be out-of-the-money for any $t>0$, while converging to the at-the-money strike as $t\to 0$. It turns out that both with and without a continuous component there exists a small maturity log-moneyness regime, depending on the order of the second order term, where the ATM asymptotic expansion can indeed be used to include close-to-the-money options. 

The rest of the article is organized as follows. First, the underlying asset price model and some useful {notations are} introduced in Section \ref{Sec:tmpstble}. Next, Sections \ref{PureJump} and \ref{StochVol} present the close-to-the-money asymptotic expansions in the pure-jump case and the case including a continuous volatility component, respectively. Finally, a numerical analysis is carried out in Section \ref{Sec:Examples}. The proofs of lemmas and other technical details are deferred to Appendices \ref{ApProf1}-\ref{ApTecLem}.

\section{The model and some relevant notation}\label{Sec:tmpstble}

Throughout, $X:=(X_t)_{t\geq{}0}$ denotes a pure-jump L\'evy process defined on a filtered probability space $(\Omega,\calF,(\calF_{t})_{t\geq{}0},\bbp)$ satisfying the usual conditions, with triplet $(0,b,\nu)$ such that: 
\begin{align}\label{NdCndTSMrt}
{{\rm (i)}\;\; \int_{1}^{\infty}e^{x}\nu(dx)
	<\infty,\quad\text{and}\quad {\rm (ii)}\;\; \bbe\left(e^{X_{1}}\right)=\exp\left(b+\int_{\bbr_{0}}\left(e^{x}-1-x{\bf 1}_{\{|x|\leq{}1\}}\right)\nu(dx)\right)=1},
\end{align}
where $\bbr_{0}:=\bbr\backslash\{0\}$, and the L\'evy triplet is given relative to the truncation function ${\bf 1}_{\{|x|\leq{}1\}}$ (see {Section 8} in \cite{Sato:1999}). Note that we are implicitly assuming that the risk-free rate $r$ is zero, and that $\bbp$ is a martingale measure for the exponential L\'evy process $S_t:=S_0e^{X_t}$. Moreover, the L\'{e}vy measure $\nu$ is assumed to admit a density $s:\bbr_{0}\to [0,\infty)$ of the form
\begin{align}\label{DstyFrm}
	s(x)= |x|^{-Y-1}C\left(\frac{x}{|x|}\right)\bar{q}(x),
\end{align}
for $Y\in(1,2)$, constants $C(1),C(-1)\in[0,\infty)$ such that $C(1)+C(-1)>0$,
and a bounded measurable function $\bar{q}:\bbr_{0}\to [0,\infty)$ such that
\begin{equation}\label{FirstCdn}
	\lim_{x\to{}0} \bar{q}(x)=1.
\end{equation}
As explained in the introduction, we also wish to incorporate an independent stochastic volatility component into the model. To that end, we assume that $(\Omega,\calF,(\calF_{t})_{t\geq{}0},\bbp)$ also carries a process $(V,Y):=(V_t,Y_t)_{t\geq 0}$, independent of $X$, such that
\begin{align}
		dV_t & = -\frac{1}{2}\sigma^2(Y_t)dt+\sigma(Y_t)\left(\rho dW_t^1+\sqrt{1-\rho^2}dW_t^2\right),\qquad V_0=0, \label{modelV}\\
		dY_t & = \alpha(Y_t)dt+\gamma(Y_t)dW_t^1,\qquad Y_0=y_0,\label{modelY}
\end{align}
and consider the asset price process $S_t:=S_0e^{X_t+V_t}$. Here, $(W_t^1)_{t\geq 0}$ and $(W_t^2)_{t\geq 0}$ are assumed to be independent standard Brownian motions relative to the filtration $(\calF_{t})_{t\geq{}0}$, $\rho\in[-1,1]$, and $\alpha(\cdot)$, $\gamma(\cdot)$ and $\sigma(\cdot)$ are assumed to be such that $e^{V_t}$ is a well defined $\bbp$-martingale. In particular, $\alpha(\cdot)$ and $\gamma(\cdot)$ are such that (\ref{modelY}) admits a unique strong solution, while $\sigma(\cdot)$ is such that the integrals in (\ref{modelV}) are well defined.

As in \cite{LopGonHou:2013}, an important ingredient in our proofs consists of some suitable probability density transformations. In order to define these {transformations}, we further assume that the filtration $(\calF_{t})_{t\geq{}0}$ is such that $\calF=\vee_{t\geq{}0}\calF_{t}$. Then, using the martingale condition {$\bbe\left(e^{X_t}\right)=1$}, a probability measure $\mathbb{P}^{*}$ on $(\Omega,\mathcal{F})$ is defined via
\begin{equation}\label{DSM}
	 \frac{d {\bbp}^{*}|_{\calF_{t}}}{d \bbp |_{\calF_{t}}}=e^{X_{t}}, \qquad t\geq{}0.
\end{equation}
Under this probability measure, $X$ is a L\'evy process with triplet $(0,b^{*},\nu^{*})$ given by
\begin{align}\label{triplet1}
\nu^{*}(dx):=e^{x}\nu(dx)=e^{x}s(x)dx,\qquad b^{*}:=b+\int_{|x|\leq{}1}x\left(e^{x}-1\right)s(x) dx,
\end{align}
(see {\cite{Sato:1999}, Theorem 33.1}). Once the measure $\bbp^*$ has been defined, another locally equivalent measure, $\widetilde{\bbp}$, is constructed, under which the L\'evy triplet $(0,\tilde b,\tilde\nu)$ of $X$ takes the form
\begin{align}\label{triplet2}
	\tilde{\nu}(dx):= {|x|^{-Y-1} C\left(\frac{x}{|x|}\right)}dx=|x|^{-Y-1}\left({C(1)}{\bf 1}_{\{x>0\}}+{C(-1)}{\bf 1}_{\{x<0\}}\right)dx,\qquad
	\tilde{b}:=b^{*}+\int_{|x|\leq{}1} x(\tilde{\nu}-\nu^{*})(dx).
\end{align}
In  particular, $X$ is a $Y$-stable L\'evy process under $\widetilde{\bbp}$. Note that condition (\ref{Eq:StndCnd1a}-vi) ensures that $\tilde{\nu}$ is equivalent to $\nu^{*}$ and, thus, by virtue of Theorem 33.1 in \cite{Sato:1999}, the measure transformation {$\bbp^{*}\to\widetilde{\bbp}$} is well defined {provided that} the following condition is satisfied:
\begin{align}\label{ChangeMeasureCond}
	\int_{\bbr_{0}}\left(e^{\varphi(x)/2}-1\right)^{2}\nu^{*}(dx)=
    \int_{\bbr_{0}}\left(1-e^{-\varphi(x)/2}\right)^{2}\tilde{\nu}(dx)<\infty,
\end{align}
where
\begin{align*}
	\varphi(x):=\ln\left(\frac{d\tilde\nu}{d\nu^*}\right)=-\ln\bar{q}(x)-x.
\end{align*}
We finish this section with some useful notation. Let us first define the centered process 
\begin{equation}\label{SSSP}
Z_{t}:=X_{t}-t\tilde{\gamma},
\end{equation}
where $\widetilde{\gamma}:=\widetilde\bbe\left(X_{1}\right)$, which is necessarily a strictly $Y$-stable process under $\widetilde{\bbp}$. Second, denoting the jump measure of the process {$X$} by $N$ and its compensated measure under $\widetilde{\bbp}$ by $\bar{N}(dt,dx)$, the following representation for the log-density process can be obtained (see Theorem 33.2 in \cite{Sato:1999}):
\begin{equation}\label{EMMN}
	{U_{t}:=\log\frac{d \widetilde{\bbp}|_{\calF_{t}}}{d \bbp^{*}|_{\calF_{t}}}=\widetilde{U}_{t}+\eta t},
\end{equation}
{with}
\begin{align}\label{Uplusminuseta}
\widetilde{U}_{t}:=\int_{0}^{t}\int_{\bbr_{0}} \varphi(x)\bar{N}(ds,dx),\qquad\eta:=\int_{\bbr_{0}}\left(e^{-\varphi(x)}-1+\varphi(x)\right)\tilde{\nu}(dx),
\end{align}
{provided that}
\begin{equation}\label{IntCndEta}
	{\int_{\bbr_{0}}\left|e^{-\varphi(x)}-1+\varphi(x)\right|\tilde{\nu}(dx)<\infty.}
\end{equation}
At this point is it worth mentioning that conditions (\ref{ChangeMeasureCond}) and (\ref{IntCndEta}) will not be assumed to be satisfied in the sequel. As explained in the introduction, the idea is to approximate the ``close-to-the-money" option prices for models satisfying the weaker condition (\ref{MinCond0}), with the corresponding option prices for models satisfying a set of stronger conditions which, in particular, implies (\ref{ChangeMeasureCond}) and (\ref{IntCndEta}).

\section{Pure-jump L\'evy model}\label{PureJump}
Consider the pure-jump exponential L\'evy model introduced in Section \ref{Sec:tmpstble}. In this section, we show that the second-order expansion of \cite{LopGonHou:2013} is valid assuming only the weakest possible conditions under which it is well defined. The following theorem is the main result of this section.
\begin{thm}\label{2ndASY}
Consider the exponential L\'evy model $S_t:=S_0e^{X_t}$, where $X:=(X_t)_{t\geq{}0}$ is a pure-jump L\'evy process whose triplet $(0,b,\nu)$ satisfies (\ref{NdCndTSMrt})-(\ref{DstyFrm}) for a bounded measurable function {$\bar{q}:\bbr\backslash\{0\}\to [0,\infty)$} such that $\displaystyle\lim_{x\to{}0} \bar{q}(x)=1$ and 
\begin{align}\label{MinCond}
	\int_{|x|\leq 1}|x|^{-Y}\left|1-\bar{q}(x)\right|dx<\infty. 
\end{align}
In that case, if the log-moneyness, $\kappa_{t}$, of the corresponding call option is of the form $\kappa_{t}:=\theta t+o(t)$, as $t\to{}0$, for some $\theta\in\bbr$, then
the second order asymptotic expansion for the close-to-the-money option price is given by
\begin{equation}\label{CL2}		 
\frac{1}{S_{0}}\mathbb{E}(S_t-{S_{0}}e^{\kappa_t})^{+}= d_{1} t^{{\frac{1}{Y}}}+d_{2} t +o(t), \qquad t\to{}0,
\end{equation}
with $d_{1}:=\widetilde{\bbe}(Z_{1}^{+})$ and $d_{2}:={\tilde\vartheta}+(\tilde{\gamma}- \theta)\widetilde{\bbp}(Z_{1}\geq 0)$, where, under $\widetilde{\bbp}$, $Z_{1}$ is a strictly stable r.v. with L\'evy measure $\tilde{\nu}(dx):=C(x/|x|)|x|^{-Y-1}dx$, and
\begin{align}\label{vartheta}
	\tilde{\vartheta} &:=C(1)\int_{0}^{\infty}\left(e^{x}\bar{q}(x)-\bar{q}(x)-x\right)x^{-Y-1}dx,\\
	\tilde\gamma &:= \widetilde{\bbe}\left(X_1\right)= b+\frac{C(1)-C(-1)}{Y-1}+C(1)\int_0^1x^{-Y}\left(1-\bar\q(x)\right)dx-C(-1)\int_{-1}^0|x|^{-Y}\left(1-\bar\q(x)\right)dx.
	\label{Deftildegamma}
\end{align}
\end{thm}
\begin{rem}\label{Remark1}
Condition (\ref{MinCond}) is the minimal condition under which the expansion (\ref{CL2}) makes sense since, otherwise, the integrals in (\ref{Deftildegamma}) are not well-defined. 
Note also that the form of the log-moneyness $\kappa_t$ chosen in Theorem \ref{2ndASY} is, in some sense, the most relevant one. As shown therein, if $\kappa_t$ converges to $0$ at a rate faster than $t$ (that is, $\theta=0$ in the theorem), the second order asymptotic expansion coincides precisely with the one in the at-the-money case. On the contrary, if $\kappa_t$ converges slower to $0$, the second order term no longer incorporates information on the tempering function $\bar\q$, and its order is determined by $\kappa_t$. More precisely, if, for instance, $\kappa_{t}=\theta\t^{\beta}+o(t^{\beta})$, with $\frac{1}{Y}<\beta<1$ and $\theta\neq{}0$ ($\beta>\frac{1}{Y}$ ensures that $d_1 t^{1/Y}$ remains the leading order term of the expansion), then 
\begin{equation*}	 
\frac{1}{S_{0}}\mathbb{E}(S_t-{S_{0}}e^{\kappa_t})^{+}= d_{1} t^{{\frac{1}{Y}}}+d_{2} t^{\beta} +o(t), \qquad t\to{}0,
\end{equation*}
where $d_1$ is as in the theorem, and $d_2:=\theta\,\widetilde{\bbp}(Z_{1}\geq 0)$, where again $Z_{1}$ is still a strictly stable random variable with L\'evy measure $\tilde{\nu}(dx):=C(x/|x|)|x|^{-Y-1}dx$. At this point, it may {also} be relevant to refer to the work of \cite{TankMij}, where the close-to-the-money asymptotic regime $\kappa_{t}:=\theta\sqrt{t\ln\left(1/t\right)}$ is considered, leading to slower rates of convergence for the option prices.
\end{rem}

As explained in the introduction, the result in Theorem \ref{2ndASY} will be obtained through two steps, the first of which consists of relaxing the conditions given in Eq.~(\ref{Eq:StndCnd1a}) to those in Eq.~(\ref{NewAssumEq}). The following proposition gives this result, whose proof is based on an approach similar to that in \cite{LopGonHou:2013} and is presented in Appendix \ref{ApProf1}.

\begin{lem}\label{2ndASYStep1}
{Let $S_t:=S_0e^{X_t}$, where $X_t$ is a pure-jump L\'evy process whose triplet $(0,b,\nu)$ satisfies (\ref{NdCndTSMrt})-(\ref{DstyFrm}) for a bounded measurable function $\bar{q}:\bbr\backslash\{0\}\to [0,\infty)$ satisfying the conditions in (\ref{NewAssumEq}). Then, the second order asymptotic expansion (\ref{CL2}) holds true.}
\end{lem}

\medskip
{\noindent\textbf{Proof of Theorem \ref{2ndASY}.}}

\noindent
Throughout, we assume without loss of generality that $S_0=1$ 
and we let $\epsilon_{0}\in (0,1)$ be a constant such that $\inf_{|x|\leq\epsilon_{0}}\bar\q(x)>0$. The existence of $\epsilon_{0}$ is guaranteed since $\bar\q(x)\to 1$ as $x\to 0$. The idea is to approximate {the} option prices corresponding to $X$ with those corresponding to a L\'evy process ${X}^{(\delta)}$ satisfying the conditions in (\ref{NewAssumEq}).   Here, $\delta$ is a parameter whose value serves to control the distance between the two models' option prices. 
In order to define $X^{(\delta)}$, let us first look at the L\'evy-It\^o decomposition of the process $X$, with truncation function ${\bf 1}_{\{|x|\leq\delta\}}$, for each $\delta\in(0,\epsilon_0)$ (see \cite{Sato:1999}). More precisely, consider the decomposition\begin{align*}
X_t & = \hat X_t^{(\delta)} + \bar{X}_t^{(\delta)} + b^{(\delta)}t 
:= \int_0^t\int_{|x|>\delta}xN(ds,dx) + \lim_{\epsilon\downarrow 0}\int_0^t\int_{\epsilon<|x|\leq\delta}x\bar N(ds,dx) -t\int_{\bbr_0}(e^x-1-x{\bf 1}_{\{|x|\leq\delta\}})\nu(dx), 
\end{align*}
for any $t\geq{}0$. 
Next, on a suitable extension $(\Omega^{(\delta)},\calF^{(\delta)},\bbp^{(\delta)})$ of the original probability space $(\Omega,\calF,\bbp)$, we define {several} independent L\'evy processes, $\tilde\X^{(\delta,1)}$, $\tilde\X^{(\delta,2)}$, $\tilde\X^{(\delta,3)}$, and $R^{(\delta)}$, such that they are also independent of the original process $X$. {Concretely, consider the L\'evy measures
\begin{align*}
	\tilde\nu_1^{(\delta)}(dx) &:= C\left(x/|x|\right){\bf 1}_{\{\bar q(x)\geq 1,|x|\leq \delta\}}|x|^{-Y-1}dx,\\
	\tilde\nu_2^{(\delta)}(dx)&:=C\left(x/|x|\right)\left(\bar q(x)-1\right)^+{\bf 1}_{\{|x|\leq \delta\}}|x|^{-Y-1}dx,\\
	\tilde\nu_3^{(\delta)}(dx)&:=C\left(x/|x|\right){\bf 1}_{\{\bar q(x)< 1,|x|\leq{}\delta\}}\bar\q(x)|x|^{-Y-1}dx,\\
	 \nu_R^{(\delta)}(dx)  &= C\left(x/|x|\right)\left((1-\bar\q(x))^+{\bf 1}_{\{|x|\leq{}\delta\}}+e^{-|x|/\delta}{\bf 1}_{\{|x|\geq{}\epsilon_0\}}\right)|x|^{-Y-1}dx.
\end{align*}
Then, with respect to the truncation ${\bf 1}_{\{|x|\leq{}\delta\}}$, the L\'evy triplet of $\tilde{X}^{(\delta,i)}$ is {set to be} $(0,0,\tilde\nu_i^{(\delta)})$, for $i=1,2,3$, while the L\'evy triplet of $R^{(\delta)}$ is given by $(0,0,\nu_R^{(\delta)})$. Let us recall that,} by the definition of a probability space extension (see {Chapter 5} in \cite{Kallenberg}), the {law} of $X$ under $\bbp^{(\delta)}$ remains unchanged. In what follows, all expected values will be {taken} with respect to the extended probability measure $\bbp^{(\delta)}$, so for simplicity we denote the expectation under $\bbp^{(\delta)}$ by $\bbe$. Now, by adding the L\'evy measures of the {involved} processes, it is clear that the law of the process 
\[
	\breve{X}^{(\delta)}_{t}:=\hat X_t^{(\delta)} + \tilde X_t^{(\delta,1)} + \tilde X_t^{(\delta,2)} + \tilde X_t^{(\delta,3)} + b^{(\delta)}t,
\]
coincides with that of the process $X$ and, thus, the price {process} $\breve{S}^{(\delta)}:=\exp(\breve{X}^{(\delta)})$ is such that
\begin{equation}\label{EqTwoMd}
	\bbe\left(\breve{S}^{(\delta)}_t-e^{{\kappa_t}}\right)^{+}=\bbe\left(S_t-e^{{\kappa_t}}\right)^{+},
\end{equation}
for any $t\geq{}0$. Next, we approximate the {law of the process} $\breve{X}^{(\delta)}$ with {that} of the following {process}, again defined on the extended probability space $(\Omega^{(\delta)},\calF^{(\delta)},\bbp^{(\delta)})$ for {each} $\delta\in(0,\epsilon_0)$:
\[	
	{X}_{t}^{(\delta)}:=\hat\X_{t}^{(\delta)}+\tilde\X_t^{(\delta,1)}+\tilde\X_t^{(\delta,3)}+R^{(\delta)}+\bar\beta^{(\delta)}t.
\]
Above, $\bar\beta^{(\delta)}$ is chosen so that {the resultant  price process, 
\[
	 {{S}}^{(\delta)}_{t}:=e^{{X}_{t}^{(\delta)}},
\]
is a martingale under $\bbp^{(\delta)}$. In turn, since the triplets of the processes in question are with respect to the truncation function ${\bf 1}_{\{|x|\leq{}\delta\}}$,}
\[
	\bar\beta^{(\delta)}:=-\int_{\bbr_0}\left(e^{x}-1-x{\bf 1}_{\{|x|\leq\delta\}}\right)C\left(x/|x|\right)\left({\bf 1}_{\{|x|\leq{}\delta\}}+\bar\q(x){\bf 1}_{\{|x|\geq{}\delta\}}+e^{-|x|/\delta}{\bf 1}_{\{|x|\geq{}\epsilon_0\}}\right)|x|^{-Y-1}dx.
\]
With $\bar\beta^{(\delta)}$ as above, the L\'evy triplet $(0,{\beta}^{(\delta)},\nu^{(\delta)})$ of ${X}^{(\delta)}$ with respect to the truncation function ${\bf 1}_{\{|x|\leq 1\}}$ is given by
\begin{align*}
	{\nu}^{(\delta)}(dx)& :=C(x/|x|)|x|^{-Y-1} {\bar{q}^{(\delta)}(x)}dx:=C(x/|x|)|x|^{-Y-1}\left({\bf 1}_{|x|\leq{}\delta}+\bar\q(x){\bf 1}_{|x|\geq{}\delta}+e^{-|x|/\delta}{\bf 1}_{|x|\geq{}\epsilon_0}\right)dx,\\
		{\beta^{(\delta)}} & = \bar\beta^{(\delta)} + \int_{\bbr_0}x{\bf 1}_{\{\delta\leq |x|\leq 1\}}{\nu}^{(\delta)}(dx)
	=-\int_{\bbr_0}\left(e^{x}-1-x{\bf 1}_{\{|x|\leq 1\}}\right)\nu^{(\delta)}(dx),
\end{align*}
so it is clear that ${\bar{q}^{(\delta)}}$ satisfies the conditions (\ref{NewAssumEq}-i) and (\ref{NewAssumEq}-iii). To show that ${\bar{q}^{(\delta)}}$ also satisfies (\ref{NewAssumEq}-ii), note that, since $\bar{q}$ is bounded, for some $B\in (0,\infty)$ and $|x|\geq\epsilon_{0}$,
\[
	-\frac{1}{\delta}\leq \frac{\ln \bar{q}^{(\delta)}(x)}{|x|}\leq \frac{B}{|x|},
\] 
which clearly implies (\ref{NewAssumEq}-ii). Since $\bar{q}^{(\delta)}$ satisfies all the conditions in (\ref{NewAssumEq}), we know from Lemma \ref{2ndASYStep1} that 
\begin{align}\label{ScnOrderApM}
\lim_{t\to0}t^{\frac{1}{Y}-1}\left(t^{-\frac{1}{Y}}\bbe\left[\left(S_t^{(\delta)}-{e^{\kappa_{t}}}\right)^+\right]-{{d_1}}\right)=d_2^{(\delta)},
\end{align}
where {{$d_{1}:=\widetilde{\bbe}(Z_{1}^{+})$}} is independent of $\delta$, and $d_{2}^{(\delta)}:=\tilde{\vartheta}^{(\delta)}+{\left(\tilde{\gamma}^{(\delta)}-\theta\right)}\widetilde{\bbp}(Z_{1}\geq 0)$ with
\begin{align}
	&\tilde\vartheta^{(\delta)} := C(1)\int_{0}^{\infty}\left(e^{x}\bar{q}^{(\delta)}(x)-\bar{q}^{(\delta)}(x)-x\right)x^{-Y-1}dx,\label{VarThetaM}\\
	&\tilde\gamma^{(\delta)} :={\beta^{(\delta)}}+\frac{C(1)-C(-1)}{Y-1}+C(1)\int_{0}^{1}x^{-Y}\left(1-\bar{q}^{(\delta)}(x)\right)dx-C(-1)\int_{-1}^{0}|x|^{-Y}\left(1-\bar{q}^{(\delta)}(x)\right)dx\label{GammaM}. 
\end{align}
Now, we proceed to compare the close-to-the-money option prices under {both prices processes $(S^{(\delta)}_{t})_{t\geq{}0}$ and $(S_{t})_{t\geq{}0}$}. To this end, let us first note the following simple {inequality}, which readily follows from 
{(\ref{EqTwoMd}) and the fact that $(a+b)^+\leq a^+ + |b|$ for any real numbers $a$ and $b$:}
\begin{align}\label{tri}
\bbe\left(S_t^{(\delta)}-e^{{\kappa_t}}\right)^+ - \bbe\left|\breve{S}^{(\delta)}_t-S_t^{(\delta)}\right| \leq \bbe\left(S_t-e^{{\kappa_t}}\right)^{+} \leq 
\bbe\left(S_t^{(\delta)}-e^{{\kappa_t}}\right)^+ + \bbe\left|\breve{S}_t^{(\delta)}-S_t^{(\delta)}\right|.
\end{align}
{From (\ref{tri}), it then} follows that
\begin{align}\label{tri2first}
t^{\frac{1}{Y}-1}\left(t^{-\frac{1}{Y}}\bbe\left(S_t^{(\delta)}-e^{{\kappa_t}}\right)^{+}-d_{1}\right) -{t^{-1}} \bbe\left|\breve{S}^{(\delta)}_t-S_t^{(\delta)}\right| &\leq 
t^{\frac{1}{Y}-1}\left(t^{-\frac{1}{Y}}\bbe\left(S_t-e^{{\kappa_t}}\right)^{+} -d_{1}\right)\\
&\leq t^{\frac{1}{Y}-1}
\left(t^{-\frac{1}{Y}}\bbe\left(S_t^{(\delta)}-e^{{\kappa_t}}\right)^{+}-d_{1}\right) + {t^{-1}}\bbe\left| \breve{S}^{(\delta)}_t-S_t^{(\delta)}\right|.\nonumber
\end{align}
We first show that $\bbe|\breve{S}^{(\delta)}_t-S_{t}^{(\delta)}|$ converges to $0$ fast enough. Indeed, using the independence of the processes involved,
\begin{align}\label{SError}
\bbe\left|\breve{S}^{(\delta)}_t-S_t^{(\delta)}\right|
& = \bbe\left(e^{\hat\X_t^{(\delta)}+\tilde\X_t^{(\delta,1)}+\tilde\X_t^{(\delta,3)}+b^{(\delta)}t}\left|e^{\tilde\X_t^{(\delta,2)}}-e^{R_t^{(\delta)}+(\bar\beta^{(\delta)}-b^{(\delta)})t}\right|\right)\nonumber\\
& = \bbe\left(e^{\hat\X_t^{(\delta)}+\tilde\X_t^{(\delta,1)}+\tilde\X_t^{(\delta,3)}+b^{(\delta)}t}\right)\bbe\left|e^{\tilde\X_t^{(\delta,2)}}-e^{R_t^{(\delta)}+(\bar\beta^{(\delta)}-b^{(\delta)})t}\right|\nonumber\\
&\leq \bbe\left|e^{\tilde\X_t^{(\delta,2)}}-1\right| + \bbe\left|e^{R_t^{(\delta)}+(\bar\beta^{(\delta)}-b^{(\delta)})t}-1\right|,
\end{align}
because {
\[
	\bbe\left(e^{\hat\X_t^{(\delta)}+\tilde\X_t^{(\delta,1)}+\tilde\X_t^{(\delta,3)}+b^{(\delta)}t}\right)=\bbe\left(e^{\tilde\X_t^{(\delta,2)}}\right)^{-1}\leq e^{-\bbe\tilde\X_t^{(\delta,2)}}= 1,
\]
by} Jensen's inequality and the fact that $\tilde\X^{(\delta,2)}$ is a martingale. Now, note that by (\ref{MinCond}), $R^{(\delta)}$ is a finite variation process and, thus, can be decomposed as
\begin{align*}
R_t^{(\delta)} 
= \sum_{s\leq t}\Delta\R_s^{(\delta)}{\bf 1}_{\{|\Delta\R_s^{(\delta)}|>\delta\}} + \left(\sum_{s\leq t}\Delta\R_s^{(\delta)}{\bf 1}_{\{|\Delta\R_s^{(\delta)}|\leq\delta\}} -t\int_{|x|\leq \delta}x\nu^{(\delta)}_{R}(dx)\right)
=: R_t^{(\delta,1)}+R_t^{(\delta,2)},
\end{align*} 
where $R_t^{(\delta,1)}$ is a compound Poisson process with L\'evy measure $\nu_R^{(\delta,1)}(dx):=C(x/|x|)|x|^{-Y-1}e^{-|x|/\delta}{\bf 1}_{\{|x|\geq{}\epsilon_0\}}dx$, and $R_t^{(\delta,2)}$ has L\'evy triplet $(0,0,\nu_R^{(\delta,2)})$, {with respect to the truncation ${\bf 1}_{\{|x|\leq{}1\}}$}, where $ \nu_R^{(\delta,2)}(dx):=C(x/|x|)(1-\bar\q(x))^+|x|^{-Y-1}{\bf 1}_{\{|x|\leq{}\delta\}}dx$. Similarly, {by (\ref{MinCond}),} $\tilde\X_t^{(\delta,2)}$ is a finite variation process and has the representation
\begin{align*}
\tilde\X_t^{(\delta,2)} = \sum_{s\leq t}\Delta\tilde\X_s^{(\delta,2)} - t\int_{|x|\leq\delta}x\tilde\nu^{(\delta,2)}(dx).
\end{align*}
For the first term in (\ref{SError}), we now have, for every $0<\delta<\epsilon_0$ and $0<t\leq{}1$,
\begin{align}\nonumber
\bbe\left|e^{\tilde\X_t^{(\delta,2)}}-1\right|
& = \bbe\left(\left|e^{\tilde\X_t^{(\delta,2)}}-1\right|{\bf 1}_{\{|\tilde\X_t^{(\delta,2)}|\leq 1\}}\right) + \bbe\left(\left|e^{\tilde\X_t^{(\delta,2)}}-1\right|{\bf 1}_{\{|\tilde\X_t^{{(\delta,2)}}|>1\}}\right)\\
\nonumber
& \leq  K_1\bbe\left|\tilde\X_t^{(\delta,2)}\right| +  K_{2}\left(\bbe\left(e^{2\tilde\X_t^{(\delta,2)}}+1\right)\bbp^{(\delta)}\left({\left|\tilde\X_t^{(\delta,2)}\right|>1}\right)\right)^{\half}\\
& \leq K_1\bbe\left|\tilde\X_t^{(\delta,2)}\right| + K_3\left(\bbp^{(\delta)}\left({\left|\tilde\X_t^{(\delta,2)}\right|>1}\right)\right)^{\half},\label{NdTrRHS}
\end{align}
where $K_1$, $K_2$, {and $K_3$}, are constants, which can be chosen independently of $\delta$, since 
\begin{align*}
	\bbe\left(e^{2\tilde\X_t^{(\delta,2)}}\right) = e^{t\int_{|x|\leq{}\delta}{C({x}/{|x|})}\left(e^{2x}-1-2x\right)\left(\bar{q}(x)-1\right)^{+}|x|^{-Y-1}dx}
	\leq e^{t\int_{|x|\leq{}1}{C({x}/{|x|})}\left(e^{2x}-1-2x\right)\left(\bar{q}(x)-1\right)^{+}|x|^{-Y-1}dx}<{\infty.}
\end{align*}
For the first term in (\ref{NdTrRHS}), we have
\begin{align*}
\bbe\left|\tilde\X_t^{(\delta,2)}\right|\leq \bbe\sum_{s\leq t}\left|\Delta\tilde\X_s^{(\delta,2)}\right|+t\int_{|x|\leq\delta}|x|\tilde\nu^{(\delta,2)}(dx)= 2t\int_{|x|\leq\delta}|x|\tilde\nu^{(\delta,2)}(dx) 
= 2t\int_{|x|<\delta}(\bar\q(x)-1)^{+}|x|^{-Y}dx, 
\end{align*}
and, for the second term in (\ref{NdTrRHS}), there exist $0<t_0\leq{}1$ such that for any {$0<t<t_{0}$ and} $0<\delta<\epsilon_0$,
\begin{align*}
\bbp^{(\delta)}\left({\left|\tilde\X_t^{(\delta,2)}\right|>1}\right)
\leq \bbp^{(\delta)}\left(\sum_{s\leq t}\left|\Delta\tilde\X_s^{(\delta,2)}\right|>1- t\int_{|x|\leq\delta}|x|\nu^{(\delta,2)}(dx)\right)
\leq \bbp^{(\epsilon_0)}\left(\sum_{s\leq t}\left|\Delta\tilde\X_s^{(\epsilon_0,2)}\right|>\half\right)={O(t^2)},
\end{align*}
by selecting $\varepsilon_{0}$ small enough (see Lemma $3.2$ in \cite{RusWorn:2002} or Remark 3.1 in \cite{FigHou:2008}), and where the last inequality follows from the fact that the distribution of $\sum_{s\leq t}\big|\Delta\tilde\X_s^{(\epsilon_{0},2)}\big|{\bf 1}_{\{|\Delta\tilde\X_s^{(\epsilon_{0},2)}|\leq{}\delta\}}$ under $\bbp^{(\epsilon_0)}$ is the same as that of $\sum_{s\leq t}\big|\Delta\tilde\X_s^{(\delta,2)}\big|$ under $\bbp^{(\delta)}$.
Therefore, for a constant $K<\infty$ independent of $\delta$,
\begin{align}\label{XError}
\limsup_{t\to 0}t^{-1}\bbe\left|e^{\tilde\X_t^{(\delta,2)}}-1\right|\leq K\int_{|x|<\delta}(\bar\q(x)-1)^+|x|^{-Y}dx. 
\end{align} 
For the second term in (\ref{SError}), {for any $0<t\leq{}1$ and $\delta\in(0,\epsilon_{0})$,}
\begin{align} 
\bbe\left|e^{R_t^{(\delta)}+(b^{(\delta)}-\bar\beta^{(\delta)})t}-1\right| 
&\leq e^{(b^{(\delta)}-\bar\beta^{(\delta)})t}\bbe\left|e^{R_t^{(\delta,1)}+R_t^{(\delta,2)}}-1\right|+\left|e^{(b^{(\delta)}-\bar\beta^{(\delta)})t}-1\right| \nonumber\\ 
&\leq e^{(b^{(\delta)}-\bar\beta^{(\delta)})t}\left(\bbe\left(e^{R_t^{(\delta,2)}}\right)\bbe\left|e^{R_t^{(\delta,1)}}-1\right| + \bbe\left|e^{R_t^{(\delta,2)}}-1\right|\right)+\left|e^{(b^{(\delta)}-\bar\beta^{(\delta)})t}-1\right|\nonumber\\
&\leq K_1\bbe\left|e^{R_t^{(\delta,1)}}-1\right| + K_2\bbe\left|e^{R_t^{(\delta,2)}}-1\right| +K_3\left|b^{(\delta)}-\bar\beta^{(\delta)}\right|t,\label{EsDcmNd}
\end{align}
where $K_1$, $K_2$, and $K_3$ are {absolute} constants that can be chosen independently of $\delta$ since, by (\ref{MinCond}),
\begin{align*}
\bbe\left(e^{R_t^{(\delta,2)}}\right) = e^{t\int_{|x|\leq\delta}C(x/|x|)(e^x-1)(1-\bar\q(x))^+|x|^{-Y-1}dx}
\leq e^{\int_{|x|\leq 1}C(x/|x|)|e^x-1||1-\bar\q(x)|^+|x|^{-Y-1}dx} <\infty, 
\end{align*}
and, for any $\delta\in(0,\epsilon_0)$,
\begin{align*}
\left|b^{(\delta)}-\bar\beta^{(\delta)}\right|& =\left|\int_{\bbr_0}\left(e^x-1-x{\bf 1}_{\{|x|\leq\delta\}}\right)(\nu-\nu^{{(\delta)}})(dx)\right|\\
&\leq \int_{\bbr_0}C(x/|x|){\left|e^x-1-x{\bf 1}_{\{|x|\leq\delta\}}\right|}|x|^{-Y-1}\left(|\bar\q(x)-1|{\bf 1}_{\{|x|\leq\delta\}} + e^{-\frac{|x|}{\delta}}{\bf 1}_{\{|x|\geq\epsilon_0\}}\right)dx\\
&\leq \int_{\bbr_0}C(x/|x|){\left(\left|e^x-1-x{\bf 1}_{\{|x|\leq\epsilon_0\}}\right|+|x|{\bf 1}_{\{|x|\leq\epsilon_0\}}\right)}|x|^{-Y-1}\left(|\bar\q(x)-1|{\bf 1}_{\{|x|\leq\epsilon_0\}} + e^{-\frac{|x|}{\delta}}{\bf 1}_{\{|x|\geq\epsilon_0\}}\right)dx<\infty.
\end{align*}
The second term in (\ref{EsDcmNd}) can be taken care of in a similar fashion as the first term in (\ref{SError}) to obtain
\begin{align}\label{R1Error}
\limsup_{t\to 0}t^{-1}\bbe\left|e^{R_t^{(\delta,2)}}-1\right| \leq K\int_{|x|<\delta}(1-\bar\q(x))^{+}|x|^{-Y}dx,
\end{align} 
where $K$ is independent of $\delta$.
For the first term in (\ref{EsDcmNd}), denoting the intensity of jumps and the first jump of $R^{(\delta,1)}$ by $\lambda^{(\delta)}$ and $\xi^{(\delta)}$, respectively, and conditioning on the number of jumps, we have
\begin{align*}
\bbe\left|e^{R_t^{(\delta,1)}}-1\right|
= e^{-\lambda^{(\delta)}t }\lambda^{(\delta)}t \bbe\left|e^{\xi^{(\delta)}}-1\right| + o(t)
\leq \lambda^{(\delta)} t\left(\bbe\left(e^{\xi^{(\delta)}}\right)+1\right)+o(t),
\end{align*}
and, thus,
\begin{align}\label{R2Error}
\limsup_{t\to 0}t^{-1}\bbe\left|e^{R_t^{(\delta,1)}}-1\right|
\leq \lambda^{(\delta)}\left(\bbe\left(e^{\xi^{(\delta)}}\right) + 1\right).
\end{align}
Combining (\ref{SError})-(\ref{R2Error}) gives
\begin{align*}
\limsup_{t\to 0}t^{-1}\bbe\left|\breve{S}^{(\delta)}_t-S_t^{(\delta)}\right| 
 \leq K\int_{|x|<\delta}|1-\bar\q(x)||x|^{-Y}dx + \lambda^{(\delta)}\left(\bbe\left(e^{\xi^{(\delta)}}\right) + 1\right) + \left|b^{(\delta)}-\bar\beta^{(\delta)}\right|,
\end{align*}
for some constant $0<K<\infty$.
Combining the above relationship with (\ref{ScnOrderApM}) {and (\ref{tri2first})}, we get
\begin{align}\label{Snwch}
 d_2^{(\delta)}-r^{(\delta)} \leq \liminf_{t\to{}0}t^{\frac{1}{Y}-1}\left(t^{-\frac{1}{Y}}\bbe\left(S_t-e^{\kappa_t}\right)^{+} -d_{1}\right)\leq \limsup_{t\to{}0}t^{\frac{1}{Y}-1}\left(t^{-\frac{1}{Y}}\bbe\left(S_t-e^{\kappa_t}\right)^{+} -d_{1}\right)\leq d_2^{(\delta)}+r^{(\delta)}.
\end{align}
It remains to show 
\begin{equation}\label{LstRelTP}
	{\rm (i)}\;\lim_{\delta\searrow{}0}d_{2}^{(\delta)}=d_{2}, \qquad {\rm (ii)}\;\lim_{\delta\searrow{}0}r^{(\delta)}=0.
\end{equation}
The first limit follows from $\tilde\vartheta^{(\delta)}\to{}\tilde\vartheta$, $\beta^{(\delta)}\to{}b$,  and $\tilde\gamma^{(\delta)}\to{}\tilde\gamma$. Indeed, 
\begin{align*}
	\tilde\vartheta^{(\delta)} &= C(1)\int_{0}^{\infty}\left(e^{x}\bar{q}^{(\delta)}(x)-\bar{q}^{(\delta)}(x)-x\right)x^{-Y-1}dx\\
	&=\tilde\vartheta+C(1)\int_{x\geq{}\epsilon_{0}}\left(e^{x}-1\right)e^{-|x|/\delta}x^{-Y-1}dx
	+C(1)\int_{0}^{\delta}\left(e^{x}-1\right)(1-\bar\q(x))x^{-Y-1}dx,
\end{align*}
which converges to $\tilde\vartheta$ by the dominated convergence theorem and (\ref{MinCond}). One can similarly show the other two limits $\beta^{(\delta)}\to{}b$ and $\tilde\gamma^{(\delta)}\to{}\tilde\gamma$. For (\ref{LstRelTP}-ii), the first part of $r^{(\delta)}$ converges to zero, again by (\ref{MinCond}), and $|b^{(\delta)}-\bar\beta^{(\delta)}|\to{}0$ is shown in a similar fashion as $\tilde\vartheta^{(\delta)}\to{}\tilde\vartheta$. Lastly,
\begin{align*}
	\lambda^{(\delta)}\left(\bbe\left(e^{\xi^{(\delta)}}\right)+1\right)
	= \int_{|x|\geq \epsilon_{0}}\left(e^{x}+1\right)C\left(x/|x|\right)|x|^{-Y-1}e^{-|x|/\delta}dx
	\leq 2\int_{|x|\geq \epsilon_{0}}C\left(x/|x|\right)|x|^{-Y-1}e^{|x|\big(1-\frac{1}{\delta}\big)}dx,
\end{align*}
and, for $\delta<1/2$, the integrand is dominated by the integrable function $\left(C(1)\vee C(-1)\right)|x|^{-Y-1}{\bf 1}_{\{|x|\geq{}\epsilon_{0}\}}$.
Hence, the dominated convergence theorem applies and we conclude that $r^{(\delta)}\to{}0$ as $\delta\to{}0$. Finally, (\ref{Snwch}) and (\ref{LstRelTP}) prove (\ref{CL2}).
\hfill\qed

\begin{rem}\label{Remark1b}
\hfill
\vspace{-0.2 cm}
\begin{enumerate}
\item 
As customary, one can map the expansion (\ref{CL2}) into an expansion for the close-to-the-money Black-Scholes implied volatility $\hat\sigma(t)$ of the model. Concretely, we have the following small-time behavior for $\hat\sigma(t)$:
\begin{align}\label{AsyIVPureCGMY}
\hat{\sigma}(t)=\sigma_{1}t^{\frac{1}{Y}-\frac{1}{2}}+\sigma_{2}t^{{\frac{1}{2}}}+o(t^{{\frac{1}{2}}}),\quad t\rightarrow 0,
\end{align}
where
\begin{align}\label{1stCoefIVPureCGMY}
\sigma_{1}:=\sqrt{2\pi}\,\widetilde{\bbe}\left(Z_{1}^{+}\right),\qquad
\sigma_{2}:=\sqrt{2\pi}\left(\tilde{\vartheta}+\left(\tilde{\gamma}-\theta\right)\widetilde{\bbp}\left(Z_{1}\geq 0\right)\right).
\end{align}
The proof of (\ref{AsyIVPureCGMY}) is identical to  the proof of Corollary 3.7 in \cite{LopGonHou:2013} and is therefore omitted. 
\item It is also worth mentioning that $\widetilde{\bbe}\left(Z_{1}^{+}\right)$ and $\widetilde{\bbp}\left(Z_{1}\geq 0\right)$ have the following explicit expressions (see \cite{LopGonHou:2013} and references therein):
\begin{align*}
\widetilde{\bbe}\left(Z_{1}^{+}\right)&=\frac{A^{\frac{1}{Y}}}{\pi}\Gamma(-Y)^{\frac{1}{Y}}\left|\cos\left(\frac{\pi Y}{2}\right)\right|^{\frac{1}{Y}}\cos\left(\frac{1}{Y}\arctan\left(\frac{B}{A}\tan\left(\frac{Y\pi}{2}\right)\right)\right)\Gamma\left(1-\frac{1}{Y}\right)\left(1+\left(\frac{B}{A}\right)^{2}\tan^{2}\left(\frac{\pi Y}{2}\right)\right)^{\frac{1}{2Y}},\\
\widetilde{\bbp}\left(Z_{1}\geq{}0\right)&=\frac{1}{2}+\frac{1}{\pi Y}\arctan\left(\frac{B}{A}
	\tan\left(\frac{Y\pi}{2}\right)\right),
\end{align*}
where $A:=C(1)+C(-1)$ and $B:=C(1)-C(-1)$. 
\item By using the expression for $b$ coming from the martingale condition (\ref{NdCndTSMrt}), the second order term in (\ref{CL2}) can be written as
\begin{align*}
d_2&=\widetilde{\bbp}(Z_{1}< 0)\,C(1)\int_{0}^{\infty}\left(e^{x}\bar{q}(x)-\bar{q}(x)-x\right)x^{-Y-1}dx\\
&\quad -\widetilde{\bbp}(Z_{1}\geq 0)\left(C(-1)\int_{-\infty}^{0}\left(e^{x}\bar{q}(x)-\bar{q}(x)-x\right)|x|^{-Y-1}dx+\theta\right),
\end{align*}
which is independent of the truncation function chosen for the L\'evy triplet of $X$.
\end{enumerate}
\end{rem}

\section{A L\'evy Jump Model With Stochastic Volatility}\label{StochVol}
A second order approximation for ATM option prices, when the asset price process includes a nonzero Brownian component, is presented in \cite{LopGonHou:2013}. In this section we will show that the continuous Brownian part can be replaced by an independent stochastic volatility process. As described in Section \ref{Sec:tmpstble}, we consider the asset price process $S_{t}:=S_{0}e^{X_{t}+V_{t}}$, where $X$ is a pure-jump tempered stable-like process satisfying the conditions of Theorem \ref{2ndASY}, and $V$ is defined as in (\ref{modelV})-(\ref{modelY}). Moreover, it is assumed that $\sigma(y_0)>0$, and that there exists an open interval $I$, containing $y_{0}$, on which $\alpha(\cdot)$ and $\gamma(\cdot)$ are uniformly bounded and $\sigma^2(\cdot)$ is Lipschitz continuous.

The following result will be used in the proof of the second order option price approximation for the process $(S_t)_{t\geq 0}$. Its proof is deferred to Appendix \ref{ApTecLem}. 
\begin{lem}\label{HittingTime}
Let $(Y_{t})_{t\geq{}0}$ be as in (\ref{modelY}), and $\tau:=\inf\{t\geq 0: Y_t\notin (a,b)\}$, where $a$ and $b$ are such that $a<y_0<b$. Then, for any $k\in\bbn$, $\bbp\left(\tau\leq t\right)=O(t^k)$, as $t\to{}0$.
\end{lem}
The second order approximation for the close-to-the-money option prices can now be stated as follows.
\begin{thm}\label{2ndASYSV}
{Consider the model $S_{t}:=S_{0}e^{X_{t}+V_{t}}$, as described above,} and let $\kappa_t := \theta t^{\frac{3-Y}{2}}+o(t^{\frac{3-Y}{2}})$, as $t\to{}0$, for some $\theta\in\bbr$. Then, the second order asymptotic expansion for the call option price is
\begin{align}\label{CL2C}
		\frac{1}{S_0}\bbe\left(S_t-S_0e^{\kappa_{t}}\right)^+=d_1t^{\half}+d_2t^{\frac{3-Y}{2}}+o\left(t^{\frac{3-Y}{2}}\right),\quad t\to 0,
\end{align}
where 
\begin{align}\label{d1d2}
	d_1 := \frac{\sigma(y_0)}{\sqrt{2\pi}}\;,\qquad 
	d_2 := \frac{\theta}{2} + \frac{2^{-\frac{Y+1}{2}}}{\sqrt{\pi}}\Gamma\left(1-\frac{Y}{2}\right)\frac{C(1)+C(-1)}{Y(Y-1)}\sigma(y_0)^{1-Y}\;.
\end{align}
\end{thm}
\begin{rem}
By the same reasoning as in Remark \ref{Remark1}, the form of the log-moneyness $\kappa_t$ chosen in Theorem \ref{2ndASYSV} is the most relevant one. The expansion reduces to the one in the at-the-money case ($\theta=0$) when $\kappa_t$ converges to $0$ at a rate faster than $t^{\frac{3-Y}{2}}$. In particular, this would be the case if $\kappa_t=\theta t + o(t)$ as in Theorem \ref{2ndASY}. On the contrary, if, for instance, $\kappa_{t}=\theta\t^{\beta}+o(t^{\beta})$ with $\half<\beta<\frac{3-Y}{2}$ and $\theta\neq{}0$, then 
\begin{align}\label{RelNdH1}
		\frac{1}{S_0}\bbe\left(S_t-S_0e^{\kappa_{t}}\right)^+=d_1t^{\half}+d_2t^{\beta}+o\left(t^{\frac{3-Y}{2}}\right),\quad t\to 0,
\end{align}
where $d_1$ is as in (\ref{d1d2}) and $d_2:=\theta/2$. The condition $\beta>\half$ is necessary for the first order term $d_1$ to be as in (\ref{d1d2}) (see, also, Theorem $3.1$ in \cite{MuhNut:2009}). Roughly, (\ref{RelNdH1}) says that if $\kappa_{t}$ converges to $0$ fast enough for the leading order term to be $d_{1}t^{1/2}$, but slower than $t^{(3-Y)/2}$, then the second order term and its order are determined by $\kappa_{t}$ and, thus, contain no additional information on the underlying financial model.
\end{rem}

\begin{rem}
	The expansion (\ref{CL2C}) leads to the following expansion for the corresponding close-to-the-money Black-Scholes implied volatility $\hat\sigma(t)$:
	\begin{align}\label{AsyIVGerCGMY} \hat{\sigma}(t)=\sigma(y_{0})+\left(\theta\sqrt{\frac{\pi}{2}}+\frac{{(C(1)+C(-1)})2^{-\frac{Y}{2}}}{Y(Y-1)}\Gamma\left(1-\frac{Y}{2}\right)\sigma(y_{0})^{1-Y}\right)t^{1-\frac{Y}{2}}+o\left(t^{1-\frac{Y}{2}}\right),\quad t\rightarrow 0.
\end{align}
The proof of (\ref{AsyIVGerCGMY}) is identical to  the proof of Corollary 4.3 in \cite{LopGonHou:2013} and is therefore omitted.
\end{rem}

\noindent\textbf{Proof of Theorem \ref{2ndASYSV}.}
Throughout, we take $S_0=1$ without loss of generality. The result will now be shown in the following three steps.

\noindent\textbf{Step 1)}
We first show that $X$ can be assumed to have a L\'evy density of the form
\begin{align}\label{LevyX}
	\tilde{s}(x) = C\left({x}/{|x|}\right)e^{-|x|}|x|^{-Y-1},
\end{align}
which, in particular, satisfies the conditions in (\ref{NewAssumEq}) (in fact, it even satisfies the stronger conditions in (\ref{Eq:StndCnd1a})).
To show that we use a procedure similar to the one in the proof of Theorem \ref{2ndASY}. Let $\delta\in (0,1)$ be a fixed constant such that $\inf_{|x|\leq\delta}\bar\q(x)>0$, and let $(0,b^{{(\delta)}},\nu)$ be the L\'evy triplet of $X$ with respect to the truncation function ${\bf 1}_{\{|x|\leq \delta\}}$. 
Then, define $(\Omega^{(\delta)},\calF^{(\delta)},\bbp^{(\delta)})$, an extension of the original probability space $(\Omega,\calF,\bbp)$, along with independent L\'evy processes, $\tilde\X^{(\delta,1)}$, $\tilde\X^{(\delta,2)}$, and $\tilde\X^{(\delta,3)}$, which are also independent of the original processes $X$ and $V$. More precisely, the L\'evy triplet of $\tilde{X}^{(\delta,i)}$, with respect to ${\bf 1}_{\{|x|\leq{}\delta\}}$, is given by $(0,0,\tilde\nu_i^{(\delta)})$, for $i=1,2,3$, where
\begin{align*}
	\tilde\nu_1^{(\delta)}(dx) &:= C\left(x/|x|\right){\bf 1}_{\{|x|\leq \delta\}}\left(e^{-|x|}{\bf 1}_{\{\bar\q(x)\geq e^{-|x|}\}}+\bar\q(x){\bf 1}_{\{\bar\q(x)< e^{-|x|}\}}\right)|x|^{-Y-1}dx,\\
	\tilde\nu_2^{(\delta)}(dx)&:=C\left(x/|x|\right)\left({\bf 1}_{\{|x|\leq \delta}\}\left(\bar q(x)-e^{-|x|}\right)^++{\bf 1}_{\{|x|\geq  \delta\}}\bar\q(x)\right)|x|^{-Y-1}dx,\\
	\tilde\nu_3^{(\delta)}(dx)&:=C\left(x/|x|\right)\left({\bf 1}_{\{|x|\leq \delta\}}\left(e^{-|x|}-\bar q(x)\right)^++{\bf 1}_{\{|x|\geq  \delta\}}e^{-|x|}\right)|x|^{-Y-1}dx.
\end{align*}
As before we will denote the expectation under $\bbp^{(\delta)}$ by $\bbe$. Now, it is clear that the law of the L\'evy process 
\[
	\breve{X}^{(\delta)}_{t}:= \tilde X_t^{(\delta,1)} + \tilde X_t^{(\delta,2)} + b^{(\delta)}t,
\]
coincides with that of the process $X$ and, thus, the price process $\breve{S}^{(\delta)}:=\exp(\breve{X}^{(\delta)}+V_t)$ is such that
\begin{equation*}
	\bbe\left(\breve{S}^{(\delta)}_t-e^{{\kappa_t}}\right)^{+}=\bbe\left(S_t-e^{{\kappa_t}}\right)^{+},
\end{equation*}
for any $t\geq{}0$. Next, define the process
\[	
	{X}_{t}^{(\delta)}:=\tilde X_t^{(\delta,1)} + \tilde X_t^{(\delta,3)}+\beta^{(\delta)}t,
\]
where $\beta^{(\delta)}$ is chosen so that {the resultant price process, 
\[
	 {S}^{(\delta)}_{t}:=e^{{X}_{t}^{(\delta)}+V_t},
\]
is a martingale under $\bbp^{(\delta)}$. Note that $X^{(\delta)}$ has a L\'evy density of the form (\ref{LevyX}). We will show that the second order term of the close-to-the-money option prices is the same under both price processes $(S^{(\delta)}_{t})_{t\geq{}0}$ and $(S_{t})_{t\geq{}0}$. As done in (\ref{tri}), it follows that
\begin{align}\label{tri2}
t^{\frac{Y-2}{2}}\left(t^{-\frac{1}{2}}\bbe\left(S_t^{(\delta)}-e^{{\kappa_t}}\right)^{+}-d_{1}\right) 
- t^{\frac{Y-3}{2}} \bbe\left|\breve{S}^{(\delta)}_t-S_t^{(\delta)}\right| &\leq 
t^{\frac{Y-2}{2}}\left(t^{-\frac{1}{2}}\bbe\left(S_t-e^{{\kappa_t}}\right)^{+} -d_{1}\right)\\
&\leq t^{\frac{Y-2}{2}}
\left(t^{-\frac{1}{2}}\bbe\left(S_t^{(\delta)}-e^{{\kappa_t}}\right)^{+}-d_{1}\right) +  t^{\frac{Y-3}{2}}\bbe\left| \breve{S}^{(\delta)}_t-S_t^{(\delta)}\right|,\nonumber
\end{align}
so, for the option prices under both price processes $(S^{(\delta)}_{t})_{t\geq{}0}$ and $(S_{t})_{t\geq{}0}$ to have the same second order term, it suffices that
\begin{align}\label{DistBtwTwoModels}
\bbe\left| \breve{S}^{(\delta)}_t-S_t^{(\delta)}\right| = \bbe\left(e^{V_{t}}\right)\bbe\left|e^{\breve\X^{(\delta)}_t}-e^{X^{(\delta)}}\right|= O(t),\quad {t\to 0}.
\end{align}
Using the independence of the processes in question, we have
\begin{align}\label{Xcond}
\bbe\left|e^{\breve\X^{(\delta)}_t}-e^{X^{(\delta)}}\right| 
& = \bbe\left(e^{\tilde X_t^{(\delta,1)}+\b^{(\delta)}t}\right)
\bbe\left|e^{\tilde X_t^{(\delta,2)}}-e^{\tilde X_t^{(\delta,3)}+(\beta^{(\delta)}-b^{(\delta)})t}\right|\nonumber\\
&\leq \bbe\left|e^{\tilde X_t^{(\delta,2)}}-1\right|
+\bbe\left|e^{\tilde X_t^{(\delta,3)}+(\beta^{(\delta)}-b^{(\delta)})t}-1\right|,
\end{align}
since, by Jensen's inequality and the fact that $\tilde\X^{(\delta,2)}$ is a martingale,
\[
	\bbe\left(e^{\tilde\X_t^{(\delta,1)}+b^{(\delta)}t}\right)=\bbe\left(e^{\tilde\X_t^{(\delta,2)}}\right)^{-1}\leq e^{-\bbe\tilde\X_t^{(\delta,2)}}= 1, 
\]
The order of the terms in (\ref{Xcond}) can then be shown to be $O(t)$, using arguments similar to the ones used to show that the terms in (\ref{SError}) were of order $O(t)$, and by noting that $\tilde X_t^{(\delta,2)}$ and $\tilde X_t^{(\delta,3)}$ are finite variation processes. Indeed,
\begin{align*}
\int_{|x|\leq\delta}\left|\bar\q(x)-e^{-|x|}\right||x|^{-Y}dx \leq 
\int_{|x|\leq\delta}\left|\bar\q(x)-1\right||x|^{-Y}dx  + \int_{|x|\leq\delta}\left|1-e^{-|x|}\right||x|^{-Y}dx <\infty.
\end{align*}
From (\ref{tri2}) and (\ref{DistBtwTwoModels}), in order to obtain (\ref{CL2C}), it suffices to show that 
\[
	\lim_{t\to{}0}t^{\frac{Y-2}{2}}\left(t^{-\frac{1}{2}}\bbe\left(S^{(\delta)}_t-e^{{\kappa_t}}\right)^{+} -d_{1}\right)=d_{2}.
\]
Hence, from the outset, we assume that $X$ has a L\'evy density as in (\ref{LevyX}).

\noindent\textbf{Step 2)}
We will now show the validity of (\ref{CL2C}) in the case where there exist constants $m$ and $M$ such that 
\begin{align}\label{SigmaMm}
0<m\leq\sigma(y)\leq M<\infty, \text{ for all } y  \text{ in the range } \calR_{Y}\text{ of } (Y_{t})_{ t\leq 1},
\end{align}
i.e., $\calR_{Y}:=\cup_{0\leq t\leq{}1}{\rm supp}(Y_{t})$, with ${\rm supp}(Y_{t})$ representing the support of $Y_{t}$.
The idea is to reduce the problem to the case where the process $Y$ is deterministic, by conditioning the option's payoff on the realization of the 
{process $W^{1}$. 
To formalize this idea, we need to introduce some notation. On a filtered probability space $(\breve\Omega,\breve\calF,(\breve\calF_{t})_{t\geq{}0},\breve\bbp)$ satisfying the usual conditions, we define independent processes $\breve{X}$ and $\breve{W}^{2}$, such that the law of $(\breve{X}_t)_{0\leq{}t\leq{}1}$ under $\breve\bbp$ is the same as the law of $(X_{t})_{0\leq{}t\leq{}1}$ under $\bbp^{}$, and $(\breve{W}_t^{2})_{0\leq{}t\leq{}1}$ is a standard Brownian motion. Also, for some deterministic functions $y:=(y_{s})_{s\in[0,1]}$ and ${q}:=({q}_{s})_{s\in[0,1]}$ belonging to $C([0,1])$, the class of continuous functions on $[0,1]$, let $(\breve{V}^{y,q}_{t})_{0\leq{}t\leq{}1}$ be defined as
\begin{align}\label{DtrmVolMod}
	\breve{V}^{y,q}_t & = -\half\int_{0}^{t}\sigma^2\left(y_u\right)du+\rho {q}_{t}+\sqrt{1-\rho^2}\int_{0}^{t}\sigma\left(y_u\right)d\breve{W}^{2}_u.
\end{align} 
With this notation at hand, we consider 
a functional $\Phi:[0,1]\times C([0,1])\times C([0,1])\to \bbr_{+}$ defined as 
\begin{align}\label{OptPricDtrmVol}
\Phi\left(t,y,{q}\right):=\breve{\bbe}\left(e^{\breve{X}_{t}+\breve\V^{y,{q}}_t}-e^{\kappa_t}\right)^+.	
\end{align}
Then it is clear that, for any $t\in[0,1]$, 
\begin{equation}\label{CndStep0}	
	\bbe\left(\left.\left(S_{t}-e^{\kappa_t}\right)_+\right|W^{1}_s,s\in[0,1]\right)=\Phi\left(t,(Y_{s})_{s\in[0,1]},\left(Q_{s}\right)_{s\in[0,1]}\right),
\end{equation}
where 
\[
	{Q}_{s}:=\int_{0}^{s}\sigma(Y_{u})dW^{1}_{u}, \qquad s\in[0,1].
\]
The following terminology will also be used in the sequel:
\begin{align}\nonumber
&\breve\V^{1,y,q}_t :=-\half\rho^2\int_0^t\sigma^2(y_s)ds + \rho{q}_t,\quad 
\breve\V^{2,y,q}_t :=-\half(1-\rho^2)\int_0^t\sigma^2(y_s)ds + \sqrt{1-\rho^2}\int_0^t\sigma(y_s)d\breve\W_s^2,\\
&\psi^{y,q,w}_{t}:=\psi_t^{1,y,q,w}+\psi_t^{2,y},\quad \psi_t^{1,y,q,w} := 
\breve\V_t^{1,y,q} - \rho\sigma(y_0)w,\quad
\psi_t^{2,y} := \frac{1-\rho^2}{2}\int_0^t\sigma^2(y_s)ds,\label{DfnPsis}
\end{align}
where $w\in\bbr$ is an auxiliary constant. Later on, we shall substitute $w$ with $W_{t}^{1}(\omega)$, which, upon conditioning on $(W^{1}_{s})_{0\leq{}s\leq{}1}$, is obviously constant. 
Note that $\breve\V_t^{y,q}=\breve\V_t^{1,y,q}+\breve\V_t^{2,y,q}$. For ease of notation, we often drop the dependence on $y$, $q$, and $w$ in the above processes, unless explicitly needed. Also, with certain abuse of notation, we sometimes use the following shorthand notation:
\begin{equation}\label{WeirdNotation}
	\breve{V}^{1}_{t}(\omega):=\breve\V^{1,Y_{\cdot}(\omega),Q_{\cdot}(\omega)}_t,\quad 
	\psi_{t}(\omega):=\psi^{Y_{\cdot}(\omega),Q_{\cdot}(\omega),W^{1}_{t}(\omega)}_{t},
	\quad \psi^{1}_{t}(\omega):=\psi^{1,Y_{\cdot}(\omega),Q_{\cdot}(\omega),W^{1}_{t}(\omega)}_{t},\quad 
	\psi^{2}_{t}(\omega):=\psi^{2,Y_{\cdot}(\omega)}_{t},
\end{equation}
where {$\omega\in\Omega$ and}, as usual, $Y_{\cdot}(\omega)$ and $Q_{\cdot}(\omega)$ are seen as random elements in $C([0,1])$.

As in the pure-jump case, we shall use two probability transformations. First, define $\breve{\bbp}^*$ on $(\breve\Omega,\breve\calF)$ by
\[
	\frac{d \breve{\bbp}^{*}|_{\calF_{t}}}{d \breve\bbp |_{\calF_{t}}}=e^{\breve{X}_{t}+\breve\V_t^{2}},\qquad t\leq{}1. 
\]
Under $\breve{\bbp}^*$, $(\breve{X})_{t\leq{}1}$ has L\'evy triplet $(0,b^*,\nu^*)$ given by (\ref{triplet1}). Similarly, by Girsanov's theorem for Brownian motions, $(\breve{V}_{t}^2)_{0\leq{}t\leq{}1}$ has the representation
\begin{align*}
	d\breve{V}^2_t 
	= \half(1-\rho^2)\sigma^2(y_t)dt + \sqrt{1-\rho^2}\sigma(y_t)d\breve\W_t^{*,2},\qquad \breve{V}^2_{0}=0,
\end{align*} 
for $0\leq{}t\leq{}1$, where $(\breve{W}^{*,2}_{t})_{0\leq t\leq{}1}$ 
is a $\breve\bbp^*$-Wiener process. Next, define the probability measure $\widetilde\bbp$ as described in (\ref{EMMN}) and (\ref{Uplusminuseta}), but replacing the jump measure $N$ of the process $X$ by the jump measure of $\breve{X}$. In particular, under $\widetilde{\bbp}$, $\breve{X}$ has L\'evy triplet given by (\ref{triplet2}). Analogously to (\ref{SSSP}), we define the centered process $\breve{Z}_{t}:=\breve{X}_{t}-t\tilde\gamma$, where $\tilde{\gamma}:=\widetilde{\bbe}\left(\breve{X}_{1}\right)$. Note that the law of $(\breve{V}^2_{t})_{t\leq{}1}$ under $\widetilde\bbp$ remains unchanged. It is also useful to point out that, under both $\breve\bbp^{*}$ and $\widetilde{\bbp}$, 
\begin{align}\label{DistV}
{\rm (i)}\;t^{-1/2}\breve{V}^{2}_t \sim \mathcal{N}\left(t^{-1/2}\psi_{t}^{2},(1-\rho^2)\bar\sigma_t^2(y)\right),\quad {\rm (ii)}\;t^{-1/2}\left(\breve{V}_t -\psi_{t}\right)\sim \mathcal{N}\left(t^{-1/2}\left(\breve{V}^{1}_{t}-\psi_{t}^{1}\right),(1-\rho^2)\bar\sigma_t^2(y)\right)
\end{align}
where, for $t\in(0,1]$, $\bar\sigma_{t}:C([0,1])\to\bbr_{+}$ is defined by
\begin{align}\label{BndBarSigma}
\bar\sigma_t(y):=\sqrt{\frac{1}{t}\int_0^t\sigma^2(y_s)ds}\in[m,M].
\end{align}
In order to find the second order term of the expansion, we investigate the limit of $t^{Y/2-1}R_t$ as $t\to{}0$, where, for $0<{}t\leq{}1$, we set 
\begin{align*}
R_t 
& := t^{-1/2}\bbe\left(S_{t}-e^{\kappa_t}\right)^{+}-d_{1},
\end{align*}
which, in terms of the functional $\Phi$, can be expressed as 
\begin{align*}
R_t 
& = \bbe\left(t^{-1/2}\bbe\left(\left.\left(S_{t}-e^{\kappa_t}\right)^+\right|W^{1}_s,s\in[0,1]\right)-d_{1}\right)\\
& = \bbe\left(t^{-1/2}\Phi\left(t,(Y_{s})_{s\in[0,1]},\left({Q}_{s}\right)_{s\in[0,1]}\right)-d_{1}\right)\\
& =: \bbe\left(\bar{\Phi}\left(t,(Y_{s})_{s\in[0,1]},\left({Q}_{s}\right)_{s\in[0,1]}\right)\right).
\end{align*}
We shall show that $\lim_{t\to{}0}t^{Y/2-1}R_{t}=d_{2}$, for the constant $d_{2}$ defined in the statement of the theorem. 
First, using (\ref{CarrMadan}), the change of probability measures to $\widetilde\bbp$,  and a change of variable, 
\begin{align}\label{2ndOADecCGMYB}
\bar{\Phi}\left(t,y, q\right)
&=t^{-1/2}e^{\breve\V_t^1}\breve\bbe\left(\left(e^{\breve{X}_{t}+\breve{V}^2_{t}}-e^{\kappa_t-\breve\V_t^1}\right)^+\right)-d_{1}\nonumber\\
&=t^{-1/2}e^{\breve\V_t^1}\int_{0}^{\infty}e^{-z}\breve\bbp^{*}\left(\breve{X}_{t}+\breve{V}^2_{t}\geq{}z+\kappa_{t}-\breve\V^1_t\right)dz
-d_{1}\nonumber\\
&=e^{\breve\V_t^1}\int_{-t^{1/2}\tilde\gamma+t^{-1/2}\kappa_{t}-t^{-1/2}\psi}^{\infty}e^{-\sqrt{t}u-\tilde\gamma t+\kappa_{t}-\psi_t} \breve\bbp^{*}\left(t^{-1/2}\left(\breve{V}_{t}-\psi_t\right)\geq{}u-t^{-1/2}\breve{Z}_{t}\right)du-d_{1}\nonumber\\
&=e^{\breve\V_t^1}\int_{-t^{1/2}\tilde\gamma+t^{-1/2}\kappa_{t}-t^{-1/2}\psi_t}^{\infty}e^{-\sqrt{t}u-\tilde\gamma t+\kappa_{t}-\psi_t}\,\widetilde{\bbe}\left(e^{-\widetilde{U}_{t}-\eta t}{\bf 1}_{\{t^{-1/2}\left(\breve{V}_{t}-\psi_t\right)\geq{}u-t^{-1/2}\breve{Z}_{t}\}}\right)du-d_{1}.\nonumber
\end{align}
Next, we decompose it as follows, for any $w\in\bbr$, 
\begin{align}
\bar{\Phi}\left(t,y,q\right)&=e^{-\tilde\eta_t}e^{\breve\V_t^1}\int_{0}^{\infty}e^{-\sqrt{t}u}\left(\widetilde{\bbe}\left(e^{-\widetilde{U}_{t}}{\bf 1}_{\{t^{-1/2}\left(\breve{V}_{t}-\psi_t\right)\geq{}u-t^{-1/2}\breve{Z}_{t}\}}\right)-\widetilde\bbe\left(e^{-\widetilde{U}_{t}}{\bf 1}_{\{t^{-1/2}\left(\breve{V}_{t}-\psi_t\right)\geq{}u\}}\right)\right)du\nonumber\\
&\quad+e^{-\tilde\eta_t}e^{\breve{V}^1_t}\int_{-t^{1/2}\tilde\gamma+t^{-1/2}\kappa_{t}-t^{-1/2}\psi_t}^{0}e^{-\sqrt{t}u}\widetilde{\bbe}\left(e^{-\widetilde{U}_{t}}{\bf 1}_{\{t^{-1/2}\left(\breve{V}_{t}-\psi_t\right)\geq{}u-t^{-1/2}\breve{Z}_{t}\}}\right)du\nonumber\\
&\quad+e^{\breve{V}^1_t}\int_{0}^{\infty}e^{-\sqrt{t}u-\tilde\gamma t+\kappa_t-\psi_t}\breve\bbp^{*}\left(t^{-1/2}\left(\breve{V}_{t}-\psi_t\right)\geq{}u\right)du-d_{1}\nonumber\\
&=: {A^1(t,y,q,w)+A^2(t,y,q,w)+A^3(t,y,q,w)},
\end{align}
where $\tilde{\eta}_t:=\tilde\eta_{t}^{y,q,w}=(\eta+\tilde{\gamma})t-\kappa_t+\psi_t^{y,q,w}$. Note that the dependence of the $A^{i}$'s on the auxiliary number $w$ is because $\psi$ depends on $w$. Analogously to (\ref{WeirdNotation}), we shall sometimes use the notation
\begin{equation}\label{WeirdNotation2}
	\tilde{\eta}_t(\omega):=\tilde\eta^{Y_{\cdot}(\omega),Q_{\cdot}(\omega),W^{1}_{t}(\omega)}_{t}.
\end{equation}
We now consider each of the three terms in (\ref{2ndOADecCGMYB}) separately.

\noindent\textbf{First term:} For $A_t^1:=A^1(t,y,q,w)$ we closely follow the steps in \cite{LopGonHou:2013}, and start with the following decomposition:
\begin{align}\label{DecomAt}
A_t^1&=e^{-\tilde\eta_t}e^{\breve\V_t^1}\widetilde{\bbe}\left(e^{-\widetilde{U}_{t}}{\bf 1}_{\{t^{-1/2}\left(\breve{V}_{t}-\psi_t\right)\geq{} 0, t^{-1/2}\left(\breve{V}_{t}-\psi_t\right)+t^{-\frac{1}{2}}\breve{Z}_{t}\geq 0\}}\int_{t^{-1/2}\left(\breve{V}_{t}-\psi_t\right)}^{t^{-1/2}\left(\breve{V}_{t}-\psi_t\right)+t^{-\frac{1}{2}}\breve{Z}_{t}}e^{-\sqrt{t}u}du\right)\nonumber\\
&\quad-e^{-\tilde{\eta}_t}e^{\breve\V_t^1}\widetilde{\bbe}\left(e^{-\widetilde{U}_{t}}{\bf 1}_{\{0\leq t^{-\half}\left(\breve{V}_t-\psi_t\right)\leq-t^{-\frac{1}{2}}\breve{Z}_{t}\}}\int_{0}^{t^{-\half}\left(\breve{V}_t-\psi_t\right)}e^{-\sqrt{t}u}du\right)\nonumber\\
&\quad+e^{-\tilde{\eta}_t}e^{\breve\V_t^1}\widetilde{\bbe}\left(e^{-\widetilde{U}_{t}}{\bf 1}_{\{0\leq-t^{-\half}\left( \breve{V}_t-\psi_t\right) \leq t^{-\frac{1}{2}}\breve{Z}_{t}\}}\int_{0}^{t^{-\half}\left(\breve{V}_t-\psi_t\right)+t^{-\frac{1}{2}}\breve{Z}_{t}}e^{-\sqrt{t}u}du\right)\nonumber\\
&=: I_{1}(t,y,q,w)-I_{2}(t,y,q,w)+I_{3}(t,y,q,w).
\end{align}
We analyze each of these terms in the following three parts. 

\noindent{\textbf{i)}}
Using (\ref{DistV}-ii),
 the first term, $I_1:=I_1(t,y,q,w)$, can be written as 
\begin{align}\label{IntJ1}
	I_1
	& = e^{-\tilde{\eta}_t}e^{\breve\V_t^1}\int_0^{\infty}{J_{11}(t,{u})e^{-\sqrt{t}u}\frac{e^{-\frac{\left(u-t^{-1/2}\left(\breve\V_t^1-\psi^1_t\right)\right)^2}{2(1-\rho^2)\bar{\sigma}_t^2(y)}}}{\sqrt{2\pi(1-\rho^2)\bar{\sigma}_t^2(y)}}du}
	+ e^{-\tilde{\eta}_t}e^{\breve\V_t^1}\int_0^{\infty}{J_{12}(t,{u})e^{-\sqrt{t}{u}}\frac{e^{-\frac{\left(u-t^{-1/2}\left(\breve\V_t^1-\psi^1_t\right)\right)^2}{2(1-\rho^2){\bar\sigma}_t^2(y)}}}{\sqrt{2\pi(1-\rho^2)\bar{\sigma}_t^2(y)}}du}\nonumber\\
	 & =: I_1^1(t,y,q,w)+I_1^2(t,y,q,w),
\end{align}
where 
\begin{align}
J_{11}(t,u)&:=\widetilde{\bbe}\left({\bf 1}_{\{\breve{Z}_{t}\geq -t^{\frac{1}{2}}u\}}\bigg(\frac{e^{-\widetilde{U}_{t}}-e^{-(\widetilde{U}_{t}+\breve{Z}_{t})}}{\sqrt{t}}-t^{-\frac{1}{2}}\breve{Z}_{t}\bigg)\right),\\
J_{12}(t,u)&:=t^{-\frac{1}{2}}\widetilde{\bbe}\left(\breve{Z}_{t}{\bf 1}_{\{\breve{Z}_{t}\geq  -t^{\frac{1}{2}}u\}}\right)=t^{-\frac{1}{2}}\widetilde{\bbe}\left((-\breve{Z}_{t}){\bf 1}_{\{-\breve{Z}_{t}\geq  t^{\frac{1}{2}}u\}}\right)=t^{\frac{1}{Y}-\frac{1}{2}}\widetilde{\bbe}\left((-\breve{Z}_{1}){\bf 1}_{\{-\breve{Z}_{1}\geq  t^{\frac{1}{2}-\frac{1}{Y}}u\}}\right)
\label{DecomJ1},
\end{align}
since $\widetilde{\bbe}(\breve{Z}_{t})=0$ and $\breve{Z}_t$ is strictly $Y$-stable. It can be shown (see (B.5) in \cite{LopGonHou:2013}) that 
there exists a constant $\lambda>0$ such that, for any $0<t\leq{}1$ and $u>0$,
\begin{equation}\label{B6FromLP}
	{\rm (i)}\;t^{\frac{Y}{2}-1}J_{12}(t,u)\leq\lambda u^{1-Y} \quad\text{ and }\quad
	{\rm (ii)}\;\lim_{t\to{}0}t^{\frac{Y}{2}-1}J_{12}(t,u)=\frac{C(-1)}{Y-1}u^{1-Y}. 
\end{equation}
From (\ref{DfnPsis}), note that 
\begin{align}
t^{-1/2}\left(\breve\V^{1}_t(\omega)-\psi^{1}_{t}(\omega)\right):=t^{-1/2}\left(\breve\V_t^{1,Y_{\cdot}(\omega),Q_{\cdot}(\omega)}-\psi^{1,Y_{\cdot}(\omega),Q_{\cdot}(\omega),W^{1}_{t}(\omega)}_t\right) = t^{-1/2}\rho\sigma(y_0)W_t^1(\omega)\sim \mathcal{N}(0,\rho^2\sigma^2(y_0)).\label{SmplDisV1}
\end{align}	
Moreover, recall that $0<m\leq\bar\sigma_t(y)\leq M<\infty$ and that for $\bbp$-a.e. $\omega\in\Omega$,
\begin{align}\label{limbarsigma}
	\lim_{t\to{}0}\bar\sigma^{2}_{t}(Y_{\cdot}(\omega))=\lim_{t\to{}0}\frac{1}{t}\int_0^t\sigma^2(Y_s)ds=\sigma^{2}(y_0)=:\sigma_{0}^{2}, 
\end{align}
which follows from the continuity of $\sigma^{2}(\cdot)$ in a neighborhood of $y_{0}$.
Using the previous relationships, we get (see Appendix \ref{AuxPrfT2} for the details) 
\begin{equation}\label{AsyInJ12}
\lim_{t\to0}t^{Y/2-1}\bbe\left(I_1^2(t, Y_{\cdot}(\omega),Q_{\cdot}(\omega),W^1_{t}(\omega))\right) = \frac{C(-1)}{Y-1}\int_0^\infty{u^{1-Y}\frac{e^{-\frac{u^2}{2\sigma^{2}_{0}}}}{\sqrt{2\pi\sigma^{2}_{0}}}}du.
\end{equation}
For the first term in (\ref{IntJ1}), $I_1^1$, it turns out that 
\begin{equation}\label{RmdTI1}
	\lim_{t\to{}0} t^{Y/2 -1}\bbe\left(I_1^1(t,Y_{\cdot}(\omega),Q_{\cdot}(\omega),W^1_{t}(\omega))\right)=0.
\end{equation}
The proof of (\ref{RmdTI1}) is also deferred to Appendix \ref{AuxPrfT2}. Together {(\ref{AsyInJ12}) and (\ref{RmdTI1})} imply that
\begin{align}\label{AsyI1}
	 \lim_{t\to0}\frac{\bbe\left(I_{1}(t,Y_{\cdot}(\omega),Q_{\cdot}(\omega),W^1_{t}(\omega))\right)}{t^{1-Y/2}} =  \frac{C(-1)}{Y-1}\int_0^\infty{u^{1-Y}\frac{e^{-\frac{u^2}{2\sigma_0^2}}}{\sqrt{2\pi\sigma_0^2}}}du.
\end{align}}
\textbf{{ii)}} Using again (\ref{DistV}-ii), the term $I_{2}:=I_2(t,y,q,w)$ in (\ref{DecomAt}) can be written as 
\begin{align}\label{DecomI2}
I_{2}
&= e^{-\tilde\eta_t}e^{\breve\V_t^1}\int_{0}^{\infty}\widetilde{\bbe}\left(\left(e^{-\widetilde{U}_{t}}-1\right){\bf 1}_{\{\breve\Z_{t}\leq-t^{\frac{1}{2}}u\}}\right)\frac{1-e^{-\sqrt{t}u}}{\sqrt{t}}\frac{e^{-\frac{\left(u-t^{-1/2}\left(\breve\V_t^1-\psi^1_t\right)\right)^2}{2(1-\rho^2)\bar\sigma_t(y)^2}}}{\sqrt{2\pi(1-\rho^2){\bar\sigma_t^2(y)}}}du\nonumber\\
&\quad +e^{-\tilde\eta_t}e^{\breve\V_t^1}\int_{0}^{\infty}\widetilde{\bbp}\left(\breve\Z_{t}\leq-t^{\frac{1}{2}}u\right)\frac{1-e^{-\sqrt{t}u}}{\sqrt{t}}\frac{e^{-\frac{\left(u-t^{-1/2}\left(\breve\V_t^1-\psi^1_t\right)\right)^2}{2(1-\rho^2)\bar\sigma_t^2(y)}}}{\sqrt{2\pi(1-\rho^2)\bar\sigma_t^2(y)}}du\nonumber\\ 
& =: I_2^1(t,y,q,w)+I_2^2(t,y,q,w).
\end{align}
Using the fact that $\big(\breve\Z_t\big)_{t\geq{}0}$ is strictly $Y$-stable, we have (see (2.18)-(2.19) in \cite{LopGonHou:2013})
\[
	\widetilde\bbp\left(\breve\Z_t\leq-t^\half u\right)\frac{1-e^{-\sqrt{t}u}}{\sqrt{t}} \leq \tilde\kappa u^{1-Y},\quad 
	\lim_{t\to{}0}t^{\frac{Y}{2}-1}\widetilde\bbp\left(\breve\Z_t\leq-t^\half u\right)=\lim_{t\to{}0}t^{\frac{Y}{2}-1}\widetilde\bbp\left(\breve\Z_1\leq-t^{\half-\frac{1}{Y}} u\right)=\frac{C(-1)}{Y} u^{-Y}{,}
\]
for any $0<t\leq{}1$ and $u>0$. Therefore, following {arguments} similar to those leading to (\ref{AsyInJ12}), it follows that
\begin{align}\label{AsyI2second}
\lim_{t\to0}t^{Y/2-1}\bbe\left( I_2^2\left(t,Y_{\cdot}(\omega),Q_{\cdot}(\omega),W^1_{t}(\omega)\right)\right)=\frac{C(-1)}{Y}\int_0^\infty{u^{1-Y}\frac{e^{-\frac{u^2}{2\sigma^2_0}}}{\sqrt{2\pi\sigma^2_0}}du}.
\end{align}
For $I_2^1$, it turns out that (see Appendix \ref{AuxPrfT2} for its verification):
\begin{align}\label{AsyI2first}
\lim_{t\to{}0}t^{\frac{Y}{2}-1}\bbe\left|I_2^1\left(t,Y_{\cdot}(\omega),Q_{\cdot}(\omega),W^1_{t}(\omega)\right)\right|=0.
\end{align}
Combining (\ref{AsyI2second}) and (\ref{AsyI2first}) then leads to 
\begin{align}\label{AsyI2}
\lim_{t\rightarrow 0}t^{\frac{Y}{2}-1}\bbe\left( I_{2}\left(t,Y_{\cdot}(\omega),Q_{\cdot}(\omega),W^1_{t}(\omega)\right)\right)=\frac{C(-1)}{Y}\int_{0}^{\infty}u^{1-Y}\frac{e^{-\frac{u^{2}}{2\sigma_0^{2}}}}{\sqrt{2\pi\sigma_0^{2}}}du.
\end{align}
\textbf{{iii)}} It remains to analyze the term {$I_{3}:=I_3(t,y,q,w)$} in (\ref{DecomAt}), which  we first decompose as follows:
\begin{align}\label{DefnJ31J32}
I_{3}
&=e^{-\tilde\eta_t}e^{\breve\V_t^1}\int_{0}^{\infty}{\widetilde{\bbe}\left(e^{-\widetilde{U}_{t}}{\bf 1}_{\{\breve\Z_{t}\geq t^{\frac{1}{2}}u\}}\right)}\frac{1-e^{\sqrt{t}u}}{\sqrt{t}}\frac{e^{-\frac{\left(u-t^{-1/2}\left(\breve\V_t^1-\psi^1_t\right)\right)^2}{2(1-\rho^2){\bar\sigma_t^2(y)}}}}{\sqrt{2\pi(1-\rho^2){\bar\sigma_t^2(y)}}}du\nonumber\\
&\quad+e^{-\tilde\eta_t}e^{\breve\V_t^1}\int_{0}^{\infty}{\widetilde{\bbe}\left(e^{-\widetilde{U}_{t}}{\bf 1}_{\{\breve\Z_{t}\geq t^{\frac{1}{2}}u\}}\frac{1-e^{-\breve\Z_{t}}}{\sqrt{t}}\right)}e^{\sqrt{t} u}\frac{e^{-\frac{\left(u-t^{-1/2}\left(\breve\V_t^1-\psi^1_t\right)\right)^2}{2(1-\rho^2){\bar\sigma_t^2(y)}}}}{\sqrt{2\pi(1-\rho^2){\bar\sigma_t^2(y)}}}du\nonumber\\
& =: I_3^1(t,y,q,w)+I_3^2(t,y,q,w).
\end{align}
To deal with the first term in (\ref{DefnJ31J32}), let us first decompose the expectation therein as follows:
\begin{align*}
J_{31}(t,u)  :=\widetilde{\bbe}\left(e^{-\widetilde{U}_{t}}{\bf 1}_{\{\breve\Z_{t}\geq t^{\frac{1}{2}}u\}}\right)
=\widetilde{\bbe}\left(\left(e^{-\widetilde{U}_{t}}-1\right){\bf 1}_{\{\breve\Z_{t}\geq t^{\frac{1}{2}}u\}}\right)+\widetilde{\bbp}\left(\breve\Z_{t}\geq t^{\frac{1}{2}}u\right)
=:J_{31}^{(1)}(t,u)+J_{31}^{(2)}(t,u).
\end{align*}
Since $(\breve{Z}_t)_{t\geq{}0}$ is $Y$-stable, we can obtain the estimate {$J_{31}^{(2)}(t,u)\leq{}\kappa t^{1-\frac{Y}{2}}u^{-Y}$}, for any $0<t\leq 1$ and $u\geq 0$. Combining that with $\left|\frac{1-e^{\sqrt{t}u}}{\sqrt{t}}\right| \leq ue^{\sqrt{t}u}$, for $0<t\leq 1$, and using steps similar to the ones in (\ref{AsyInJ12}), we can show that
\begin{align}\label{AsyInJ312}
\lim_{t\rightarrow 0}t^{\frac{Y}{2}-1}\bbe\left(e^{\breve\V_t^1(\omega)-\tilde\eta_t(\omega)}\int_{0}^{\infty}J_{31}^{(2)}(t,u)\frac{1-e^{\sqrt{t}u}}{\sqrt{t}}\frac{e^{-\frac{\left(u-t^{-1/2}\left( \breve\V_t^1(\omega)-\psi^1_t(\omega)\right)\right)^2}{2(1-\rho^2){\bar\sigma_t^2(Y(\omega))}}}}{\sqrt{2\pi(1-\rho^2){\bar\sigma_t^2(Y(\omega))}}}du\right)=-\frac{C(1)}{Y}\int_{0}^{\infty}u^{1-Y}\frac{e^{-\frac{u^{2}}{2\sigma_0^{2}}}}{\sqrt{2\pi\sigma_0^{2}}}du.
\end{align}
The same procedure as in (\ref{DocomFirIntI2}-\ref{DocomFirIntI23}) {below} can be used to show that $0\leq J_{31}^{(1)}(t,z) \leq f(t)$ for $0<t\leq t_0<1$, where $f(t)=o(t^{1-Y/2})$, so for $0<t\leq t_0$:
\begin{align}\label{AsyInJ311}
0 &\,\leq  t^{\frac{Y}{2}-1}\bbe\left(e^{\breve\V_t^1(\omega)-\tilde\eta_t(\omega)}\int_{0}^{\infty}J_{31}^{(1)}(t,u)\frac{\left|1-e^{\sqrt{t}u}\right|}{\sqrt{t}}\frac{e^{-\frac{\left(u-t^{-1/2}\left( \breve\V_t^1(\omega)-\psi^1_t(\omega)\right)\right)^2}{2{(1-\rho^2)\bar\sigma_t^2(Y(\omega))}}}}{\sqrt{2\pi(1-\rho^2){\bar\sigma_t^2(Y(\omega))}}}du\right)\nonumber\\
&\, \leq f(t)t^{\frac{Y}{2}-1}\bbe\left(e^{\breve\V_t^1(\omega)-\tilde\eta_t(\omega)}\int_{0}^{\infty}ue^u\frac{e^{-\frac{\left(u-t^{-1/2}\left( \breve\V_t^1(\omega)-\psi^1_t(\omega)\right)\right)^2}{2(1-\rho^2){\bar\sigma_t^2(Y(\omega))}}}}{\sqrt{2\pi(1-\rho^2){\bar\sigma_t^2(Y(\omega))}}}du\right)\;\stackrel{t\to{}0}{\longrightarrow}\; 0.
\end{align}
To deal with the second term in (\ref{DefnJ31J32}), note that, by the self-similarity of $(\breve\Z_t)_{t\geq 0}$,
\begin{align*}
J_{32}(t,u) & :=\widetilde{\bbe}\left(e^{-\widetilde{U}_{t}}{\bf 1}_{\{\breve\Z_{t}\geq t^{\frac{1}{2}}u\}}\frac{1-e^{-\breve\Z_{t}}}{\sqrt{t}}\right)\\
&=\widetilde{\bbe}\left({\bf 1}_{\{\breve\Z_{t}\geq t^{\frac{1}{2}}u\}}\left(\frac{e^{-\widetilde{U}_{t}}-e^{-(\breve\Z_{t}+\widetilde{U}_{t})}}{\sqrt{t}}-t^{-\frac{1}{2}}\breve\Z_{t}\right)\right)+\widetilde{\bbe}\left(t^{\frac{1}{Y}-\frac{1}{2}}\breve\Z_{1}{\bf 1}_{\{\breve\Z_{1}\geq t^{\frac{1}{2}-\frac{1}{Y}}u\}}\right)\nonumber\\
&=:J_{32}^{(1)}(t,u)+J_{32}^{(2)}(t,u).
\end{align*}
Note that $J_{32}^{(2)}(t,u)$ is similar to $J_{12}(t,u)$ in $(\ref{DecomJ1})$ and, thus, the asymptotic {behavior} is similar to $(\ref{AsyInJ12})$. Concretely, 
\begin{align}\label{AsyInJ32sec}
\lim_{t\rightarrow 0}t^{\frac{Y}{2}-1}\bbe\left(e^{\breve\V_t^1(\omega)-\tilde\eta_t(\omega)}\int_{0}^{\infty}\widetilde{\bbe}\left(t^{\frac{1}{Y}-\frac{1}{2}}\breve\Z_{1}{\bf 1}_{\{\breve\Z_{1}\geq t^{\frac{1}{2}-\frac{1}{Y}}u\}}\right)e^{\sqrt{t}u}\frac{e^{-\frac{\left(u-t^{-1/2}\left( \breve\V_t^1(\omega)-\psi^1_t(\omega)\right)\right)^2}{2(1-\rho^2){\bar\sigma_t^2(Y(\omega))}}}}{\sqrt{2\pi(1-\rho^2){\bar\sigma_t^2(Y(\omega))}}}du\right)
=\frac{C(1)}{Y-1}\int_{0}^{\infty}u^{1-Y}\frac{e^{-\frac{u^{2}}{2\sigma_0^{2}}}}{\sqrt{2\pi\sigma_0^{2}}}du.
\end{align}
Next, decompose $J_{32}^{(1)}(t,u)$ as:
\begin{align*}
J_{32}^{(1)}(t,u)&=t^{-\frac{1}{2}}\widetilde{\bbe}\left({\bf 1}_{\{\breve\Z_{t}\geq t^{\frac{1}{2}}u\}}\int_{\widetilde{U}_{t}}^{\breve\Z_{t}+\widetilde{U}_{t}}\left(e^{-x}-1\right)dx\right)\\
&=t^{-\frac{1}{2}}\int_{-\infty}^{0}(e^{-x}-1)\widetilde{\bbp}\left(\breve\Z_{t}\geq t^{\frac{1}{2}}u,\widetilde{U}_{t}\leq x\leq \breve\Z_{t}+\widetilde{U}_{t}\right)dx+t^{-\frac{1}{2}}\int_{0}^{\infty}(e^{-x}-1)\widetilde{\bbp}\left(\breve\Z_{t}\geq t^{\frac{1}{2}}u,\widetilde{U}_{t}\leq x\leq \breve\Z_{t}+\widetilde{U}_{t}\right)dx\\
& \leq t^{-\frac{1}{2}}\int_{-\infty}^{0}(e^{-x}-1)\widetilde{\bbp}\left( \widetilde{U}_{t}\leq x\right)dx+t^{-\frac{1}{2}}\int_{0}^{\infty}(e^{-x}-1)\widetilde{\bbp}\left(x\leq \breve\Z_{t}+\widetilde{U}_{t}\right)dx.
\end{align*}
Using similar arguments as when dealing with the terms involving (\ref{DcIntNd}) {below} gives 
\begin{align}\label{AsyInJ321}
\lim_{t\rightarrow{}0}t^{\frac{Y}{2}-1}\bbe\left(e^{\breve\V_t^1(\omega)-\tilde\eta_t(\omega)}\int_{0}^{\infty}J_{32}^{(1)}(t,u)e^{\sqrt{t}u}\frac{e^{-\frac{\left(u-t^{-1/2}\left( \breve\V_t^1(\omega)-\tilde\eta_t(\omega)\right)\right)^2}{2(1-\rho^2){\bar\sigma_t^2(Y(\omega))}}}}{\sqrt{2\pi(1-\rho^2){\bar\sigma_t^2(Y(\omega))}}}du\right)=0.
\end{align}
Combining the above gives,
\begin{align}\label{AsyI3}
\lim_{t\to0}t^{Y/2-1} \bbe\left(I_3(t,Y_{\cdot}(\omega),Q_{\cdot}(\omega),W_{t}^1(\omega))\right)=\frac{C(1)}{Y(Y-1)}\int_{0}^{\infty}u^{1-Y}\frac{e^{-\frac{u^{2}}{2\sigma_0^{2}}}}{\sqrt{2\pi\sigma_0^{2}}}du.
\end{align}
Finally, (\ref{DecomAt}), (\ref{AsyI1}), (\ref{AsyI2}), and (\ref{AsyI3}) yield
\begin{align}\label{AsyA1}
\lim_{t\to0}\frac{\bbe\left( A_t^1(t,Y_{\cdot}(\omega),Q_{\cdot}(\omega),W_{t}^1(\omega))\right)}{t^{1-Y/2}}=\frac{C(1)+C(-1)}{2Y(Y-1)}\,\sigma_0^{1-Y}\,
\bbe\left(|W_{1}^{1}|^{1-Y}\right).
\end{align}
\textbf{Second term:}
For the term $A_t^2:=A^2(t,y,q,w)$ in (\ref{2ndOADecCGMYB}), by the density transformation $\widetilde\bbp\rightarrow\breve\bbp^*$: 
\begin{align}\label{A2decomp}
A_t^2 &= e^{-\tilde\eta_t}e^{\breve{V}^1_t}\int_{-t^{1/2}\tilde\gamma+t^{-1/2}\kappa_{t}-t^{-1/2}\psi_t}^{0}e^{-\sqrt{t}z}\widetilde{\bbe}\left(e^{-\widetilde{U}_{t}}{\bf 1}_{\{t^{-1/2}\left(\breve{V}_{t}-\psi_t\right)\geq{}z-t^{-1/2}\breve{Z}_{t}\}}\right)dz\nonumber\\ 
& = e^{-\tilde{\gamma}t+\kappa_t-\psi^2_t}e^{\breve{V}^1_t-\psi^1_t}\int_{-t^{1/2}\tilde{\gamma}+t^{-1/2}\kappa_t-t^{-1/2}\psi_t}^{0}{e^{-\sqrt{t}z}\breve\bbp^*\left(t^{-\half}\left(\breve\V_t-\psi_t\right)\geq z-t^{-\half}\breve\Z_t\right)}dz\nonumber\\
& = e^{-\tilde{\gamma}t+\kappa_t-\psi^2_t}e^{\breve{V}^1_t-\psi^1_t}\int_{-t^{1/2}\tilde{\gamma}-t^{-1/2}\psi_t}^{0}{e^{-\sqrt{t}z}\breve\bbp^*\left(t^{-\half}\left(\breve\V_t-\psi_t\right)\geq z-t^{-\half}\breve\Z_t\right)}dz\nonumber\\
&\quad+ e^{-\tilde{\gamma}t+\kappa_t-\psi^2_t}e^{\breve{V}^1_t-\psi^1_t}\int_{-t^{1/2}\tilde{\gamma}+t^{-1/2}\kappa_t-t^{-1/2}\psi_t}^{-t^{1/2}\tilde{\gamma}-t^{-1/2}\psi_t}{e^{-\sqrt{t}z}\breve\bbp^*\left(t^{-\half}\left(\breve\V_t-\psi_t\right)\geq z-t^{-\half}\breve\Z_t\right)}dz\nonumber\\
&=:A_t^{2,1}(y,q,w)+A_t^{2,2}(y,q,w).
\end{align}
For the first term,
\begin{align*}
\left|A_t^{2,1}\right| \leq Ke^{\breve\V_t^1-\psi^1_t}\left|-t^{1/2}\tilde{\gamma}-t^{-1/2}\psi_t\right|e^{\sqrt{t}\left|-t^{1/2}\tilde{\gamma}-t^{-1/2}\psi_t\right|}
 \leq Ke^{\breve\V_t^1-\psi^1_t+\left|\psi_t\right|}\left(t^{1/2}|\tilde{\gamma}|+t^{-1/2}|\psi_t|\right),
\end{align*}
for some constant $K$. Using (\ref{SigmaMm}), (\ref{DfnPsis}), Cauchy's inequality, and the {inequality} $e^{|x|}\leq{}e^{x}+e^{-x}$, we have, for any $p>0$ and $0<t\leq{}1$,
\begin{align*}
\bbe\left(e^{p\left(\breve\V_t^1(\omega)-\psi^1_t(\omega)+\left|\psi_t(\omega)\right|\right)}\right)
&\leq K\bbe\left(e^{p\rho\sigma(y_0)\left(W_t^1+\left|W_t^1\right|\right)+p\rho\left|\int_0^t\sigma(Y_s)dW_s^1\right|}\right)\\
&\leq K\left(1+\bbe\left(e^{4p\rho\sigma(y_0)W_t^1}\right)\right)^{1/2}\left(\bbe\left(e^{2p\rho\int_0^t\sigma(Y_s)dW_s^1}\right)+\bbe\left(e^{-2p\rho\int_0^t\sigma(Y_s)dW_s^1}\right)\right)^{1/2}\leq B,
\end{align*}
for a constant $B<\infty$ independent of $t$. Indeed, by Novikov's condition and Girsanov's theorem, 
\[
	\bbe\left(e^{\pm 2p\rho\int_0^t\sigma(Y_s)dW_s^1}\right)\leq	
	e^{2p^{2}\rho^{2}M^{2}t}\bbe\left(e^{\pm 2p\rho\int_0^t\sigma(Y_s)dW_s^1-2p^{2}\rho^{2}\int_0^t\sigma^{2}(Y_s)ds}\right)=e^{2p^{2}\rho^{2}M^{2}t}<\infty.
\]
Therefore,
\begin{align}\label{AsyA21}
\bbe\left|A_t^{2,1}(Y_{\cdot}(\omega),Q_{\cdot}(\omega),W^1_{t}(\omega))\right| &= O(t^{1/2}) + Kt^{-1/2}\bbe\left(e^{\breve\V_t^1(\omega)-\psi^1_t(\omega)+\left|\psi_t(\omega)\right|}|\psi_t(\omega)|\right) \nonumber\\
&
\leq O(t^{1/2}) + Kt^{-1/2}\sqrt{\bbe\left(\psi_t(\omega)\right)^2} = O(t^{1/2}), 
\end{align}
because $\bbe\left(\psi_t(\omega)\right)^2=O({t^2})$. Indeed, $0\leq \left(\psi_t^2(\omega)\right)^2\leq Kt^2$ for some $K<\infty$, due to (\ref{SigmaMm}), and
\begin{align*}
\bbe\left(\psi_t^1(\omega)\right)^2 & = \bbe\left(\breve\V_t^1(\omega) - \rho\sigma(y_0)W_t^1(\omega)\right)^2\nonumber\\
& = \bbe\left(\rho\int_0^t\left(\sigma(Y_s)-\sigma(y_0)\right)dW_s^1 -\frac{\rho^2}{2}\int_0^t\sigma^2(Y_s)ds\right)^2\nonumber\\
& \leq K\left(\int_0^t\bbe\left(\sigma(Y_s)-\sigma(y_0)\right)^2ds +t\int_0^t\bbe\left(\sigma^4(Y_s)\right)ds\right)\nonumber\\
& \leq K\left(\int_0^t\bbe\left(\sigma^2(Y_s)-\sigma^2(y_0)\right)^2ds +t\int_0^t\bbe\left(\sigma^4(Y_s)\right)ds\right)
\end{align*}
The second term is clearly $O(t^2)$ due to (\ref{SigmaMm}). The first term is also of order $O(t^2)$. Indeed, let $I$ be an interval containing $y_0$, where $\sigma^2(\cdot)$ is Lipschitz with constant $L$ and $\alpha(\cdot)$, and $\gamma(\cdot)$ are bounded, and let $\tau:=\inf\left\{s:Y_s\notin I\right\}$. Then,
\begin{align}\label{sigma2order}
\bbe\left(\sigma^2(Y_t)-\sigma^2({y_0})\right)^2 & = \bbe\left(\left(\sigma^2(Y_t)-\sigma^2({y_0})\right)^2{\bf 1}_{\{\tau>t\}}\right)  + \bbe\left(\left(\sigma^2(Y_t)-\sigma^2({y_0})\right)^2{\bf 1}_{\{\tau\leq t\}}\right)\nonumber\\
& \leq L\bbe\left(\left(Y_t-{y_0}\right)^2{\bf 1}_{\{\tau>t\}}\right) + (M^2-m^2)^2\bbp\left(Y_t\notin I\right)= O(t), \quad {t\to 0},
\end{align} 
where the order of the second term follows from Lemma \ref{HittingTime}, and
\begin{align*}
\bbe\left(\left(Y_t-{y_0}\right)^2{\bf 1}_{\{\tau>t\}}\right) &= \bbe\left(\left(\int_0^t\alpha(Y_s)ds+\int_0^t\gamma(Y_s)dW_s^1\right)^2{\bf 1}_{\{\tau>t\}}\right)\\
&\leq 2\bbe\left(\int_0^t\alpha(Y_s){\bf 1}_{\{\tau>t\}}ds\right)^2+2\bbe\left(\int_0^t\gamma^2(Y_s){\bf 1}_{\{\tau>t\}}ds\right)=O(t), \quad {t\to 0}.
\end{align*} 
Hence, $\bbe\left(\psi_t(\omega)\right)^2=O({t^2})$ and, thus, from (\ref{AsyA21}), it follows that
\begin{align}\label{AsyA2}
\lim_{t\to0}t^{Y/2-1}\bbe\left(A_t^{2,1}(Y_{\cdot}(\omega),Q_{\cdot}(\omega),W^1_{t}(\omega))\right)= 0.
\end{align}
For the second part of (\ref{A2decomp}), a change of variables gives
\begin{align*} 
A_t^{2,2}& = e^{-\tilde{\gamma}t+\kappa_t-\psi^2_t}e^{\breve{V}^1_t-\psi^1_t}\int_{-t^{1/2}\tilde\gamma+t^{-1/2}\kappa_{t}-t^{-1/2}\psi_t}^{-t^{1/2}\tilde\gamma-t^{-1/2}\psi_t}{e^{-\sqrt{t}z}\breve\bbe^*\left({\bf 1}_{\{t^{-1/2}\left(\breve{V}_{t}-\psi_t\right)\geq{}z-t^{-1/2}\breve{Z}_{t}\}}\right)}dz\nonumber\\
& = e^{-\tilde{\gamma}t+\kappa_t}e^{\breve{V}^1_t}\int_{-t^{1/2}\tilde\gamma+t^{-1/2}\kappa_{t}}^{-t^{1/2}\tilde\gamma}{e^{-\sqrt{t}u} \breve\bbe^*\left({\bf 1}_{\{t^{-1/2}\breve{V}_{t}\geq{}u-t^{-1/2}\breve{Z}_{t}\}}\right)}du\\
&=e^{-\tilde{\gamma}t+\kappa_t}e^{\breve{V}^1_t}\int_{-t^{1/2}\tilde\gamma+t^{-1/2}\kappa_{t}}^{-t^{1/2}\tilde\gamma}e^{-\sqrt{t}u}B^{2,2}_{t}(y,q,u)du.
\end{align*}
By the change probability measures $\breve\bbp^{*}\to\breve\bbp$, 	
\[
	{B^{2,2}_{t}(y,q,u)}=\breve\bbe\left(e^{\breve{V}^{y,q}_t+\breve\X_t}{\bf 1}_{\{t^{-1/2}\breve{V}^{y,q}_{t}\geq{}u-t^{-1/2}\breve{Z}_{t}\}}\right),
\]
in particular, reversing the argument in (\ref{CndStep0}), 
\begin{align*}
	\bbe\left(B_{t}^{2,2}(Y_{\cdot}(\omega),Q_{\cdot}(\omega),u)\right)&=
	\bbe\left(\bbe\left(\left.e^{{V}_t+\X_t}{\bf 1}_{\{t^{-1/2}{V}_{t}\geq{}u-t^{-1/2}{Z}_{t}\}}\right|W^{1}_s,s\in[0,1]\right)\right)\\
	&=
	\bbe\left(e^{{V}_t+\X_t}{\bf 1}_{\{t^{-1/2}{V}_{t}\geq{}u-t^{-1/2}{Z}_{t}\}}\right)\\
	&=\bbp^{*}\left(t^{-1/2}{V}_{t}\geq{}u-t^{-1/2}{Z}_{t}\right),
\end{align*}
where for the last equality we used the change of probability measures 
\begin{equation}\label{DSM2b}
	 {\frac{d {\bbp}^{*}|_{\calF_{t}}}{d \bbp |_{\calF_{t}}}}=e^{V_{t}+X_{t}}, \qquad t\geq{}0.
\end{equation}
Hence, by Fubini's theorem and the condition $\kappa_t := \theta t^{\frac{3-Y}{2}}+o(t^{\frac{3-Y}{2}})$, 
\begin{align}
\lim_{t\to 0}t^{\frac{Y}{2}-1}\bbe\left(A_t^{2,2}(Y_{\cdot}(\omega),Q_{\cdot}(\omega),W_{t}^{1}(\omega))\right) &= 
\lim_{t\to 0}e^{-\tilde{\gamma}t+\kappa_t}\kappa_{t}t^{\frac{Y-3}{2}}\frac{1}{t^{-1/2}\kappa_{t}}\int_{-t^{1/2}\tilde\gamma+t^{-1/2}\kappa_{t}}^{-t^{1/2}\tilde\gamma}e^{-\sqrt{t}u}\bbe\left(B_{t}^{2,2}(Y_{\cdot}(\omega),Q_{\cdot}(\omega),u)\right)du\nonumber\\
&= \theta\,\lim_{t\to 0}\frac{1}{t^{-1/2}\kappa_{t}}\int_{-t^{1/2}\tilde\gamma+t^{-1/2}\kappa_{t}}^{-t^{1/2}\tilde\gamma}{e^{-\sqrt{t}u} \bbp^*\left(t^{-1/2}{V}_{t}\geq{}z-t^{-1/2}{Z}_{t}\right)}du\nonumber\\
&= \frac{\theta}{2},
\end{align}
where we have used that $t^{-1/2}\kappa_{t}\to{}0$ and the fact that
\begin{align}\label{GaussConv}
	\bbp^*\left(t^{-1/2}{V}_{t}\geq{}z-t^{-1/2}{Z}_{t}\right)\;\stackrel{t\to{}0}{\longrightarrow}\; \bbp^*\left(\Lambda\geq{}z\right)\;\stackrel{z\to{}0}{\longrightarrow}\;\frac{1}{2},
\end{align}
where $\Lambda$ is a centered Gaussian variable. Indeed, as verified in  Appendix \ref{AuxPrfT2}, under $\bbp^*$,
\begin{equation}\label{CnvDistV}
	t^{-1/2}\V_t\,\ld\, \Lambda,
\end{equation}
and $t^{-1/Y}\Z_{t}$ converges in distribution to a $Y$-stable random variable $\Z$ since, due to the independence of $X$ and $V$, the distribution of $Z$ under $\bbp^*$ is that of a tempered stable process  (see Proposition $1$ in \cite{rosenbaum.tankov.10}).

\noindent\textbf{Third term:} {By letting $\sigma_0:=\sigma(y_0)$,} noting that $d_{1}=\sigma_{0}\bbe(\Lambda)^{+}=\int_{0}^{\infty}\bbp(\sigma_{0}\Lambda\geq{}z)dz$, where $\Lambda$ is a standard normal variable,  and using (\ref{DistV}-ii), the term $A_t^3$ in (\ref{2ndOADecCGMYB}) can trivially be decomposed as follows:
\begin{align}\label{A3decomp}
A_t^3
&=\int_0^{\infty}\left(e^{\breve{V}^1_t}e^{-\sqrt{t}z-\tilde\gamma t+\kappa_t-\psi_t}\int_{\frac{z-t^{-1/2}\left(\breve{V}_{t}^1-\psi_t^1\right)}{\sqrt{1-\rho^2}\bar\sigma_t(y)}}^{\infty}\phi(x)dx-\int_{z/\sigma_{0}}^{\infty}{\phi(x)dx}\right)dz\nonumber\\	    
& = \int_0^{\infty}{(e^{\breve{V}^1_t}e^{-\sqrt{t}z-\tilde\gamma t+\kappa_t-\psi_t}-1)\int_{z/\sigma_{0}}^{\infty}{\phi(x)dx}dz} 	    
+\int_0^{\infty}{e^{\breve{V}^1_t}e^{-\sqrt{t}z-\tilde\gamma t+\kappa_t-\psi_t}\int_{\frac{z-t^{-1/2}\left(\breve{V}_{t}-\psi_t^1\right)}{\sqrt{1-\rho^2}\bar\sigma_t(y)}}^{z/\sigma_{0}}{\phi(x)dx}dz}\nonumber\\	 
& =: J^{1}_{t}(y,q,w) + J^{2}_{t}(y,q,w),
\end{align}
where $\phi(x):=\frac{e^{-\half x^2}}{\sqrt{2\pi}}$. For $J^{1}_{{t}}$, using the notation $\hat\Phi(z):=\int_{z}^{\infty}\phi(x)dx$, we obtain: 
\begin{align}\label{AsyJ1}
\bbe\left|J^{1}_{t}(Y_{\cdot}(\omega),Q_{\cdot}(\omega),W^1_{t}(\omega))\right| &\leq \bbe\int_0^{\infty}{\left|e^{\breve{V}^1_t(\omega)-\psi_t^1(\omega)}e^{-\tilde\gamma t+\kappa_t-\psi_t^2(\omega)}e^{-\sqrt{t}z}-1\right|\hat\Phi\left(\frac{z}{\sigma_{0}}\right)}dz\nonumber\\
&\quad+ \bbe\int_0^{\infty}
e^{-\tilde\gamma t+\kappa_t-\psi^2_t(\omega)}\left|e^{\breve{V}^1_t(\omega)-\psi^1_t(\omega)}-1\right|\hat\Phi\left(\frac{z}{\sigma_{0}}\right)dz\nonumber\\
&\quad +\bbe\int_{0}^{\infty}\left|e^{-\tilde\gamma t+\kappa_t-\psi^2_t(\omega)}-1\right|\hat\Phi\left(\frac{z}{\sigma_{0}}\right)dz\nonumber\\
&= O(\sqrt{t}), 
\end{align}
{as $t\to 0$,} which can be shown using that $|\tilde\gamma t|+|\kappa_t|+|\psi^2_t(\omega)|\leq K \sqrt{t}$, a.s., for a constant $K$, and that $\breve{V}^1_t(\omega)-\psi^1_t(\omega)\sim N\left(0,\rho^2\sigma^{2}_{0}t\right)$. Then, for $J^{2}_{{t}}$, 
\begin{align}\label{AsyJ2}
J^{2}
&= e^{-\tilde\gamma t+\kappa_t}e^{-\psi^2_t}e^{\breve{V}^{1}_{t}-\psi_t^1}\int_0^{\infty}{e^{-\sqrt{t}z}\int_{\frac{z-t^{-1/2}\left(\breve{V}^{1}_{t}-\psi_t^1\right)}{\sqrt{1-\rho^2}\bar\sigma_t(Y)}}^{\frac{z-t^{-1/2}\left(\breve{V}^{1}_{t}-\psi_t^1\right)}{\sqrt{1-\rho^2}\sigma_{0}}}{\phi(x)dx}dz}\nonumber\\
&\quad + e^{-\tilde\gamma t+\kappa_t}e^{-\psi^2_t}e^{\breve{V}^{1}_{t}-\psi_t^1}\int_0^{\infty}{e^{-\sqrt{t}z}\int_{\frac{z-t^{-1/2}\left( \breve{V}^{1}_{t}-\psi_t^1\right)}{\sqrt{1-\rho^2}\sigma_{0}}}^{z/\sigma_{0}}{\phi(x)dx}dz}\nonumber\\
&=:J^{2,1}_{t}(y,q,w) + J^{2,2}_{t}(y,q,w)
\end{align}
For the first term, recall from (\ref{DfnPsis})-(\ref{WeirdNotation}) that $\breve{V}^{1}_{t}(\omega)-\psi^{1}_t(\omega)=\rho\sigma_{0}W_t^1$, so
\begin{align*}
\left|J^{2,1}_{t}(Y_{\cdot}(\omega),Q_{\cdot}(\omega),W^1_{t}(\omega))\right| &\leq K_{1}e^{\rho\sigma_{0}W_t^1}\int_{0}^{\infty}\left|z-\rho\sigma_{0}\frac{W_t^1}{\sqrt{t}}\right|{\left|\frac{1}{\sigma_{0}}-\frac{1}{\bar\sigma_t(Y)}\right|}\frac{1}{{\sqrt{2\pi}}}e^{- \half\left(\frac{z-\rho\sigma_{0}W_t^1/\sqrt{t}}{\sqrt{1-\rho^2}M}\right)^2}dz\nonumber\\
&\leq K_{2}e^{\rho\sigma_{0}W_t^1}\left(1+\left|\frac{W_t^1}{\sqrt{t}}\right|\right)\left|\frac{1}{\sigma_{0}}-\frac{1}{\bar\sigma_t(Y)}\right|,
\end{align*}
for some constant $K_{1},K_{2}$. Hence, $\bbe\left(J^{2,1}_{t}(Y_{\cdot}(\omega),Q_{\cdot}(\omega),W^1_{t}(\omega))\right)=O(t^{\half})$, because 
\begin{align}\label{finalstep}
\bbe\left({\bar\sigma_t(Y)}-\sigma_0\right)^2 & \leq\frac{1}{(2m)^2}\bbe\left({\bar\sigma_t^2(Y)}-\sigma_0^2\right)^2
\leq \frac{1}{(2m)^2}\frac{1}{t}\int_0^t\bbe\left(\sigma^2(Y_s)-\sigma^2({y_0})\right)^2ds,
\end{align}
is of order $O(t)$ by (\ref{sigma2order}). 
Finally, for $J^{2,2}_{{t}}$, we have
\begin{align}\label{J22}
\bbe\left(J^{2,2}_{t}(Y_{\cdot}(\omega),Q_{\cdot}(\omega),W^1_{t}(\omega))\right) &=
e^{-\tilde\gamma t+\kappa_t}\bbe\int_0^{\infty}{\left(e^{-\sqrt{t}z}\left(e^{\rho\sigma_{0}W_t^1-\psi^2_t(\omega)}-1\right)+\left( e^{-\sqrt{t}z}-1\right)+1\right)\int_{\frac{z-{\rho\sigma_{0}W_t^1}/{\sqrt{t}}}{\sqrt{1-\rho^2}\sigma_{0}}}^{z/\sigma_{0}}{\phi(x)dx}dz}.
\end{align}
The second term is $O(\sqrt{t})$ since $0\leq{}1-e^{-\sqrt{t}z}\leq\sqrt{t}z$ for $z>0$ and
\begin{align}\label{intbound} \sup_{0<t\leq{}1}\bbe\left|\int_0^{\infty}z\int_{\frac{z-{\rho\sigma_{0}W_t^1}/{\sqrt{t}}}{\sqrt{1-\rho^2}\sigma_{0}}}^{z/\sigma_{0}}{\phi(x)dx}dz\right|<\infty,
\end{align}
as verified in Appendix \ref{AuxPrfT2}. Cauchy's inequality can be used to show that the first term is also $O(\sqrt{t})$, since, due to the fact that $\psi_t^{2}>0$,  
\begin{align*}
\bbe\left(e^{-\psi^2_t(\omega)}e^{\rho\sigma_{0}W_t^1}-1\right)^2 
&\leq 2\bbe\left(e^{-2\psi^2_t(\omega)}\left(e^{\rho\sigma_{0}W_t^1}-1\right)^2\right) + 2\bbe\left(e^{-\psi^2_t(\omega)}-1\right)^2\\
&\leq 2\bbe\left(e^{\rho\sigma_{0}W_t^1}-1\right)^2 + 2\bbe\left(1-e^{-\psi^2_t(\omega)}\right)^2\\
&\leq 2\left(e^{2\rho^{2}\sigma_{0}^{2}t}-1\right)-4\left(e^{\frac{\rho^{2}\sigma_{0}^{2}}{2}t}-1\right)
+ 2\bbe\left(\psi^2_t(\omega)\right)^2
= O(t), \quad {t\to 0},
\end{align*}
and 
\begin{align*}
\sup_{0<t\leq{}1} \bbe\left(\int_0^{\infty}{e^{-\sqrt{t}z}\int_{\frac{z-{\rho\sigma_{0}W_t^1}/{\sqrt{t}}}{\sqrt{1-\rho^2}\sigma_{0}}}^{z/\sigma_{0}}{\phi(x)dx}dz}\right)^2 < \infty,
\end{align*} 
which can be shown similarly to (\ref{intbound}). Finally, the third term can be shown to be zero. Indeed, by Fubini's theorem, we have
\begin{align}\label{WRel}
\bbe\left(\int_0^{\infty}\int_{\frac{z-{\rho\sigma_{0}W_t^1}/{\sqrt{t}}}{\sqrt{1-\rho^2}\sigma_{0}}}^{z/\sigma_{0}}{\phi(x)dx}dz\right)  
= \int_{z=0}^{\infty}\int_{u=-\infty}^{\infty}\int_{x=\frac{z-{\rho\sigma_{0}u}}{\sqrt{1-\rho^2}\sigma_{0}}}^{z/\sigma_{0}}{\phi(u)\phi(x)dx}dudz.
\end{align}
Now, for an arbitrary $K\in (0,\infty)$, note that
\begin{align*}
\int_{z=0}^{K}\int_{u=-\infty}^{\infty}\int_{x=\frac{z-{\rho\sigma_{0}u}}{\sqrt{1-\rho^2}\sigma_{0}}}^{z/\sigma_{0}}{\phi(u)\phi(x)dx}dudz 
&= \int_{z=0}^{K}\int_{u=-\infty}^{\infty}\int_{x=-\infty}^{z/\sigma_{0}}{\phi(u)\phi(x)dx}dudz\nonumber\\ 
&\quad - \int_{z=0}^{K}\int_{u=-\infty}^{\infty}\int_{x=-\infty}^{\frac{z-{\rho\sigma_{0}u}}{\sqrt{1-\rho^2}\sigma_{0}}}{\phi(u)\phi(x)dx}dudz\nonumber\\
&= \int_{z=0}^{K}\int_{x=-\infty}^{z/\sigma_{0}}{\phi(x)dx}dz \nonumber\\
&\quad- \int_{z=0}^{K}\int_{u=-\infty}^{\infty}\int_{s=-\infty}^{{z}/(\sqrt{1-\rho^2}\sigma_{0})}\phi(u)\phi\left(s-\frac{\rho u}{\sqrt{1-\rho^2}}\right)dsdudz.
\end{align*} 
The second term in the last expression can further be manipulated as follows: 
\begin{align*}
&\int_{z=0}^{K}\int_{u=-\infty}^{\infty}\int_{s=-\infty}^{{z}/(\sqrt{1-\rho^2}\sigma_{0})}\phi(u)\phi\left(s-\frac{\rho u}{\sqrt{1-\rho^2}}\right)dsdudz\nonumber\\  &=\int_{z=0}^{K}\int_{u=-\infty}^{\infty}\int_{s=-\infty}^{{z}/\left(\sqrt{1-\rho^2}\sigma_{0}\right)}\frac{e^{-\frac{\left(u-\rho\s\sqrt{1-\rho^2}\right)^2}{2(1-\rho^2)}}}{\sqrt{2\pi(1-\rho^2)}}\sqrt{\frac{1-\rho^2}{2\pi}}e^{-\frac{\left(1-\rho^2\right)s^2}{2}}dsdudz\nonumber\\ &=\int_{z=0}^{K}\int_{s=-\infty}^{{z}/\left(\sqrt{1-\rho^2}\sigma_{0}\right)}\sqrt{\frac{1-\rho^2}{2\pi}}e^{-\frac{\left(1-\rho^2\right)s^2}{2}}dsdz=\int_{z=0}^{K}\int_{x=-\infty}^{z/\sigma_{0}}{\phi(x)dx}dz.
\end{align*}
Therefore, for any $K\in(0,\infty)$, 
\[
	\int_{z=0}^{K}\int_{u=-\infty}^{\infty}\int_{x=\frac{z-{\rho\sigma_{0}u}}{\sqrt{1-\rho^2}\sigma_{0}}}^{z/\sigma_{0}}{\phi(u)\phi(x)dx}dudz=0,
\]
which shows that the expectation in (\ref{WRel}) is equal {to} $0$.

Thus, (\ref{A3decomp})-(\ref{J22}) show that
\begin{align}\label{AsyA3}
\lim_{t\to0}\frac{\bbe\left(A_t^3(Y_{\cdot}(\omega),Q_{\cdot}(\omega),W^1_{t}(\omega))\right)}{t^{1-Y/2}}=0. 
\end{align}

\noindent 
Finally, combining (\ref{2ndOADecCGMYB}), (\ref{AsyA1}), (\ref{AsyA2}), and (\ref{AsyA3}), gives the following second order term for the option prices of $(S_t)_{t\geq{}0}$:
\begin{align*}
\lim_{t\to0}t^{Y/2-1}R_t = \frac{\theta}{2}+\frac{C(1)+C(-1)}{2Y(Y-1)}\,\sigma(y_0)^{1-Y}\,\bbe\left(|W_{1}^{1}|^{1-Y}\right)
=\frac{\theta}{2} + \frac{2^{-\frac{Y+1}{2}}}{\sqrt{\pi}}\Gamma\left(1-\frac{Y}{2}\right)\frac{C(1)+C(-1)}{Y(Y-1)}\sigma(y_0)^{1-Y}.
\end{align*}
where the last step follows from the well known formula for the centered moment of a Gaussian random variable (see, e.g., (25.6) in \cite{Sato:1999}).
This shows that the second order expansion (\ref{CL2C}) holds under condition (\ref{SigmaMm}).

\noindent\textbf{Step 3)}
We will now show that the expansion extends to the case when $\sigma(\cdot)$ is no longer assumed to satisfy (\ref{SigmaMm}). To that end, define a process $(\bar\S_t)_{t\leq{}1}$ of the form 
\[
	\bar{S}_{t}:=e^{X_t+\bar\V_t},
\] 
where $\bar\V$ is defined as in (\ref{modelV})-(\ref{modelY}), but replacing $\sigma(y)$ with $\bar\sigma(y):=\sigma(y){\bf 1}_{\{m<\sigma(y)< M\}}+m{\bf 1}_{\{\sigma(y)\leq m\}}+M{\bf 1}_{\{M\leq\sigma(y)\}}$. Here $m$ and $M$ are such that $0<m<\sigma(y_0)<M<\infty$. In that case, we know, by Step 2 above, that
\begin{align*}
\lim_{t\rightarrow{}0}t^{\frac{Y}{2}-1}\left(t^{-\frac{1}{2}}{\bbe\left(\bar\S_{t}-{e^{\kappa_{t}}}\right)^{{+}}}-d_1\right)=d_2,
\end{align*}
with $d_1$ and $d_2$ as in (\ref{d1d2}). {On the hand, using the identity $(a+b)^+\leq a^+ + |b|$, it follows that}
\begin{align*}
\bbe\left(\bar\S_t-e^{{\kappa_t}}\right)^+ - \bbe\left|{S}_t-\bar\S_t\right| \leq \bbe\left(S_t-e^{{\kappa_t}}\right)^{+} \leq 
\bbe\left(\bar\S_t-e^{{\kappa_t}}\right)^+ + \bbe\left|S_t-\bar\S_t\right|.
\end{align*}
{Therefore, for the close-to-the-money option prices under both processes $(S_t)_{t\geq{}0}$ and $(\bar\S_t)_{t\geq{}0}$ to have the same second-order term, it suffices that}
\begin{align*}
	\lim_{t\to{}0}t^{\frac{Y-3}{2}}\,\bbe\left|S_t-\bar\S_t\right|=0.
\end{align*}
To show the latter, first note that 
\[
	\bbe\left|S_t-\bar\S_t\right|=\bbe\left(e^{X_t}\right)\bbe\left|e^{V_t}-e^{\bar\V_t}\right| = \bbe\left|e^{V_t}-e^{\bar\V_t}\right|,
\]
where we have used the independence of $X$ and $V$. Since $1<Y<2$, it suffices to show that
\begin{align}\label{barVtoVcond}
\bbe\left|e^{V_t}-e^{\bar\V_t}\right| = O(t), \quad {t\to 0}.
\end{align}
Define a probability measure $\widehat\bbp$ on $(\Omega,\calF)$ by $\frac{d {\widehat\bbp}|_{\calF_{t}}}{d \bbp |_{\calF_{t}}}=e^{V_t},\, 0\leq{}t\leq{}1$. By Girsanov's theorem, the following representations hold under $\widehat\bbp$:
\begin{align*}
dV_t & = \half\sigma^2(Y_t)dt + \sigma(Y_t){\left(\rho d\widehat\W_t^1+\sqrt{1-\rho^2}d\widehat\W_t^2\right)},\\
d\bar\V_t & = -\half\bar\sigma^2(Y_t)dt + \bar\sigma(Y_t)\sigma(Y_t)dt + \bar\sigma(Y_t){\left(\rho d\widehat\W_t^1+\sqrt{1-\rho^2}d\widehat\W_t^2\right)},\\
dY_t & = (\alpha(Y_t)+\rho\sigma(Y_t)\gamma(Y_t))dt + \gamma(Y_t)d{\widehat\W_t^1}=:\widehat\alpha(Y_t) + \gamma(Y_t)d{\widehat\W_t^1},
\end{align*}
where $(\widehat\W^1_t)_{t\geq 0}$ and $(\widehat\W^2_t)_{t\geq 0}$ are independent $\widehat\bbp$-Brownian motions. Then, since $\bbe\left(e^{V_{t}}\right)=\bbe\big(e^{\bar{V}_{t}}\big)=1$,
\begin{align*}
\half\bbe\left|e^{V_t}-e^{\bar\V_t}\right| & = \bbe\left(e^{V_t}-e^{\bar\V_t}\right)_{+}-\half\bbe\left(e^{V_t}-e^{\bar\V_t}\right) = \bbe\left(e^{V_t}-e^{\bar\V_t}\right)_{+}= \widehat\bbe\left(1-e^{\bar\V_t-V_t}\right)_{+}= 
\int_0^{\infty}e^{-y}\widehat\bbp\left(V_t-\bar\V_t\geq y\right)dy.
\end{align*}
Noting that $V_t-\bar\V_t=0$ for $t< T:=\inf\{t\geq 0:\sigma(Y_t)\leq m \text{ or }\sigma(Y_t)\geq M\}$ gives: 
\begin{align*}
\half\bbe\left|e^{V_t}-e^{\bar\V_t}\right| 
 \leq \widehat\bbp\left(T\leq t\right)\int_0^{\infty}e^{-y}dy = \widehat\bbp\left(T\leq t\right).
\end{align*}
Due to the continuity of $\sigma(\cdot)$ at $y_0$, we can select $m^Y$ and $M^Y$ such that $m^Y<y_0<M^Y$ and
\begin{align*}
T \geq T^Y := \inf\{t\geq0:Y_t\leq m^Y \text{ or } Y_t\geq M^Y\},
\end{align*}
and, thus, by Lemma \ref{HittingTime},
\begin{align*}
\half\bbe\left|e^{V_t}-e^{\bar\V_t}\right| \leq \widehat\bbp\left(T^Y\leq t\right) = O(t),\quad {t\to 0}.
\end{align*}
Therefore, (\ref{barVtoVcond}) is satisfied, which in turn, as explained above, implies
\begin{align*}
\lim_{t\rightarrow{}0}t^{\frac{Y}{2}-1}\left(t^{-\frac{1}{2}}\frac{1}{S_{0}}\,\bbe\left(S_{t}-S_{0}\right)_{+}-d_1\right)=d_2,
\end{align*}
with $d_1$ and $d_2$ as in (\ref{d1d2}).
\hfill\qed

\section{Numerical Examples}\label{Sec:Examples}
In this section we perform a numerical analysis for the pure-jump CGMY model with stochastic volatility. Under the CGMY model, the L\'evy measure is given by
\begin{equation}\label{LevyMeasCGMY2}
\nu(dx)=|x|^{-Y-1}q(x)dx=|x|^{-Y-1}\left(Ce^{-Mx}\,{\bf 1}_{\{x> 0\}}+Ce^{Gx}\,{\bf 1}_{\{x<0\}}\right)dx,
\end{equation}
with corresponding parameters $C, G, M>0$ and $Y\in(1,2)$. The martingale condition (\ref{NdCndTSMrt}) implies that $M>1$, and it will be useful to note that the constants $\eta$ and $\tilde\gamma$ from Section \ref{Sec:tmpstble} can be written as (see \cite{LopGonHou:2013})
\begin{align*}
\eta = C\Gamma(-Y)\left((M-1)^{Y}+(G+1)^{Y}\right),\qquad
\tilde\gamma = -C\Gamma(-Y)\left((M-1)^{Y}+(G+1)^{Y}-M^{Y}-G^{Y}\right).
\end{align*}
For the continuous component we will consider the Heston stochastic volatility model for which the drift and volatility parameters of (\ref{modelV})-(\ref{modelY}) are given by 
\begin{align*}
	\sigma(y) = \sqrt{y}, \qquad
	\alpha(y) = \kappa(\theta-y), \qquad
	\gamma(y) = \epsilon\sqrt{y},
\end{align*}
with all coefficients strictly positive, and satisfying the Feller condition $2\kappa\theta-\epsilon^{2}>0$. 
In \cite{LopGonHou:2013} it is shown that the second order term in the pure-jump CGMY model is given by
\[
	d_{2}=\frac{C\Gamma(-Y)}{2}\left((M-1)^{Y}-M^{Y}-(G+1)^{Y}+G^{Y}\right),
\]
while in the CGMY model with a stochastic volatility component, {$C(1)=C(-1)=C$} and, thus,
\[
	d_{2}=\frac{C\sigma(y_0)^{1-Y}2^{\frac{1-Y}{2}}}{Y(Y-1)\sqrt{\pi}}\Gamma\left(1-\frac{Y}{2}\right).
\]
To estimate the option prices we will use a Monte Carlo based method. First, using the two measure transformations introduced in Section \ref{Sec:tmpstble}, we have
\begin{align}\label{MC}
\bbe\left(e^{X_t+V_t}-1\right)^+=\bbe^*\left(e^{-X_t}\left(e^{X_t+V_t}-1\right)^+\right)=\widetilde\bbe\left(e^{-U_t}\left(e^{V_t}-e^{-X_t}\right)^+\right).
\end{align}
Then, using (\ref{SSSP})-(\ref{Uplusminuseta}) and (\ref{LevyMeasCGMY2}), $U_t$ and $X_t$ can be seen to have the following representations under $\widetilde\bbp$,
\begin{align*}
U_t = (M-1)\bar\U_t^++(G+1)\bar\U_t^-+\eta t,\qquad X_t = \bar\U_t^+-\bar\U_t^-+\tilde\gamma t,
\end{align*}
where $\bar\U_t^+$ and $\bar\U_t^{-}$ are independent $Y$-stable random variables with scale, skewness, and location parameters given by $t^{1/Y}C|\cos(\pi Y/2)|\Gamma(-Y)$, $1$ and $0$, respectively. The expression (\ref{MC}) can then easily be computed by Monte Carlo, using traditional simulation schemes for the continuous component $V_t$. 
Overall, the second order approximation offers an improvement over the first order approximation, but it seems to be much more accurate in the pure-jump model than in the model with a continuous component. Figure \ref{Figure1} shows its performance for the pure-jump CGMY model {for} different values of $Y$, the index of jump activity. The second order approximation seems to improve for decreasing $Y$, i.e. decreasing jump activity. A similar sensitivity analysis for the other three CGMY parameters is carried out in \cite{LopGonHou:2013}. 

In the case of a stochastic volatility component, we have seen that the second order approximation is the same as in the case of a Brownian component, it only incorporates information on the spot volatility of the continuous component. It is also apparent that for small maturities, the simulated option prices are very similar for the models with either stochastic volatility or a Brownian component, the difference being negligible compared to the difference between the prices and the approximations. Figure \ref{Figure2} compares the first and second order approximations to the simulated option prices under a CGMY model with either Heston stochastic volatility or a Brownian component, and varying degree of jump activity. Again, \cite{LopGonHou:2013} offers a sensitivity analysis with respect to the other CGMY parameters. 

\begin{figure}
\hspace{-1 cm}
    \includegraphics[width=20.0cm,height=6.0cm]{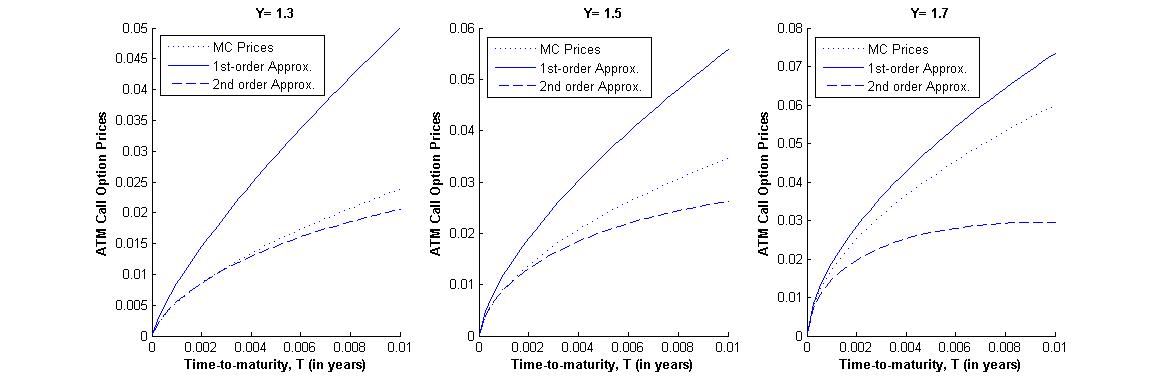}
    \caption{Comparison of the ATM call option prices computed by Monte Carlo, and the first- and second-order approximations. The model is pure-jump CGMY with parameters $C=0.5, G=2, M=3.6$.}\label{Figure1}
\end{figure}

\begin{figure}
	\hspace{-1 cm}
    \includegraphics[width=20.0cm,height=6.0cm]{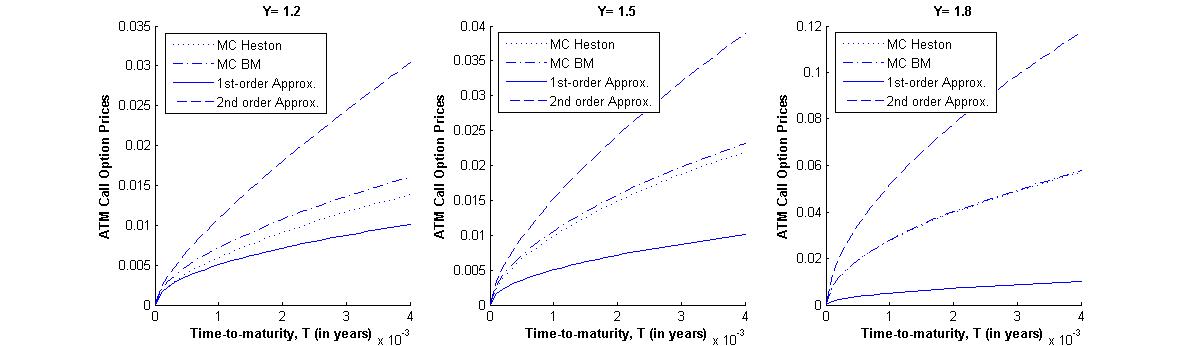}
    \caption{Comparison of the ATM call option prices computed by Monte Carlo, and the first- and second-order approximations. The pure-jump part is CGMY with parameters $C=0.5, G=2, M=3.6,$ and the continuous part is a Heston stochastic volatility model with $y_0=\theta=0.4^2,\kappa=\epsilon=1$, or a Brownian motion with $\sigma=0.4$.}\label{Figure2}
\end{figure}

\appendix

\section{Proof of Lemma \ref{2ndASYStep1}}\label{ApProf1}

We start by giving some needed technical lemmas, whose proofs are deferred to Appendix \ref{ApTecLem}. The following result shows that the conditions in (\ref{NewAssumEq}) suffice for both the change of probability measure  from $\bbp^{*}$ to $\widetilde{\bbp}$ and the representation (\ref{EMMN}) to hold true.
\begin{lem}\label{FirstLem}
Under the conditions (\ref{FirstCdn}) and (\ref{NewAssumEq}), both (\ref{ChangeMeasureCond}) and (\ref{IntCndEta}) hold true.
\end{lem}

\noindent
We will also make use of the following two lemmas, the first of which is an extension of Lemma $3.3$ in \cite{LopGonHou:2013}.
\begin{lem}\label{KLTCTBR}
Under {(\ref{NewAssumEq})}, the following two assertions hold true:
\begin{enumerate}
\item For any $v>0$,
\begin{align}
{\rm (i)}\;\;&{\lim_{t\to 0}\frac{1}{t}\widetilde{\bbp}\left(Z_{t}^{+}+\widetilde{U}_{t}\geq v\right)=\int_{\bbr_{0}}{\bf 1}_{\{{x^{-}}-\ln\bar{q}(x)\geq v\}} \tilde{\nu}(dx)},\label{SmplLmtSLJ1}\\ 
{\rm (ii)}\;\;&{\lim_{t\to 0}\frac{1}{t}\widetilde{\bbp}\left(Z_{t}^{+}+\widetilde{U}_{t}\leq -v\right)=\int_{\bbr_{0}}{\bf 1}_{\{{x^{-}}-\ln\bar{q}(x)\leq -v\}} \tilde{\nu}(dx)}\label{SmplLmtSLJ2}
\end{align}
\item There exist constants $\tilde{\kappa}<\infty$ and $t_{0}>0$ such that
\begin{equation}\label{TIITI}
{\rm (i)}\;\; \frac{1}{t}\widetilde{\bbp}\left(|\widetilde{U}_{t}|\geq v\right)\leq\tilde{\kappa}v^{-Y},\quad\quad
{{\rm (ii)}\;\;\frac{1}{t}\widetilde{\bbp}\left({|Z_{t}|}+|\widetilde{U}_{t}|\geq v\right)\leq\tilde{\kappa}v^{-Y}},
\end{equation}
for any $0<t\leq{}t_{0}$ and  $v>0$.
\end{enumerate}
\end{lem}
\begin{lem}\label{L2}
Let $(\xi_t)_{t\geq0}$ be a centered L\'evy process with a L\'evy measure $\rho$ such that $R:=\inf\{r:\rho(x:|x|>r)=0\}<\infty$. {Then, given a fixed arbitrary $k\in\bbn$}, there exist constants $\tilde\kappa<\infty$ and $v_0>0$ such that
\begin{align}\label{ExpIneqBS}
\frac{1}{t}{\bbp}\left(|\xi_t|\geq v\right)\leq \tilde\kappa e^{-kv},
\end{align} 
for any {$0<t\leq1$} and $v>v_0$.
\end{lem}
 
\noindent
We are now ready to show the result. 

\medskip
{\noindent\textbf{Proof of Lemma \ref{2ndASYStep1}.}}

\noindent
Throughout, we assume without loss of generality that $S_0=1$.
 The proof follows the arguments in \cite{LopGonHou:2013}. 
 First, we utilize the following representation obtained in \cite{CM09},
\begin{align}\label{CarrMadan}
\mathbb{E}(S_t-e^{\kappa_t})^{+}
=\int_{0}^{\infty}e^{-x}\mathbb{P}^{*}(X_{t}>\kappa_t+x)dx,
\end{align}
together with the density transformation $\bbp^{*}\to\widetilde{\bbp}$ introduced in Eq.~(\ref{EMMN}), to rewrite the scaled option price in {the form}
\begin{align*}	
t^{-1/Y}\mathbb{E}(S_t-e^{{\kappa_t}})^{+}
& = e^{{\kappa_t}-(\tilde{\gamma}+\eta)t}\int_{-\tilde{\gamma}t^{1-1/Y}+t^{-1/Y}{\kappa_t}}^{\infty}e^{-t^{1/Y}v}\,
\widetilde{\bbe}\left(e^{-\widetilde{U}_{t}}{\bf 1}_{\{t^{-1/Y}Z_{t}\geq{}v\}}\right)dv.
\end{align*}	
{Lemma $3.1$ in \cite{LopGonHou:2013} proved the formula {above} for $\kappa_t=0$. The general case is proved analogously.} Then, the error $D(t):=t^{-1/Y}\mathbb{E}(S_t-e^{{\kappa_t}})^{+}-\widetilde{\bbe}(Z_{1}^{+})$ can be decomposed as follows
\begin{align}\label{DecompD}
{\D(t)}: & = {e^{\kappa_t-(\tilde{\gamma}+\eta)t}}\left({\int_{0}^{\infty}}e^{-t^{1/Y}v}\,
\widetilde{\bbe}\left(e^{-\widetilde{U}_{t}}{\bf 1}_{\{t^{-1/Y}Z_{t}\geq{}v\}}\right)dv-\widetilde{\bbe}(Z_{1}^{+})\right)\nonumber\\ 
& \quad+ {(e^{\kappa_t-(\tilde\gamma+\eta)t}-1)}\widetilde{\bbe}(Z_{1}^{+})\nonumber\\
& \quad-e^{{\kappa_t}-(\tilde{\gamma}+\eta)t}\int^{-\tilde{\gamma}t^{1-1/Y}+t^{-1/Y}{\kappa_t}}_{0}e^{-t^{1/Y}v}\,
\widetilde{\bbe}\left(e^{-\widetilde{U}_{t}}{\bf 1}_{\{t^{-1/Y}Z_{t}\geq{}v\}}\right)dv\nonumber\\
& =:\D_1(t) + \D_2(t) - \D_3(t).
\end{align}
Using that $\kappa_{t}=\theta t+o(t)$ and $Y>1$, it is easy to see that 
 \begin{equation}\label{DD2}
 	t^{\frac{1}{Y}-1}D_2(t)=o(1),\qquad t\to{}0.
\end{equation}
To handle the term $D_{3}(t)$, we proceed as in Lemma {A.1.} in \cite{LopGonHou:2013}. Concretely, using the density transformation $\widetilde{\bbp}\to\bbp^{*}$ and then change of variables $u=t^{1/Y-1}v$,
\begin{align}\label{AsympD3}\nonumber
t^{\frac{1}{Y}-1}D_3(t) 
&=e^{\kappa_t-(\tilde{\gamma}+\eta)t}\int_{0}^{-\tilde{\gamma}+t^{-1}\kappa_t}e^{-ut}\,
\bbp^{*}\left(t^{-1/Y}Z_{t}\geq{}t^{1-1/Y}u\right)du\\
& \longrightarrow\, \left(-\tilde\gamma+\theta\right)\bbp^*(\widetilde\Z\geq 0)=\left(-\tilde\gamma+\theta\right)\widetilde\bbp(Z_1\geq 0), 
\end{align}
as $t\to 0$, since $t^{-1}\kappa_{t}\to{}\theta$, $t^{1-1/Y}u\to{}0$, and $t^{-1/Y}Z_{t}\to \widetilde\Z$ in distribution under $\bbp^*$, where $\widetilde\Z$ is a centered $Y$-stable random variable (see Proposition $1$ in \cite{rosenbaum.tankov.10}), which is the same as the distribution of $Z_1$ under $\widetilde\bbp$.
For the first term {in (\ref{DecompD})}, we further decompose it as follows:
\begin{align}\label{DecompD1}
D_1(t)&=\int_{0}^{\infty}e^{-t^{1/Y}v}\widetilde{\bbe}\left(e^{-\widetilde{U}_{t}}{\bf 1}_{\{t^{-1/Y}Z_{t}\geq{}v\}}\right)dv-\widetilde{\bbe}(Z_{1}^{+})\nonumber\\
&\quad\,+\left({e^{\kappa_t-(\tilde{\gamma}+\eta)t}}-1\right)\widetilde{\bbe}(Z_{1}^{+})\nonumber\\
&\quad\,+({e^{\kappa_t-(\tilde{\gamma}+\eta)t}}-1)\left(\int_{0}^{\infty}e^{-t^{1/Y}v}\widetilde{\bbe}\left(e^{-\widetilde{U}_{t}}{\bf 1}_{\{t^{-1/Y}Z_{t}\geq{}v\}}\right)dv-{\widetilde{\bbe}}(Z_{1}^{+})\right)\nonumber\\
&=:\bar\D_{1}(t)+\bar\D_{2}(t)+\bar\D_{3}(t), 
\end{align}
where it is clear that $\bar\D_{3}(t)=o(\bar\D_{1}(t))$ and that $t^{1/Y-1}\bar\D_{2}(t)=o(1)$, as $t\to{}0$. Next, using Fubini's theorem as well as the identities $\widetilde\bbe\big(\widetilde{U}_{t}\big)=0$, $t^{-1/Y}\widetilde\bbe\left(Z^{+}_t\right)=\widetilde\bbe\left(Z^{+}_1\right)$, and $\widetilde\bbe\big(e^{-\widetilde{U}_{t}}\big)=e^{\eta t}$,
\begin{align}\label{DecomD12}
t^{1/Y-1}\bar\D_{1}(t)&=t^{1/Y-1}\left(\frac{e^{\eta t}-1}{t^{1/Y}}+\frac{1-\widetilde{\bbe}\left(e^{-(Z_{t}^{+}+\widetilde{U}_{t})}\right)-{\widetilde{\bbe}}\left(Z_{t}^{+}+\widetilde{U}_{t}\right)}{t^{1/Y}}\right)\nonumber\\
&=\frac{e^{\eta t}-1}{t}+\frac{1}{t}\widetilde{\bbe}\left(\int_{0}^{Z_{t}^{+}+\widetilde{U}_{t}}\left(e^{-v}-1\right)dv{\bf 1}_{\{Z_{t}^{+}+\widetilde{U}_{t}\geq 0\}}\right)-\frac{1}{t}\widetilde{\bbe}\left(\int_{Z_{t}^{+}+\widetilde{U}_{t}}^{0}\left(e^{-v}-1\right)dv{\bf 1}_{\{Z_{t}^{+}+\widetilde{U}_{t}\leq 0\}}\right)\nonumber\\
&=\frac{e^{\eta t}-1}{t}+{\frac{1}{t}\int_{0}^{\infty}\left(e^{-v}-1\right)\widetilde{\bbp}(Z_{t}^{+}+\widetilde{U}_{t}\geq v)dv}-{\frac{1}{t}\int_{0}^{\infty}\left(e^{v}-1\right)\widetilde{\bbp}(Z_{t}^{+}+\widetilde{U}_{t}\leq -v)dv}\nonumber\\
&=: {\bar{D}_{11}(t)+\bar{D}_{12}(t)-\bar{D}_{13}(t)}, 
\end{align}
where {clearly $\bar{D}_{11}(t)\rightarrow\eta$, as $t\to 0$}.
For $\bar{D}_{12}$, (\ref{TIITI}-ii) allows to pass the limit inside the integral, {and so (\ref{SmplLmtSLJ1}) implies that}
\begin{align}\label{D12Domi}
\lim_{t\to 0}\bar{D}_{12}(t)=\int_{0}^{\infty}(e^{-v}-1)\int_{\bbr_{0}}{\bf 1}_{\{x^{-}-\ln\bar{q}(x)\geq v\}}\tilde{\nu}(dx)dv=:\vartheta^1.
\end{align}
{In \cite{LopGonHou:2013}, it was proved that $\lim_{t\to{}0}\bar{D}_{13}(t)=0$ under the assumptions in (\ref{Eq:StndCnd1a}). The idea therein was to change variable to $u=t^{-1/Y}v$ and then dominate the resulting probability inside the integral with $u^{-1}\widetilde{\bbe}\left(\exp\left(-t^{-1/Y}(Z_{t}+\widetilde{U}_{t})\right)\right)$, which is $O(u^{-1})$ as $t\to{}0$ under the conditions in (\ref{Eq:StndCnd1a}). However, it turns out that if (\ref{Eq:StndCnd1a}-iii) does not hold, then $\widetilde{\bbe}\left(\exp\left(-t^{-1/Y}(Z_{t}+\widetilde{U}_{t})\right)\right)$ will {diverge as $t\to{}0$}. Instead, here we justify that the limit can be passed into the integration so that, in light of (\ref{SmplLmtSLJ2}),}
\begin{align}\label{D13Domi}
\lim_{t\to 0}{\bar{D}_{13}(t)}=\int_{0}^{\infty}\left(e^{v}-1\right)\lim_{t\to 0}\frac{1}{t}\widetilde{\bbp}(Z_{t}^{+}+\widetilde{U}_{t}\leq -v)dv=\int_{0}^{\infty}(e^{v}-1)\int_{\bbr_{0}}{\bf 1}_{\{x^{-}-\ln\bar{q}(x)\leq -v\}}\tilde{\nu}(dx)dv=:\vartheta^2.
\end{align}
Note that $\vartheta^2=0$ under (\ref{Eq:StndCnd1a}-iii)-(\ref{Eq:StndCnd1a}-iv) since $x^{-}-\ln\bar{q}(x)=-\ln\bar{q}(x)\geq x>0$ for $x>0$ and $x^{-}-\ln\bar{q}(x)=-x-\ln\bar{q}(x)\geq -x>0$ for $x<0$. 

\noindent 
To show that the dominated convergence theorem can be applied in (\ref{D13Domi}), fix $v_0>0$ and split $D_{13}(t)$ into the sum of two integrals
\begin{align*}
\bar{D}_{13}(t)={\int_{0}^{v_0}\left(e^{v}-1\right)\frac{1}{t}\widetilde{\bbp}(Z_{t}^{+}+\widetilde{U}_{t}\leq -v)dv}
+{\int_{v_0}^{\infty}\left(e^{v}-1\right)\frac{1}{t}\widetilde{\bbp}(Z_{t}^{+}+\widetilde{U}_{t}\leq -v)dv}
=:\bar{D}_{13}^1(t)+\bar{D}_{13}^2(t).
\end{align*}
For the term $\bar{D}_{13}^1(t)$, note that, by (\ref{TIITI}-ii), {for $t\leq t_0$}
\begin{align}\label{D13DC1}
{\frac{1}{t}\left(e^v-1\right)\widetilde{\bbp}(Z_t^+ + \widetilde U_t \leq-v)} \leq \tilde\kappa{\left(e^v-1\right)v^{-Y}\in \bbl^{1}((0,v_{0}))}.
\end{align}
For $\bar{D}_{13}^2(t)$, let
\begin{align*}
Q_t := Z_t+\widetilde U_t = \int_0^t\int(-\ln\bar q(x))\bar\N(ds,dx),
\end{align*}
and fix $\varepsilon>0$ to define
\begin{align*}
Q_t^{(\varepsilon)} := Q_t - \int_0^t\int_{\{-\ln\bar\q(x)>\varepsilon\}}(-\ln\bar q(x))N(ds,dx)=
{\int_0^t\int_{-\ln \bar{q}(x)\leq \varepsilon}(-\ln\bar q(x))\bar\N(ds,dx)- t\int_{-\ln \bar{q}(x)> \varepsilon}(-\ln\bar q(x))\tilde{\nu}(dx)},
\end{align*}
which has bounded jumps, because $\bar\q(x)$ is bounded. Also note that $Q_{t}\geq{}Q^{(\varepsilon)}_{t}$ and $-\infty<\mu:=\widetilde\bbe\big(Q_1^{(\varepsilon)}\big)<\widetilde\bbe\big(Q_t^{(\varepsilon)}\big)<0$, for $0<t\leq 1$. {Using these identities and Lemma \ref{L2}, we can select $v_0$ such that
\begin{align}\label{D13DC2}
	\frac{1}{t}\left(e^{v}-1\right)\widetilde{\bbp}(Z_{t}^{+}+\widetilde{U}_{t}\leq -v)&\leq\frac{1}{t}\left(e^{v}-1\right)\widetilde{\bbp}\left(Q_t\leq-v\right)\nonumber \\
	&\leq \frac{1}{t}
	\left(e^{v}-1\right)\widetilde{\bbp}\left(Q_t^{(\varepsilon)}-\widetilde\bbe\, Q_t^{(\varepsilon)}\leq -v-\mu\right)\\
	&\leq 
	\tilde\kappa e^{-2\mu} (e^v-1)e^{-2v}\in \bbl^{1}([v_{0},\infty)).\nonumber 
\end{align}
Together} (\ref{D13DC1}) and (\ref{D13DC2}) justify the use of the dominated convergence theorem in (\ref{D13Domi}). Combining (\ref{DecompD})-(\ref{D13Domi}), gives {{
\begin{equation}\label{DExLmN}
	{\lim_{t\to{}0}t^{1/Y-1}\left(t^{-\frac{1}{Y}}\mathbb{E}(S_t-e^{K_t})^{+}-\widetilde{\bbe}(Z_{1}^{+})\right)={\left(\tilde\gamma-\theta\right)}\widetilde\bbp(Z_1\geq 0) + {\vartheta^1}-\vartheta^2+\eta=:{(\tilde\gamma-\theta)}\widetilde\bbp(Z_1\geq 0) + \vartheta+\eta}.
\end{equation}}}
Finally, applying Fubini's theorem to the right-hand sides of (\ref{D12Domi}) and (\ref{D13Domi}) gives
\begin{align*}
\vartheta &={C(1)}\int_{0}^{\infty}\left(\int_{0}^{{(-\ln\bar{q}(x))\vee 0}}(e^{-v}-1)dv-\int_{0}^{{\ln\bar{q}(x)\vee0}}(e^{v}-1)dv\right)x^{-Y-1}dx\\
&\quad+ {C(-1)}\int_{-\infty}^{0}\left(\int_{0}^{{(-x-\ln\bar{q}(x))\vee{}0}}(e^{-v}-1)dv-\int_{0}^{{(x+\ln\bar{q}(x))\vee0}}(e^{v}-1)dv\right) x^{-Y-1}dx\\
&={C(1)}\int_{0}^{\infty}\left(1-e^{\ln\bar{q}(x)}+\ln\bar{q}(x)\right)x^{-Y-1}dx+{C(-1)}\int_{-\infty}^{0}\left(1-e^{x+\ln\bar{q}(x)}+x+\ln\bar{q}(x)\right)|x|^{-Y-1}dx.
\end{align*}
One can similarly show that the constant $\eta$ defined in (\ref{Uplusminuseta}) can be written as:
\begin{align*}
	\eta=
	{C(1)}\int_{0}^{\infty}\left(e^{x+\ln \bar{q}(x)}-1-\ln\bar{q}(x)-x\right)x^{-Y-1}dx+{C(-1)}\int_{-\infty}^{0}\left(e^{x+\ln \bar{q}(x)}-1-\ln\bar{q}(x)-x\right)|x|^{-Y-1}dx.
\end{align*}
Combining the {expressions} for $\vartheta$ and $\eta$ yields (\ref{vartheta}). The expression for $\tilde{\gamma}$ in (\ref{Deftildegamma}) follows from
\[
	\tilde\gamma= \widetilde{\bbe} X_{1}=\tilde{b}+\int_{\{|x|>{}1\}} x\tilde{\nu}(dx)=b^{*}+\int_{|x|\leq 1}x(\tilde{\nu}-\nu^{*})(dx)+\int_{\{|x|>1\}}x\tilde{\nu}(dx),
\]
and standard simplifications. \hfill\qed

\section{Proof of auxiliary results related to Theorem \ref{2ndASYSV}}\label{AuxPrfT2}

{\noindent\textbf{Proof of (\ref{AsyInJ12}).}}

\noindent 
{By Fubini's Theorem and the terminology in (\ref{DfnPsis}), we can write
\[
	t^{Y/2-1}\bbe\left(I_1^2(t, Y_{\cdot}(\omega),Q_{\cdot}(\omega),W^1_{t}(\omega))\right) = e^{-(\eta+\tilde\gamma)t+\kappa_{t}}
	\int_{0}^{\infty} t^{Y/2-1}J_{12}(t,u)e^{-\sqrt{t}u}\bbe\left(\Xi_{t}(u,\omega)\right)du,
\]
where
\[	
	\Xi_{t}(u,\omega)=e^{-\frac{1-\rho^{2}}{2}t\bar\sigma_{t}^{2}(Y_{\cdot}(\omega))}
	e^{\rho\sigma_{0}W_{t}^{1}(\omega)}\frac{e^{-\frac{\left(u-\rho\sigma_{0}t^{-1/2}W_{t}^{1}(\omega)\right)^2}{2(1-\rho^2)\bar{\sigma}_t^2(Y_{\cdot}(\omega))}}}{\sqrt{2\pi(1-\rho^2)\bar{\sigma}_t^2(Y_{\cdot}(\omega))}}.
\]
We first show that 
\begin{equation}\label{LmtExp}
	\lim_{t\to{}0}\bbe\left(\Xi_{t}(u,\omega)\right)= \bbe\left(\frac{e^{-\frac{\left(u-\rho\sigma_{0}\Lambda\right)^2}{2(1-\rho^2){\sigma}_0^2}}}{\sqrt{2\pi(1-\rho^2){\sigma}_0^2}}\right)= \frac{e^{-\frac{u^2}{2{\sigma}_0^2}}}{\sqrt{2\pi{\sigma}_0^2}},
\end{equation}
where $\Lambda\sim\mathcal{N}(0,1)$. Indeed, using (\ref{SmplDisV1})-(\ref{limbarsigma}) together with Slutsky's theorem and the continuous mapping theorem,
\[
	\Xi_{t}(u,\omega)\ld\frac{e^{-\frac{\left(u-\rho\sigma_{0}\Lambda\right)^2}{2(1-\rho^2){\sigma}_0^2}}}{\sqrt{2\pi(1-\rho^2){\sigma}_0^2}},
\]
as $t\to{}{}0$. It is also easy to see that the collection $\left(\Xi_{t}(u,\omega)\right)_{0<t\leq{}1}$ is uniformly integrable since, due to (\ref{BndBarSigma}), there exists a constant $K$ such that 
$\Xi_{t}(u,\omega)\leq{}K\exp(\rho\sigma_{0}W^{1}_{t}(\omega))$ and clearly $\sup_{0<t\leq{}1}\bbe\left(e^{2\rho\sigma_{0}W^{1}_{t}}\right)<\infty$. Therefore, we conclude the first equality in (\ref{LmtExp}). The second equality therein holds since
\begin{align}\label{CmplSqrArg}
{\bbe}\left(\frac{e^{-\frac{\left(u-\rho\sigma_{0}\Lambda\right)^2}{2(1-\rho^2){\sigma}_0^2}}}{\sqrt{2\pi(1-\rho^2){\sigma}_0^2}}\right)
= \int_{\bbr}\frac{e^{-\frac{\left(u-x\right)^2}{2(1-\rho^2){\sigma}_0^2}}}{\sqrt{2\pi(1-\rho^2){\sigma}_0^2}}\frac{e^{-\frac{x^2}{2\rho^2\sigma_0^2}}}{\sqrt{2\pi\rho^2\sigma_0^2}}dx
= \frac{e^{-\frac{u^2}{2{\sigma}_0^2}}}{\sqrt{2\pi{\sigma}_0^2}},
\end{align}
by completing the square. Next, we show that 
\begin{equation}\label{SmplRelNd}
	\lim_{t\to{}0}\int_{0}^{\infty} t^{Y/2-1}J_{12}(t,u)e^{-\sqrt{t}u}\bbe\left(\Xi_{t}(u,\omega)\right)du=
	\int_{0}^{\infty} \lim_{t\to{}0}t^{Y/2-1}J_{12}(t,u)e^{-\sqrt{t}u}\bbe\left(\Xi_{t}(u,\omega)\right)du.
\end{equation}
suffices to show
{By the dominated convergence theorem, together with (\ref{B6FromLP}-i),} it suffices to show the existence of a bound $B_{t}(u)$ such that $\bbe\left(\Xi_{t}(u,\omega)\right)\leq{}B_{t}(u)$ and 
\[
	\lim_{t\to{}0}\int_{0}^{\infty} u^{1-Y}B_{t}(u)du=
	\int_{0}^{\infty} \lim_{t\to{}0}u^{1-Y}B_{t}(u)du<\infty. 
\]
To this end, note that, in view of (\ref{BndBarSigma}), there exists a constant $K$ such that 
\[
	\Xi_{t}(u,\omega)\leq K
	e^{\rho\sigma_{0}W_{t}^{1}(\omega)}e^{-\frac{1}{2(1-\rho^2)M^{2}}\left(u-t^{-1/2}\rho\sigma_{0}W_{t}^{1}(\omega)\right)^2}.
\]
Hence, by Cauchy's inequality, 
\[
	\bbe\left(\Xi_{t}(u,\omega)\right)\leq Ke^{\rho^{2}\sigma^{2}_{0}t}
	\bbe\left(e^{-\frac{1}{(1-\rho^2)M^{2}}\left(u-\rho\sigma_{0}\Lambda\right)^2}\right)^{\frac{1}{2}}\leq 
	K_{1}e^{\rho^{2}\sigma^{2}_{0}t}e^{-K_{2}u^{2}},
\]
for some positive constants $K_{1},K_{2}$, where the last inequality above can be obtained by a procedure similar to that leading to (\ref{CmplSqrArg}).
 Finally, together (\ref{LmtExp}), (\ref{SmplRelNd}), and (\ref{B6FromLP}-ii) implies (\ref{AsyInJ12}). 
\hfill\qed

{\noindent\textbf{Proof of (\ref{RmdTI1}).}}

\noindent 
Below, $\tilde{\kappa}$ denotes a generic constant whose value may change from line to line. Let us first note the decomposition
\begin{align}\nonumber
J_{11}(t,u)&=t^{-\frac{1}{2}}\widetilde{\bbe}\left({\bf 1}_{\{\breve{Z}_{t}\geq 0\}}\int_{\widetilde{U}_{t}}^{\widetilde{U}_{t}+\breve{Z}_{t}}\left(e^{-x}-1\right)dx\right)-t^{-\frac{1}{2}}\widetilde{\bbe}\left({\bf 1}_{\{-t^{\frac{1}{2}}u\leq \breve{Z}_{t}\leq 0\}}\int_{\widetilde{U}_{t}+\breve{Z}_{t}}^{\widetilde{U}_{t}}\left(e^{-x}-1\right)dx\right)\nonumber\\
&=t^{-\frac{1}{2}}\int_{\bbr}\left(e^{-x}-1\right)T_{1}(t,x)dx -t^{-\frac{1}{2}}\int_{\bbr}\left(e^{-x}-1\right)T_{2}(t,x,u)dx,\label{DcIntNd}
\end{align}
where, for $t>0$ and $u>0$, we set 
\begin{align}\label{T12}
T_{1}(t,x):=\widetilde{\bbp}\left(\breve{Z}_{t}\geq 0,\widetilde{U}_{t}\leq x\leq\widetilde{U}_{t}+\breve{Z}_{t}\right),\quad T_{2}(t,x,u):=\widetilde{\bbp}\left(-t^{\frac{1}{2}}u\leq \breve{Z}_{t}\leq 0,\widetilde{U}_{t}+\breve{Z}_{t}\leq x\leq\widetilde{U}_{t}\right).
\end{align}
For the first integral in (\ref{DcIntNd}) on the domain of integration $x\in (0,\infty)$, we use (\ref{TIITI}-ii) to conclude that $T_1(t,x)\leq\tilde\kappa t x^{-Y}$, for $t\leq t_0<1$. Hence, recalling the notation introduced in (\ref{DfnPsis}), (\ref{WeirdNotation}), and (\ref{WeirdNotation2}) and using (\ref{BndBarSigma}), we have: 
\begin{align}\label{AsyInJ11}
	0 & \leq  t^{\frac{Y}{2}-1}\bbe\left(e^{-\tilde\eta_{t}(\omega)+\breve{V}_{t}^{1}(\omega)}\int_0^{\infty}{\int_0^\infty{\frac{1-e^{-x}}{\sqrt{t}}T_1(t,x)dx}e^{-\sqrt{t}u}\frac{e^{-\frac{\left(u-t^{-1/2}(\breve{V}_{t}^{1}(\omega)-\psi_{t}^{1}(\omega))\right)^2}{2(1-\rho^2)\bar{\sigma}_t^2(Y(\omega))}}}{\sqrt{2\pi(1-\rho^2)\bar{\sigma}_t^2(Y(\omega))}}du}\right) \nonumber\\
	& =  e^{-(\eta+\tilde\gamma)t+\kappa_{t}}t^{\frac{Y}{2}-1}\bbe\left(e^{-\frac{1-\rho^{2}}{2}t\bar\sigma_{t}^{2}(Y(\omega))}e^{\rho\sigma_{0}W_{t}^{1}(\omega)}\int_0^{\infty}{\int_0^\infty{\frac{1-e^{-x}}{\sqrt{t}}T_1(t,x)dx}e^{-\sqrt{t}u}\frac{e^{-\frac{\left(u-t^{-1/2}\rho\sigma_{0}W_{t}^{1}(\omega)\right)^2}{2(1-\rho^2)\bar{\sigma}_t^2(Y(\omega))}}}{\sqrt{2\pi(1-\rho^2)\bar{\sigma}_t^2(Y(\omega))}}du}\right) \nonumber\\
	& \leq  \tilde\kappa t^{\frac{Y-1}{2}}\bbe\left(e^{\rho\sigma_{0}W_{t}^{1}(\omega)}\int_0^{\infty}{\int_0^\infty{(1-e^{-x})x^{-Y}dx}\frac{e^{-\frac{\left(u-t^{-1/2}\rho\sigma_{0}W_{t}^{1}(\omega)\right)^2}{2(1-\rho^2)\bar{\sigma}_t^2(Y(\omega))}}}{\sqrt{2(1-\rho^2)\pi\bar{\sigma}_t^2(Y(\omega))}}du}\right) \nonumber\\
	& \leq{}  \tilde\kappa t^{\frac{Y-1}{2}}\int_0^\infty{(1-e^{-x})x^{-Y}dx}\stackrel{t\to{}0}{\longrightarrow} 0,
\end{align}
where for the last inequality we used that the integral with respect to $u$ is bounded by $1$ and that $\sup_{0<t\leq{}1}\bbe\left( e^{\rho\sigma_{0}W_{t}^{1}}\right)<\infty$.
The second integral in (\ref{DcIntNd}) on the domain of integration $x\in (0,\infty)$ can be handled similarly. For the first integral in (\ref{DcIntNd}) on the domain of integration $x\in (-\infty,0)$, note that \begin{align}
0 & \leq  t^{\frac{Y}{2}-1}\bbe\left(e^{-\tilde\eta_{t}(\omega)+\breve{V}_{t}^{1}(\omega)}\int_0^{\infty}{\int_{-\infty}^{0}{\frac{e^{-x}-1}{\sqrt{t}}T_1(t,x)dx}e^{-\sqrt{t}u}\frac{e^{-\frac{\left(u-t^{-1/2}(\breve{V}_{t}^{1}(\omega)-\psi_{t}^{1}(\omega))\right)^2}{2(1-\rho^2)\bar{\sigma}_t^2(Y(\omega))}}}{\sqrt{2\pi(1-\rho^2)\bar{\sigma}_t^2(Y(\omega))}}du}\right) \nonumber\\
 &= e^{-(\eta+\tilde\gamma)t+\kappa_{t}}t^{\frac{Y}{2}-1}\bbe\left(e^{-\frac{1-\rho^{2}}{2}t\bar\sigma_{t}^{2}(Y(\omega))}e^{\rho\sigma_{0}W_{t}^{1}(\omega)} \int_{0}^{\infty}\int_{-\infty}^{0}\frac{e^{-x}-1}{\sqrt{t}}T_{1}(t,x)dxe^{-\sqrt{t}u}\frac{e^{-\frac{\left(u-t^{-1/2}\rho\sigma(y_0)W_t^1(\omega)\right)^2}{2(1-\rho^2)\bar{\sigma}_t^2(Y(\omega))}}}{\sqrt{2\pi(1-\rho^2)\bar{\sigma}_t^2(Y(\omega))}}du\right)\nonumber\\
 & \leq\tilde{\kappa}t^{\frac{Y-1}{2}}\int_{-\infty}^0\left(e^{-x}-1\right)\frac{1}{t}\widetilde\bbp\left(\widetilde U_t\leq x\right)dx\,\stackrel{t\to{}0}{\longrightarrow}\, 0.
\label{AsyDecomJ113}
\end{align}
To see the validity of the last limit,
fix $x_0>0$ and split the integral therein, which we denote $V(t)$, into two parts
\[
	V(t)=\int_{-\infty}^{-x_0}\left(e^{-x}-1\right)\frac{1}{t}\widetilde\bbp\left(\widetilde U_t\leq x\right)dx 
+\int_{-x_0}^0\left(e^{-x}-1\right)\frac{1}{t}\widetilde\bbp\left(\widetilde U_t\leq x\right)dx=:V_1(t)+V_2(t).
\]
By virtue of (\ref{TIITI}-ii), there exist $\tilde\kappa<\infty$ and $t_0>0$ such that 
\begin{align}\label{T1DC1}
V_2(t) = \int_{-x_0}^0\left(e^{-x}-1\right)\frac{1}{t}\widetilde\bbp\left(\widetilde U_t\leq x\right)dx
\leq \tilde\kappa\int_{-x_0}^0\left(e^{-x}-1\right)x^{-Y}dx < \infty,
\end{align}
for all $0<t<t_0$. For $V_1(t)$, fix an $\varepsilon>0$ and write
\begin{align*} \widetilde\U_t=\int_0^t\int_{|x|\leq\varepsilon}\varphi(x)\bar{N}(ds,dx)
+\int_0^t\int_{|x|>\varepsilon}\varphi(x)\bar{N}(ds,dx)
=:\widetilde\U_t^1+\widetilde\U_t^2,
\end{align*} 
where $N$ is the jump measure of $\breve{X}$. Note that
\begin{equation}\label{UpBndLm}
V_1(t) \leq \int_{-\infty}^{x_0}\left(e^{-x}-1\right)\frac{1}{t}\widetilde\bbp\left(\widetilde U_t^1\leq x/2\right)dx
+ \int_{-\infty}^{x_0}\left(e^{-x}-1\right)\frac{1}{t}\widetilde\bbp\left(\widetilde U_t^2\leq x/2\right)dx.
\end{equation}
For the first integral, (\ref{ExpIneqBS}) allows to select $x_0$ such that
\begin{align}\label{T1DC2}
\int_{-\infty}^{x_0}\left(e^{-x}-1\right)\frac{1}{t}\widetilde\bbp\left(\widetilde U_t^1\leq x/2\right)dx
\leq \tilde\kappa\int_{-\infty}^{x_0}\left(e^{-x}-1\right)e^{2x}dx<\infty,
\end{align} 
for all $0<t\leq 1$, where $\tilde\kappa<\infty$ is a constant. For the second integral, note that $ \widetilde\U_t^{2}=\alpha t+\sum_{i=1}^{N_{t}^{(\varepsilon)}}{\xi_i^{\varepsilon}}$, where $\alpha\in\bbr$, and $(N_{t}^{(\varepsilon)})_{t\geq{}0}$ and $(\xi_{i}^{(\varepsilon)})_{i\geq{}1}$ are the counting process and jump sizes of $(\widetilde\U_t^2)_{t\geq{}0}$. Denote by $\lambda_{\varepsilon}:=\bbe(N_{1}^{(\varepsilon)})$ the jump intensity. Then, by conditioning on $N_{t}^{(\varepsilon)}$, we can find $0<t_0<1$ such that, for all $t\leq t_0$,
\begin{align}
\int_{-\infty}^{x_0}\left(e^{-x}-1\right)\frac{1}{t}\widetilde\bbp\left(\widetilde U_t^2\leq x/2\right)dx
	& = \int_{-\infty}^{x_0}{(e^{-x}-1)\frac{1}{t}\sum_{k=1}^{\infty}e^{-\lambda_{\varepsilon} t}\frac{(\lambda_{\varepsilon} t)^k}{k!}\tilde{\bbp}\left(\alpha t + \sum_{i=1}^k{\xi_i^{(\varepsilon)}}\leq \frac{x}{2}\right)dx} \nonumber\\
	& \leq \int_{-\infty}^{x_0}{e^{-x}\sum_{k=1}^{\infty}\frac{\lambda_{\varepsilon}^k}{k!}\tilde\bbe[e^{-4\sum_{i=1}^k{\xi_i^{(\varepsilon)}}}]e^{2x+4|\alpha| }dx} \nonumber \\
	& \leq \tilde\kappa\int_{-\infty}^{x_0}{e^{x}e^{\lambda_{\varepsilon}\mu}dx} < \infty,
\label{T1DC3}
\end{align}
where we have employed Markov's inequality, and $0 \leq \mu:=\tilde\bbe(e^{-4{\xi_1^{(\varepsilon)}}}) < \infty$, because
\begin{align*}
\tilde\bbe[e^{-4\xi_1^{(\varepsilon)}}]
=\frac{1}{\lambda_{\varepsilon}}\int_{|x|>\varepsilon}{e^{-4\varphi(x)}\tilde\nu(dx)}
= \frac{C(1)}{\lambda_{\varepsilon}}\int_{\varepsilon}^\infty{e^{4x}(\bar{q}(x))^{4}x^{-Y-1}dx}
+ \frac{C(-1)}{\lambda_{\varepsilon}}\int_{-\infty}^{-\varepsilon}{e^{4x}(\bar{q}(x))^4|x|^{-Y-1}dx}
 < \infty,
\end{align*}
the integrals being finite for the L\'evy density in (\ref{LevyX}). Combined, (\ref{T1DC1})-(\ref{T1DC3}), imply the limit in (\ref{AsyDecomJ113}) since $1<Y<2$.
Finally, for the second integral in (\ref{DcIntNd}) on the domain of integration $x\in (-\infty,0)$, note that \begin{align}
0 & \leq  t^{\frac{Y}{2}-1}\bbe\left(e^{-\tilde\eta_{t}(\omega)+\breve{V}_{t}^{1}(\omega)}\int_0^{\infty}{\int_{-\infty}^{0}{\frac{e^{-x}-1}{\sqrt{t}}T_2(t,x)dx}e^{-\sqrt{t}u}\frac{e^{-\frac{\left(u-t^{-1/2}(\breve{V}_{t}^{1}(\omega)-\psi_{t}^{1}(\omega))\right)^2}{2(1-\rho^2)\bar{\sigma}_t^2(Y(\omega))}}}{\sqrt{2\pi(1-\rho^2)\bar{\sigma}_t^2(Y(\omega))}}du}\right) \nonumber\\
 & = {e^{-(\eta+\tilde\gamma)t+\kappa_{t}}t^{\frac{Y}{2}-1}\bbe\left(e^{-\frac{1-\rho^{2}}{2}t\bar\sigma_{t}^{2}(Y(\omega))}e^{\rho\sigma_{0}W_{t}^{1}(\omega)}\int_{0}^{\infty}\int_{-\infty}^{0}\frac{e^{-x}-1}{\sqrt{t}}T_{2}(t,x,u)dxe^{-\sqrt{t}u}\frac{e^{-\frac{\left(u-t^{-1/2}\rho\sigma(y_0)W_t^1(\omega)\right)^2}{2(1-\rho^2)\bar{\sigma}_t^2(Y(\omega)}}}{\sqrt{2\pi(1-\rho^2){\bar\sigma_t^2(Y(\omega))}}}du\right)}\nonumber\\
& \leq\tilde{\kappa}t^{\frac{Y-1}{2}}\int_{-\infty}^0\left(e^{-x}-1\right)\frac{1}{t}\widetilde\bbp(\widetilde U_t+Z_t\leq x)dx
\stackrel{t\to{}0}{\longrightarrow} 0,
\label{AsyDecomJ114}
\end{align}
since the dominated convergence theorem can be applied as in (\ref{D13Domi}), and $Y>1$.
\hfill\qed

\medskip
{\noindent\textbf{Proof of (\ref{AsyI2first}).}}

\noindent Let us start by decomposing the expectation appearing in the term $I_2^1$ defined in (\ref{DecomI2}) as follows:
\begin{align}\label{DocomFirIntI2}
\left|\widetilde{\bbe}\left(\left(e^{-\widetilde{U}_{t}}-1\right){\bf 1}_{\{\breve\Z_{t}\leq-t^{\frac{1}{2}}u\}}\right)\right|
& = \left|\widetilde{\bbe}\left(\left(e^{-\widetilde{U}_{t}}-1+\widetilde\U_t\right){\bf 1}_{\{\breve\Z_{t}\leq-t^{\frac{1}{2}}u\}}\right)
- \widetilde{\bbe}\left(\widetilde\U_t{\bf 1}_{\{\breve\Z_{t}\leq-t^{\frac{1}{2}}u\}}\right)\right|\nonumber\\
& \leq \widetilde{\bbe}\left(e^{-\widetilde{U}_{t}}-1+\widetilde\U_t\right) + \widetilde{\bbe}\left(|\widetilde\U_t|\right)\nonumber\\
& =: J_{21}(t)+J_{22}(t).
\end{align}
For $J_{21}(t)$, there exist $\tilde\kappa<\infty$ and $t_0>0$ such that,
\begin{align}\label{DocomFirIntI22}
J_{21}(t) = \widetilde{\bbe}\left(e^{-\widetilde{U}_{t}}-1+\widetilde\U_t\right)
=\int_{0}^{\infty}(1-e^{-w})\widetilde{\bbp}\left(\widetilde{U}_{t}\geq w\right)dw
-\int_{-\infty}^{0}(1-e^{-w})\widetilde{\bbp}\left(\widetilde{U}_{t}\leq w\right)dw \leq \tilde\kappa\t,
\end{align}
for all $0<t\leq t_0$, where the last inequality follows from (\ref{TIITI}-i), for the first integral, and (\ref{AsyDecomJ113}), where the second integral is dealt with. For $J_{22}(t)$, using (\ref{TIITI}-i),
\begin{align}\label{DocomFirIntI23}
0\leq J_{22}(t) & =t^{1/Y}\int_{0}^{\infty}\widetilde\bbp\left(t^{-1/Y}|\widetilde{U}_{t}|\geq u\right) du\nonumber\\
\nonumber
&\leq t^{1/Y}\left(1+\int_{1}^{\infty}\widetilde\bbp\left(t^{-1/Y}|\widetilde{U}_{t}|\geq u\right) du\right)\nonumber\\
&\leq t^{1/Y}\left(1+\int_{1}^{\infty}\tilde{\kappa}t(t^{1/Y}u)^{-Y}du\right) =o(t^{1-\frac{Y}{2}}),\quad {t\to 0},
\end{align}
since $1/Y>1-Y/2$ for $1<Y<2$. It follows that 
\begin{align*}
0 & \,\leq t^{\frac{Y}{2}-1}\bbe\left|I_2^1\left(t,Y_{\cdot}(\omega),Q_{\cdot}(\omega),W^1_{t}(\omega)\right)\right| \nonumber\\
& \leq  t^{\frac{Y}{2}-1}\bbe\left(e^{-\tilde\eta_t(\omega)}e^{\breve\V_t^1(\omega)}\int_{0}^{\infty}\left|\widetilde{\bbe}\left(\left(e^{-\widetilde{U}_{t}}-1\right){\bf 1}_{\{\breve\Z_{t}\leq-t^{\frac{1}{2}}u\}}\right)\right|\frac{1-e^{-\sqrt{t}u}}{\sqrt{t}}\frac{e^{-\frac{\left(u-t^{-1/2}\left( \breve\V_t^1(\omega)-\psi^1_t(\omega)\right)\right)^2}{2(1-\rho^2){\bar\sigma_t^2(Y(\omega))}}}}{\sqrt{2\pi(1-\rho^2){\bar\sigma_t(Y(\omega))}}}du\right)\nonumber\\
&\,\leq  t^{\frac{Y}{2}-1}\left(J_{21}(t)+J_{22}(t)\right)\bbe\left(e^{-\tilde\eta_t}\int_{0}^{\infty}u\frac{e^{-\frac{\left(u-t^{-1/2}\left( \breve\V_t^1(\omega)-\psi^1_t(\omega)\right)\right)^2}{2(1-\rho^2){\bar\sigma_t^2(Y(\omega))}}}}{\sqrt{2\pi(1-\rho^2){\bar\sigma_t^2(Y(\omega))}}}du\right)\,\stackrel{t\to{}0}{\longrightarrow}\, 0,
\end{align*}
by the fact that $0<m\leq\bar\sigma\left(Y(\omega)\right)\leq M<\infty$.
\hfill\qed

\medskip
{\noindent\textbf{Proof of (\ref{CnvDistV}).}}

\noindent Under the probability measure $\bbp^*$ defined in (\ref{DSM2b}), $V_t$ has the representation 
\begin{align*} 
V_t=\rho\int_0^t\sigma(Y_s)dW_s^1+\sqrt{1-\rho^2}\int_0^t\sigma(Y_s)dW_s^2+\half\int_0^t\sigma^2(Y_s)ds =: \xi_t+\half\int_0^t\sigma^2(Y_s)ds,
\quad V_0=0.
\end{align*}
We will show $t^{-\half}\xi_t\ld\Lambda\sim \mathcal{N}(0,\sigma_0^2)$, as $t\to 0$, which implies $t^{-\half}V_t\ld\Lambda$ by (\ref{SigmaMm}) and Slutsky's theorem. For $\theta\in\bbc$, define
$M^{\theta}_t=e^{\theta \xi_t-\frac{\theta^2}{2}\int_0^t\sigma^2(Y_s)ds}$, which is a martingale since $\sigma(\cdot)$ is bounded. Let $\theta=iu/\sqrt{t}$ so
\begin{align*}
\bbe\left(e^{\frac{iu}{\sqrt{t}}\xi_t+\frac{u^2}{2t}\int_0^t\sigma^2(Y_s)ds}\right)=
\bbe\left(e^{\frac{iu}{\sqrt{t}}\xi_t}\left(e^{\frac{u^2}{2t}\int_0^t\sigma^2(Y_s)ds}-
e^{\frac{u^2}{2}\sigma_0^2}\right)
+e^{\frac{iu}{\sqrt{t}}\xi_t+\frac{u^2}{2}\sigma_0^2}\right)=1.
\end{align*}
Taking the limit as $t\to 0$ on both sides now gives
\begin{align*}
\lim_{t\to 0}\bbe\left(e^{\frac{iu}{\sqrt{t}}\xi_t}\right)=e^{-\frac{u^2}{2}\sigma_0^2},
\end{align*}
by noticing that
\begin{align*}
\bbe\left(e^{\frac{iu}{\sqrt{t}}\xi_t}\left(e^{\frac{u^2}{2t}\int_0^t\sigma^2(Y_s)ds}-e^{\frac{u^2}{2}\sigma_0^2}\right)\right)
&\leq \left(\bbe\left(e^{\frac{2iu}{\sqrt{t}}\xi_t}\right)\bbe\left(e^{\frac{u^2}{2t}\int_0^t\sigma^2(Y_s)ds}-e^{\frac{u^2}{2}\sigma_0^2}\right)^2\right)^{\half}\\
&\leq\left(\bbe\left(e^{\frac{u^2}{2t}\int_0^t\sigma^2(Y_s)ds}-e^{\frac{u^2}{2}\sigma_0^2}\right)^2\right)^{\half}\to 0,\quad t\to 0,
\end{align*}
where the last step follows from (\ref{BndBarSigma}), the dominated convergence theorem, and the fact that $\sigma(\cdot)$ is bounded in a neighborhood of $y_0$.
\hfill\qed

\medskip
{\noindent\textbf{Proof of (\ref{intbound}).}}

\noindent Since $W_t^1/\sqrt{t}\sim \mathcal{N}(0,1)$,
\begin{align*}
&\bbe\left|\int_0^{\infty}z\int_{\frac{z-{\rho\sigma_{0}W_t^1}/{\sqrt{t}}}{\sqrt{1-\rho^2}\sigma_{0}}}^{z/\sigma_{0}}{\phi(x)dx}dz\right|  
= \left(\int_{u=-\infty}^{0}+\int_{u=0}^{\infty}\right)\phi(u)\left|\int_{z=0}^{\infty}z\int_{x=\frac{z-{\rho\sigma_{0}u}}{\sqrt{1-\rho^2}\sigma_{0}}}^{z/\sigma_{0}}{\phi(x)dx}dz\right|du=:I_1+I_2.
\end{align*}
For $u<0$, we have:
\begin{align*} 
I_1 &\leq \int_{u=-\infty}^{0}\phi(u)\int_{z=0}^{\infty}z\left|\frac{z}{\sigma_{0}}-\frac{z-{\rho\sigma_{0}u}}{\sqrt{1-\rho^2}\sigma_{0}}\right|\phi\left(\frac{z}{\sigma_0}\right)dz\\
&\leq \int_{u=-\infty}^{0}\phi(u)\int_{z=0}^{\infty}\left(K_1z^2+K_2z|u|\right)\phi\left(\frac{z}{\sigma_0}\right)dz<\infty,
\end{align*}
where the $K$'s are positive constants. For $u>0$,
\begin{align*}
I_2 &\leq \int_{u=0}^{\infty}\phi(u)\left(\int_{z=\rho\sigma_0u}^{\infty}z\left|\frac{z}{\sigma_{0}}-\frac{z-{\rho\sigma_{0}u}}{\sqrt{1-\rho^2}\sigma_{0}}\right|\phi\left(\frac{z-{\rho\sigma_{0}u}}{\sqrt{1-\rho^2}\sigma_{0}}\right)dz
+\int_{z=0}^{\rho\sigma_0u}z\left|\frac{z}{\sigma_{0}}-\frac{z-{\rho\sigma_{0}u}}{\sqrt{1-\rho^2}\sigma_{0}}\right|\phi(0)dz\right)du\\
&\leq \int_{u=0}^{\infty}\phi(u)\left(\int_{x=0}^{\infty}(x+\rho\sigma_0u)\left|\frac{x+\rho\sigma_0u}{\sigma_{0}}-\frac{x}{\sqrt{1-\rho^2}\sigma_{0}}\right|\phi\left(\frac{x}{\sqrt{1-\rho^2}\sigma_{0}}\right)dx
+\int_{z=0}^{\rho\sigma_0u}z\left|\frac{z}{\sigma_{0}}-\frac{z-{\rho\sigma_{0}u}}{\sqrt{1-\rho^2}\sigma_{0}}\right|\phi(0)dz\right)du\\
&\leq \int_{u=0}^{\infty}\phi(u)\left(\int_{x=0}^{\infty}\left(K_1x^2+K_2u^2+K_3xu\right)\phi\left(\frac{x}{\sqrt{1-\rho^2}\sigma_{0}}\right)dx
+\left(K_4u^3+K_5u^4\right)\right)du<\infty.
\end{align*}
\hfill\qed

\section{Proofs of other technical lemmas}\label{ApTecLem}

{\noindent\textbf{Proof of Lemma \ref{FirstLem}.}}

\noindent 
{For (\ref{ChangeMeasureCond}), first note} that around the origin we have
\[
	\left(1-e^{-\varphi(x)/2}\right)^{2} \sim {\frac{1}{4}\varphi(x)^{2}}={\frac{1}{4}}\left(x+\ln\bar\q(x)\right)^2\leq {\frac{1}{2}\left(x^{2}+\left(\ln\bar{q}(x)\right)^{2}\right)}\sim {\frac{1}{2}\left(x^2 + (1-\bar\q(x))^2\right)},
\]
{and, thus, (\ref{NewAssumEq}-i) suffices} for the integral to be finite on any interval containing the origin. Away from the origin, the integral is finite in view of the martingale condition (\ref{NdCndTSMrt}-i) for $x>0$, and the fact that {$q(x)$} is bounded for $x<0$.
{For (\ref{IntCndEta}), note that, around the origin,} we have
\begin{align*}
	e^{-\varphi(x)}-1+\varphi(x)\sim {\frac{\varphi(x)^{2}}{2}}, 
\end{align*}
{and} the situation is the same as for the integral in (\ref{ChangeMeasureCond}). Outside the origin we have
\begin{align*}
	\int_{|x|> 1}\left|e^{-\varphi(x)}-1+\varphi(x)\right| |x|^{-Y-1}dx
	\leq C_{1} \int_{|x|> 1}\left|e^{-\varphi(x)}-1\right| |x|^{-Y-1}dx
	+ C_{2}\int_{|x|> 1}|\varphi(x)| |x|^{-Y-1}dx < \infty
\end{align*}
for some constants $C_{1},C_{2}<\infty$. The first integral above is finite by the {same arguments as those used in the} integral (\ref{ChangeMeasureCond}), and a sufficient condition for the second one to be finite is {(\ref{NewAssumEq}-ii)}.
\hfill\qed

\medskip
{\noindent\textbf{Proof of Lemma \ref{KLTCTBR}.}}

\noindent {\rm (1)} 
\noindent The proof is identical to the proof of the first part of Lemma $3.3$ in \cite{LopGonHou:2013}.
\smallskip\\
\noindent 
{\rm (2)}
Throughout, $\tilde{\kappa}>0$ denotes a generic {finite} constant that may vary from line to line. First, using the fact that $Z_t$ is strictly $Y$-stable, and therefore self-similar, we obtain
\begin{equation}
	\frac{1}{t}\widetilde{\bbp}\left({|Z_{t}|}\geq{}v\right)\leq \tilde{\kappa}v^{-Y},\label{TIIStabl}
\end{equation}
for any $0<t\leq{}1$ and  $v>0$ (see Equation $(2.18)$ in \cite{LopGonHou:2013}). It therefore suffices to show the analog inequality for $|\widetilde{U}_t|$. For that we use the following decomposition, with $\varepsilon=\alpha v$, with $\alpha>0$:
\begin{equation}\label{uDecomp}
	\bar{\widetilde{U}}_{t}^{(\varepsilon)}:=\int_{0}^{t}\int_{|\varphi(x)|\geq\varepsilon} \varphi(x)N(ds,dx),
	\qquad \widetilde{U}_{t}^{(\varepsilon)}:=\widetilde{U}_{t}-\bar{\widetilde{U}}_{t}^{(\varepsilon)}.
\end{equation}
Since $\bar{q}(x)\to{}1$ as $x\to{}0$, $\{x:|\varphi(x)|\geq{}\varepsilon\}\subset\{x:|x|\geq{}\delta\}$ for some $\delta>0$ small enough. Thus, {{$\bar{\widetilde{U}}_{t}^{(\varepsilon)}$}} is a compound Poisson process.
Let $\bar{C}:={C(1)+C(-1)}$, and note that $|\varphi(x)|=|-x-\ln\bar q(x)|\leq|x|\left(1+\left|\frac{\ln\bar q(x)}{x}\right|\right)\leq |x|(1+K)$, where $K<\infty$ because of (\ref{NewAssumEq}). Hence, denoting $N_{t}^{\varepsilon}:=N\left(\{(s,x): s\leq{}t, |\varphi(x)|\geq{}\varepsilon\}\right)$ and $\lambda_{\varepsilon}:=\tilde{\nu}\left(\{x:|\varphi(x)|\geq{}\varepsilon\}\right)$, 
\begin{align*}
\frac{1}{t}\widetilde{\bbp}\left(\Big|\bar{\widetilde{U}}_{t}^{(\varepsilon)}\Big|\geq{}v/2\right)&\leq\frac{1}{t}\widetilde{\bbp}\left(N_{t}^{(\varepsilon)}\neq 0\right)=\frac{1}{t}\left(1-e^{-\lambda_{\varepsilon}t}\right)\leq\lambda_{\varepsilon}\leq \bar{C} \int_{|\varphi(x)|\geq{}\alpha v}|x|^{-Y-1}dx\leq \bar{C} \int_{|x|\geq{}\frac{\alpha v}{1+K}}|x|^{-Y-1}dx\leq \tilde{\kappa}v^{-Y}.
\end{align*}
For ${\widetilde{U}}_{t}^{(\varepsilon)}$, we divide the positive real axis into three parts. First, by Lemma \ref{L2}, there exists $v_1<\infty$ such that 
$t^{-1}\widetilde{\bbp}\left(\Big|{\widetilde{U}}_{t}^{(\varepsilon)}\Big|\geq{}v/2\right)\leq \tilde{\kappa}v^{-Y}$ for $v>v_1$, some $\tilde\kappa<\infty$, and all $t<1$. Next, let $v_0>0$ and note that for $v_0\leq v\leq v_1$, by taking $\alpha$ small enough, 
\begin{align*}
\frac{1}{t}\widetilde{\bbp}\left(\Big|\widetilde{U}_{t}^{(\varepsilon)}\Big|\geq v/2\right)
\leq \frac{1}{t}\widetilde{\bbp}\left(\Big|\widetilde{U}_{t}^{(\varepsilon)}\Big|\geq v_0/2\right)
\leq v_1^{-Y}\leq v^{-Y},
\end{align*}
for $t$ small enough. By defining $\varphi(0):=0$, $\varphi(x)$ becomes continuous at the origin, so there exists $\delta>0$ such that $B(0,\delta)\subseteq\{\varphi(x):x\in\bbr\}$. Now consider $v<v_0:=\delta/\alpha$, using the decomposition (\ref{uDecomp}) with $\alpha=1/4$, and note that the smallest $r$ such that $(-r,r)$ supports the L\'evy measure of $\widetilde{U}_{t}^{(\varepsilon)}$ is $\varepsilon=v/4$. Then,
\begin{align*}
\left|\widetilde{\bbe}\, {\widetilde{U}}_{t}^{(\varepsilon)}\right|&\leq t\int_{|\varphi(x)|>v/4}|\varphi(x)|\tilde{\nu}(dx)
=t\bar{C}(1+K)\int_{|x|>\frac{v/4}{1+K}}|x| |x|^{-Y-1}dx=Ctv^{-Y+1},
\end{align*}
where $C$ is a positive constant. Thus, whenever $t$ and $v$ are such that $Ctv^{-Y+1}\leq v/4$,
\begin{align*}
\widetilde{\bbp}\left(\Big|\widetilde{U}_{t}^{(\varepsilon)}\Big|\geq v/2\right)
=\widetilde{\bbp}\left(-\widetilde{U}_{t}^{(\varepsilon)}\geq v/2\right)
+\widetilde{\bbp}\left(\widetilde{U}_{t}^{(\varepsilon)}\geq v/2\right)
\leq\widetilde{\bbp}\left(-\widetilde{U}_{t}^{(\varepsilon)}+\bbe\,\widetilde{U}_{t}^{(\varepsilon)}\geq v/4\right)
+\widetilde{\bbp}\left(\widetilde{U}_{t}^{(\varepsilon)}-\bbe\,\widetilde{U}_{t}^{(\varepsilon)}\geq v/4\right)
\end{align*}
Next, using a concentration inequality for centered random variables (see, e.g., \cite{Hou:2002}, Corollary 1),
\begin{align*}
\widetilde{\bbp}\left(\left|\widetilde{U}_{t}^{(\varepsilon)}\right|\geq v/2\right)\leq 2e^{\frac{v}{4\varepsilon}-\left(\frac{v}{4\varepsilon}+\frac{tV_{\varepsilon}^2}{\varepsilon^2}\right)
\log\left(1+\frac{\varepsilon v}{4tV_{\varepsilon}^2}\right)}\leq 2\left(\frac{4eV_{\varepsilon}^2}{\varepsilon v}\right)^{\frac{v}{4\varepsilon}}t^{\frac{v}{4\varepsilon}}\leq \frac{{32}eV_{v/4}^2}{v^{2}}t,
\end{align*}
where $V_{\varepsilon}^2:={\rm Var}(\widetilde{U}_{1}^{(\varepsilon)})$ and in the last inequality, $\varepsilon=v/4$. Now,
\begin{align*}
	V_{v/4}^2 = {\rm Var}\left(\widetilde{U}_{1}^{(v/4)}\right)
	&=\int_{|\varphi(x)|\leq{}v/4}\varphi^2(x)\tilde{\nu}(dx)\\
	&=\int_{\{|x|\leq{}v,\,|\varphi(x)|\leq{}v/4\}}\varphi^2(x)\tilde{\nu}(dx)
	+\int_{\{|x|\geq{}v,\,|\varphi(x)|\leq{}v/4\}}\varphi^2(x)\tilde{\nu}(dx)\\
	&\leq
	M^2\bar\C\int_{|x|\leq{}v}|x|^{-Y+1}dx +\bar\C\left(\frac{v}{4}\right)^2\int_{|x|\geq{}v}|x|^{-Y-1}dx
	\leq\tilde{\kappa} v^{2-Y},
\end{align*}
for some $\tilde{\kappa}<\infty$, where above, we set $\M:=\sup_{|x|>0}|\varphi(x)|/|x|$, which is finite in {view} of (\ref{NewAssumEq}). Therefore, whenever $Ctv^{-Y+1}\leq v/4$ (or equivalently, $4Ctv^{-Y}\leq 1$),
\begin{align*}
\frac{1}{t}\widetilde{\bbp}\left(\Big|\widetilde{U}_{t}^{(\varepsilon)}\Big|\geq v/2\right)\leq\frac{eV_{v/4}^2}{v^{2}}\leq\tilde{\kappa}v^{-Y}.
\end{align*}
Moreover, for any $t>0$ and $v>0$,
\begin{align*}
\frac{1}{t}\widetilde{\bbp}\left(\Big|\widetilde{U}_{t}^{(\varepsilon)}\Big|\geq v/2\right)
=\frac{1}{t}\widetilde{\bbp}\left(\Big|\widetilde{U}_{t}^{(\varepsilon)}\Big|\geq v/2\right){\bf 1}_{\{4Ctv^{-Y}\leq 1\}}+
\frac{1}{t}\widetilde{\bbp}\left(\Big|\widetilde{U}_{t}^{(\varepsilon)}\Big|\geq v/2\right){\bf 1}_{\{4Ctv^{-Y}>1\}}
\leq\tilde{\kappa}v^{-Y}+4Cv^{-Y}\leq \tilde{\kappa}v^{-Y},
\end{align*}
Combining the previous estimates, we finally have
\begin{equation} \label{UsflTlEstTldU}
\frac{1}{t}\widetilde{\bbp}\left(\Big|\widetilde{U}_{t}\Big|\geq v\right)\leq\frac{1}{t}\widetilde{\bbp}\left(\Big|\bar{\widetilde{U}}_{t}^{(\varepsilon)}\Big|\geq{}v/2\right)
+\frac{1}{t}\widetilde{\bbp}\left(\Big|\widetilde{U}_{t}^{(\varepsilon)}\Big|\geq v/2\right)\leq \tilde{\kappa}v^{-Y}.
\end{equation}
for all $v>0$ and $t>0$ and some constant $\tilde{\kappa}<\infty$.
\hfill\qed

\medskip
{\noindent\textbf{Proof of Lemma \ref{L2}.}}

\noindent
Using a concentration inequality for centered random variables (see, e.g., \cite{Hou:2002}, Corollary 1) gives
\begin{equation*}
	\tilde\bbp\left(|\xi_{t}|\geq v\right) \leq 2e^{\frac{v}{R}-\left(\frac{v}{R}+\frac{tV^2_{R}}{R^2}\right)\log\left(1+\frac{R v}{tV^2_{R}}\right)} 
	 = 2e^{\frac{v}{R}\left(1-\half\log\left(1+\frac{R v}{tV^2_{R}}\right)\right)}e^{-\left(\half\frac{v}{R}+\frac{tV^2_{R}}{R^2}\right)\log\left(1+\frac{R v}{tV^2_{R}}\right)},
\end{equation*}
where $V_R^2=Var(\xi_1)$, and
\begin{align*}
	e^{-\left(\half\frac{v}{R}+\frac{tV^2_{R}}{R^2}\right)\log\left(1+\frac{R v}{tV^2_{R}}\right)} 
	\leq e^{-\half\frac{v}{R}\log\left(1+\frac{R v}{tV^2_{R}}\right)}
	\leq t^{\half\frac{v}{R}}\left(\frac{R v}{V_{R}^2}\right)^{-\half\frac{v}{R}}.
\end{align*}
The relation then follows by choosing $v_0$ big enough for $\half\frac{v_0}{R}\geq 1$ {and $\frac{1}{2R}\log\left(\frac{Rv_{0}}{V_{R}^{2}}\right)>k$}
to hold.
\hfill\qed

\medskip
{\noindent\textbf{Proof of Lemma \ref{HittingTime}.}}

\noindent
Following the lines of \cite{Abu:2008}, define $Z_t=u(Y_t)$ where
\begin{align*}
u(y)=\int_{{y_0}}^ye^{-\int_{Y_0}^x\frac{2\alpha(z)}{\gamma^2(z)}dz}dx.
\end{align*}
$(Z_t)_{t\geq 0}$ is a local martingale with quadratic variaton
\begin{align*}
\langle\Z\rangle_t  =\int_0^t\left(u'(Y_s)\gamma(Y_s)\right)^2ds
 = \int_0^te^{-4\int_{Y_0}^{Y_s}\frac{\alpha(z)}{\gamma^2(z)}dz}\gamma^2(Y_s)ds,
\end{align*}
so there exists a Brownian motion $(\B_t)_{t\geq 0}$ starting at $Z_0$, such that $Z_t=\B_{{\langle\Z\rangle}_t}$. Also define the stopped processes $\bar\Y_t=Y_{t\wedge {\tau}}$ and $\bar\Z_t=Z_{t\wedge {\tau}}$. Then,
\begin{align}\label{final}
\bbp\left({\tau}\leq t\right)
& = \bbp\left(\inf_{s\leq t}u(Y_s)\leq u(a) \text{ or } \sup_{s\leq t}u(Y_s)\geq u(b)\right)\nonumber\\
& = \bbp\left(\inf_{s\leq t}Z_s\leq u(a) \text{ or } \sup_{s\leq t}Z_s\geq u(b)\right)\nonumber\\
& = \bbp\left(\inf_{s\leq t}\bar\Z_s\leq u(a) \text{ or } \sup_{s\leq t}\bar\Z_s\geq u(b)\right)\nonumber\\
& = \bbp\left(\inf_{s\leq {\langle\bar\Z\rangle}_t}\B_{s}\leq u(a)\text{ or } \sup_{s\leq {\langle\bar\Z\rangle}_t}\B_{s}\geq u(b)\right)\nonumber\\
& \leq 2\bbp\left(\sup_{s\leq \beta(t)}\B^0_{s}\geq M\right) \leq 4\Psi\left(\frac{M}{\sqrt{\beta(t)}}\right),
\end{align}
where $(\B^0_t)_{t\geq 0}$ is a standard Brownian motion, $M:=\min\left(u(b)-Z_0,Z_0-u(a)\right)$, $\beta(t)$ a deterministic function such that ${\langle\bar\Z\rangle}_t\leq\beta(t)$, and the last line follows from the fact that
\begin{align*}
\bbp\left(S_t>z\right) = 2\Psi\left(\frac{z}{\sqrt{t}}\right),\;z>0,
\end{align*}
where $S_t$ is the running supremum of a standard Brownian motion, and
\begin{align*}
\Psi(z) := \int_{{z}}^{\infty}\frac{1}{\sqrt{2\pi}}e^{-\frac{u^2}{2}}du.
\end{align*}
To show the existence of $\beta(t)$, note that
\begin{align*}
\langle\bar\Z\rangle_t 
= \int_0^te^{-4\int_{Y_0}^{\bar\Y_t}\frac{\alpha(z)}{\gamma^2(z)}dz}\gamma^2(\bar\Y_s)ds,
\end{align*}
First, assume $\gamma(y_0)\neq 0$, and use the continuity of $\gamma(\cdot)$ at $y_0$ to find $\bar\a,\bar\b$, and {$\epsilon>0$}, such that ${y_0}\in(\bar\a,\bar\b)$ and {$|\gamma(y)|>\epsilon$} for $y\in(\bar\a,\bar\b)$. Then, using the fact that $\alpha(\cdot)$ and $\gamma(\cdot)$ are locally bounded,
\begin{align*}
\langle\bar\Z\rangle_t 
\leq K_1\int_0^te^{K_2\int_{\bar\a}^{\bar\b}\frac{1}{\gamma^2(z)}dz}ds < K_3t  =: \beta(t),
\end{align*}
where $K_1,K_2,$ and $K_3$ are positive constants. As a result, (\ref{final}), and therefore also $\bbp\left({\tau}\leq t\right)$, is of order $O(t^k)$ for any $k\in\bbn$. If $\gamma(y_0)=0$, let $y'\in(a,b)$ be such that $\gamma(y')\neq 0$ (if no such $y'$ exists, $Y_t{\bf 1}_{\{a<Y_t<b\}}$ is deterministic and $\bbp({\tau}\leq t)=0$ for $t$ small enough), and define ${\tau'}:=\inf\{t\geq 0:Y_t=y'\}$. Then
\begin{align*}
\bbp\left({\tau}\leq t\right) & =\bbp\left({\tau}\leq t,{\tau'}\leq {\tau}\right)+\bbp\left({\tau}\leq t,{\tau'}> {\tau}\right)\\
& \leq \bbp\left({\tau}\leq t|{\tau'}\leq {\tau}\right) + \bbp\left({\tau}\leq t,{\tau'}> {\tau}\right)\\
& \leq \bbp\left(\bar {\tau}\leq t\right) + \bbp\left({\tau}\leq t,{\tau'}> {\tau}\right)\\
& = O(t^k), \quad {t\to 0},
\end{align*}
where $\bar \tau:=\inf\{t\geq 0:Y_t\notin (a,b)|Y_0=y'\}$, so the order of the first term follows from the case when $\gamma(y_0)\neq 0$ case, and the order of the second term follows from the fact that $(Y_s)_{s\leq t}$ is deterministic on $\{{\tau'}> t\}$, so $\bbp\left({\tau}\leq t,{\tau'}>{\tau}\right)=0$ for $t$ small enough.
\hfill\qed

\medskip
\noindent
{\textbf{Acknowledgments:} 
The authors gratefully acknowledge the constructive and helpful comments provided by two anonymous referees, which significantly contributed to improve the quality of the manuscript.}

\bibliographystyle{plain}

\end{document}